\documentclass[12pt]{article}
\usepackage{epsf}
\usepackage{color}
\usepackage{graphicx}
\usepackage{amsmath}
\usepackage{amssymb}
\usepackage{latexsym}

\begin{document}

\newcommand{\mathcalpl}{M_{\mathcalathrm{Pl}}}
\setlength{\baselineskip}{18pt}
\begin{titlepage}
\begin{flushright}
OU-HET 628/2009 \\
YITP-09-31
\end{flushright}

\vspace*{1.2cm}
\begin{center}
{\Large\bf Top Yukawa Deviation in Extra Dimension}
\end{center}
\lineskip .75em
\vskip 1.5cm

\begin{center}
{\large 
Naoyuki Haba$^{a,}$\footnote[1]{E-mail:
\tt haba@phys.sci.osaka-u.ac.jp}, 
Kin-ya Oda$^{a,}$\footnote[2]{E-mail:
\tt odakin@phys.sci.osaka-u.ac.jp}, 
and 
Ryo Takahashi$^{a,b,}$\footnote[3]{E-mail: 
\tt rtakahas@yukawa.kyoto-u.ac.jp}
}\\

\vspace{1cm}

$^a${\it Department of Physics, Graduate School of Science, Osaka
University, \\
Toyonaka, Osaka 560-0043, Japan}\\
$^b${\it Yukawa Institute for Theoretical Physics, Kyoto University,\\
 Kyoto 606-8502, Japan}\\

\vspace*{10mm}
{\bf Abstract}\\[5mm]
{\parbox{13cm}{\hspace{5mm}
%%%%%%%%%%%%%%%%%%%%%%%%%%%%%%%%%%%%%%%%%%%%%%%%%%%%%%%%%%%%%%%%
%             ABSTRACT                                         %
%%%%%%%%%%%%%%%%%%%%%%%%%%%%%%%%%%%%%%%%%%%%%%%%%%%%%%%%%%%%%%%%
We suggest a simple one-Higgs-doublet model living in the bulk of five-dimensional spacetime compactified on $S^1/Z_2$, in which the top Yukawa coupling can be smaller than the naive standard-model expectation, i.e.\ the top quark mass divided by the Higgs vacuum expectation value. 
If we find only single Higgs particle at the LHC and also observe the top Yukawa deviation, our scenario becomes a realistic candidate beyond the standard model.
The Yukawa deviation comes from the fact that the wave function profile of the free physical Higgs field can become different from that of the vacuum expectation value, due to the presence of the brane-localized Higgs potentials.
In the Brane-Localized Fermion scenario, we find sizable top Yukawa deviation, which could be checked at the LHC experiment, with a dominant Higgs production channel being the $WW$ fusion.
We also study the Bulk Fermion scenario with brane-localized Higgs potential, which resembles the Universal Extra Dimension model with a stable dark matter candidate.
We show that both scenarios are consistent with the current electroweak precision measurements.

}}
\end{center}
\end{titlepage}

%%%%%%%%%%%%%%%%%%%%%%
\section{Introduction} 
%%%%%%%%%%%%%%%%%%%%%% 

The Large Hadron Collider (LHC) experiment is just being started. One of the main missions of the LHC is the discovery of the Higgs particle. In the Standard Model (SM), the Higgs field is the origin of masses of fermions and plays a crucial role in the electroweak (EW) symmetry breaking. The SM will be completed as a renormalizable theory when the Higgs particle is discovered. However, a potential theoretical problem still exists which is so-called the gauge hierarchy problem about the big energy desert between the EW and Planck scales. The extra dimension scenario is one of fascinating approaches toward solving the problem (large extra dimensions~\cite{ArkaniHamed:1998rs} and warped extra dimension~\cite{Randall:1999ee}). 
Generally, extra dimensional theories lead to rich phenomenologies, for example, new heavy particles with compactification-scale masses, called the Kaluza-Klein (KK) particles. In the Universal Extra Dimensions (UED) model~\cite{ued}, the lightest KK particle (LKP) with an odd parity is stable, and can be a dark matter. Although the compactification scale is generically constrained to be larger than a few TeV by the EW precision measurements for Brane-Localized Fermion (BLF) scenario~\cite{kk1}--\cite{kk5}, the UED can weaken it to few hundred GeV due to the five-dimensional Lorentz symmetry, that is the KK number conservation~\cite{ued,App,Gogo}.
The TeV scale KK resonances might be discovered at the LHC as an existence of proof of the extra-dimension.

%The large hadron collider (LHC) experiment just has been started. It is well 
%known that a main mission of the LHC is discovery of the Higgs particle. The 
%Higgs is thought as origin of masses of the standard model (SM) particles, and 
%the Higgs sector plays a crucial role of the electroweak (EW) symmetry 
%breaking. The SM will be completed as low energy effective theory by 
%discovering the Higgs particle. However, a serious theoretical problem still 
%exists, which is so-called gauge hierarchy problem as a big energy desert 
%between the EW and Planck scales. An extra dimensions is one of fascinating 
%approaches toward solving the problem (large extra dimension 
%\cite{ArkaniHamed:1998rs} and warped extra dimension 
%\cite{Randall:1999ee,Randall:1999vf}.) Generally, extra dimensional theories 
%lead to rich phenomenologies. For examples, new heavy particles with 
%compactification masses, called Kaluza-Klein (KK) particles. In a universal 
%extra dimensions (UED) model~\cite{ued}, the lightest KK particle (LKP) with an
% odd parity is stable, and can be a dark matter. Although the compactification 
%scale is usually constrained to a few TeV by EW precision measurements, UED 
%constraint can be weaken it to ~500GeV due to the 5-dimensional Lorentz 
%symmetry. TeV scale KK resonances might be discovered at the LHC as an 
%existence of proof of extradimension.

In this paper, we suggest a simple one-Higgs-doublet model in which the Yukawa coupling between top and physical Higgs fields deviates from the naive SM expectation, that is the top quark mass divided by the Higgs vacuum expectation value. Our setup is a five-dimensional flat spacetime, with one SM Higgs doublet field existing in the bulk. The wave function profile of the free physical Higgs field becomes different from its Vacuum Expectation Value (VEV) when brane potentials are introduced.\footnote{In this paper, we concentrate on the case of the vanishing bulk potential and will present the more general treatment in a separate publication~\cite{HOT}.}
We consider two scenarios depending on the location of matter fermions. One scenario is the BLF scenario and the other is Bulk Fermion (BF) scenario.

In the BLF scenario, we find that the coupling between top and physical Higgs fields becomes small because of the brane potentials with large couplings and that it could be checked at the LHC experiment. In this case dominant Higgs production channel at the LHC becomes $WW$ fusion. 
%Since the wave-function profile of the free Higgs field vanishes at the boundaries, the four-dimensional effective Higgs self-couplings do not become huge even in the large coupling limit, therefore the Higgs field can be regarded as a particle (as long as the compactification scale is not higher than a few TeV). 
Even in the limit of large brane-localized Higgs coupling, the four-dimensional effective Higgs self-couplings remain finite, since the wave function of the Higgs field vanishes at the boundary in the limit. 
Therefore Higgs can be regarded as a particle (as long as a compactification scale is not higher than a few TeV). 
Reminding that the top Yukawa deviation can occur also in multi-Higgs models, if we find only one Higgs particle as well as top Yukawa deviation at LHC, our scenario become a realistic candidate beyond the SM.~\footnote{The warped gauge-Higgs unification model is another candidate~\cite{hk}.} (The top Yukawa deviation can occur in multi-Higgs models such as supersymmetric SM.) Note that our setup is similar to the UED, but the latter has no top Yukawa deviation. 

In the BF scenario, the top Yukawa deviation is not significant even with the large boundary potentials. We show that a stable dark matter exists, like in the universal extra dimension model, due to the existence of an accidental reflection symmetry, the Kaluza-Klein parity.
%In the BF scenario, the top Yukawa deviation is not significant even in the large boundary coupling limit, while a stable dark matter exists due to the existence of an accidental reflection symmetry, the KK parity. 
We also show that these two scenarios are consistent with the present electroweak precision measurements.

The paper is organized as follows: in Section 2, we present an extra-dimensional setup with brane localized Higgs potentials, and discuss general phenomenological aspects. In Sections 3 and 4, we investigate more detailed phenomenological implications of the setup with the BLF and BF scenarios, respectively. We comment on the dark matter in Section 5. The last section is devoted to the summary and discussions.
In Appendix A, we comment on the radion stabilization. In Appendix B we spell out the free part of the action of the Higgs sector.
In Appendix C, we present gauge interactions of the Higgs fields.
In Appendix D, we list the Feynman rules in our notation.

%The paper is organized as follows: in the section 2, we show an 
%extradimensional setup with brane localized Higgs potentials, and discuss 
%general phenomenological aspects. In the section 3 and 4, we investigate more 
%detailed phenomenological implications of the setup to the BLF and BF 
%scenarios, respectively. We comment on the dark matter in the section 5. The 
%section 6 is devoted to the summary and discussions.

%%%%%%%%%%%%%%%%%%%%%%%%%%%%%%%%%%%%%
\section{Bulk Higgs and its profiles}
%%%%%%%%%%%%%%%%%%%%%%%%%%%%%%%%%%%%%

In our setup, the SM Higgs doublet exists in the 5-dimensional flat space-time. 
The Higgs has brane-localized potentials, and we investigate their effects on 
the Higgs wave-function profiles, masses and couplings in this section. Though we will concentrate on the case where EW symmetry is broken through the brane-localized Higgs potentials, our analysis can be applicable in more general cases.

%%%%%%%%%%%%%%%%%%%%%%%%
\subsection{VEV profile}
%%%%%%%%%%%%%%%%%%%%%%%%
In this subsection, we show the wave function profiles of classical mode of a
bulk scalar field in a flat extradimensional setup with brane-localized Higgs 
potentials.

Let us start with the following action of a bulk field, $\Phi$, including brane-localized potentials. We take flat five dimensional spacetime compactified on $S^1/Z_2$,
\begin{eqnarray}
 S&=&\int d^4x\int_0^Ldy[-|\partial_M\Phi|^2
     -\mathcal{V}-\delta(y-L)V_L-\delta(y)V_0],\label{p1}
% \\
%   &=&\int d^4x\int_0^Ldy[\Phi^\dagger\mathcal{P}\Phi
%      -\mathcal{V}+\delta(y-L)[-\Phi^\dagger\partial_y\Phi-V_L]
%      +\delta(y)[\Phi^\dagger\partial_y\Phi-V_0]],
\end{eqnarray}
where we write five dimensional coordinates as $X^M=(x^\mu,y)$, when 
$M,N,\ldots=0\sim4$ and $\mu,\nu,\ldots=0\sim3$. We define $y\equiv X^4$. The 
brane localized potentials $V_0$ and $V_L$ are introduced at $y=0$ and $L$ 
branes, respectively.\footnote{Similar brane localized action has been 
discussed in \cite{Flacke:2008ne}, where brane localized kinetic terms and 
their effect on the EW symmetry breaking via the five dimensional Higgs 
mechanism in the UED model have been considered.} For simplicity, we assume that the potentials solely depend on $|\Phi|^2$ so that potentials can be written as $V(|\Phi|^2)$. When we write
\begin{eqnarray}
 \Phi=\frac{\Phi_R+i\Phi_I}{\sqrt{2}},
\end{eqnarray}
we obtain
\begin{align}
	{\partial V\over\partial\Phi_X}
		&=	\Phi_XV', &
	{\partial^2V\over{\partial\Phi_X^2}}
		&=	V'+\Phi_X^2V'', &
	{\partial^2V\over\partial\Phi_R\partial\Phi_I}
		&=	\Phi_R\Phi_IV'',
			\label{pot_deri}
\end{align}
where $X$ stands for $R$ and $I$, while $V$ stands for $\mathcal{V}$, $V_L$, and 
$V_0$, and we have written $V'=dV/d\,|\Phi|^2$ etc. The variation of the action
is given by
\begin{eqnarray}
 \delta S&=&\int d^4x\int_0^Ldy\bigg[\delta\Phi_X\left(\mathcal{P}\Phi_X
	     -{\partial \mathcal{V}\over\partial\Phi_X}\right)
            +\delta(y-L)\delta\Phi_X\left(-\partial_y\Phi_X
            -{\partial V_L\over\partial\Phi_X}\right)\nonumber\\
         & &\hspace{7.4cm}+\delta(y)\delta\Phi_X\left(+\partial_y\Phi_X
            -{\partial V_0\over\partial\Phi_X}\right)\bigg],
  \label{S_variation_for_complex_scalar}
\end{eqnarray}
where 
$\mathcal{P}\equiv\Box+\partial_y^2\equiv-\partial_0^2+\nabla^2+\partial_y^2$. 
The VEV of the scalar field, $\Phi^c$, is determined
by action principle, $\delta S=0$, that is,
\begin{eqnarray}
 &&\mathcal{P}\Phi_X^c-\frac{\partial\mathcal{V}}{\partial\Phi_X}^c=0,
   \label{l6}\\
 &&\left(\mp\partial_y\Phi_X^c
   -\left.\frac{\partial V_\eta}{\partial\Phi_X}^c\right)\right|_{y=\eta}=0,
   \label{l7}
\end{eqnarray}
where 
\begin{align}
 {\partial V\over\partial\Phi_X}^c(x,y)
 &\equiv\left.{\partial V\over\partial\Phi_X}\right|_{\Phi=\Phi^c(x,y)}.
 \label{shorthand_notation}
\end{align}
Here signs above and below in Eq.~(\ref{l7}) are for $\eta=L$ and $0$, 
respectively. The Eq.~(\ref{l6}) is the bulk equation of motion for the scalar 
field. Equation~(\ref{l7}) corresponds to the Boundary Condition (BC) on each 
brane. Solving the equation of motion with the BCs determines the profile of 
VEV in the bulk. For later convenience, we rewrite 
Eq.~(\ref{S_variation_for_complex_scalar})--(\ref{l7}) in a $z$ coordinate 
defined as $z\equiv y-L/2$, 
\begin{eqnarray}
 \delta S&=&\int d^4x\int_{-L/2}^{+L/2}dz
            \bigg[\delta\Phi_X\left(\mathcal{P}\Phi_X
	     -{\partial \mathcal{V}\over\partial\Phi_X}\right)
            +\delta(z-L/2)\delta\Phi_X\left(-\partial_z\Phi_X
            -{\partial V_{+L/2}\over\partial\Phi_X}\right)\nonumber\\
         & &\hspace{6.4cm}+\delta(z+L/2)\delta\Phi_X\left(+\partial_z\Phi_X
            -{\partial V_{-L/2}\over\partial\Phi_X}\right)\bigg],
\end{eqnarray}
and 
\begin{eqnarray}
 &&\mathcal{P}_z\Phi_X^c-\frac{\partial\mathcal{V}}{\partial\Phi_X}^c=0,
   \label{l6z}\\
 &&\left(\mp\partial_z\Phi_X^c
   -\left.\frac{\partial V_\xi}{\partial\Phi_X}^c
   \right)\right|_{z=\xi}=0,\label{l7z}
\end{eqnarray}
where $\mathcal{P}_z\equiv\Box+\partial_z^2$, and signs above and below in 
Eq.~(\ref{l7z}) are for $z=+L/2$ and $-L/2$, respectively.
If we take the bulk and boundary potential as
\begin{eqnarray}
 \mathcal{V}&=&m^2|\Phi|^2,\label{bup}\\
 V_\xi&=&\frac{\lambda_\xi}{4}(|\Phi|^2-v_\xi^2)^2,\label{bp}
\end{eqnarray}
respectively, the general solution of (\ref{l6z}) is written as 
\begin{eqnarray}
 \Phi^c(z)\equiv v(z)=A\cosh(mz)+B\sinh(mz).
\end{eqnarray}
When we take the same boundary potentials for both the $z=+L/2$ and $-L/2$ branes, 
the BCs (\ref{l7z}) determine $A$ and $B$ as\footnote{There are other 
solutions. Complete discussions for solving the equation of motion and a radion
stabilization are given in Appendix A.}
\begin{eqnarray}
 A &=& \pm\frac{\sqrt{\lambda v^2c_h-ms_h}}{c_h^{3/2}\sqrt{\lambda}}, 
       \label{Aeven}\\
 B &=& 0, \label{Beven}
\end{eqnarray}
or
\begin{eqnarray}
 A &=& 0,                                                        \label{Aodd}
 \\
 B &=& \pm\sqrt{\frac{\lambda v^2s_h-mc_h}{\lambda s_h(c_h-2)}}, \label{Bodd}
\end{eqnarray}
where we define 
\begin{eqnarray}
 \lambda &\equiv& \lambda_{+L/2}=\lambda_{-L/2},  \label{lambdadef}\\
 v       &\equiv& v_{+L/2}=v_{-L/2},              \label{vv}\\
 s_h     &\equiv& \sinh\left(\frac{mL}{2}\right), \label{ssh}\\
 c_h     &\equiv& \cosh\left(\frac{mL}{2}\right). \label{cch}
\end{eqnarray}
It is easy to see that solutions (\ref{Aeven}) and (\ref{Beven}) make the VEV 
profiles even under the reflection between $z=\pm L$, that is, 
$\Phi^c(L/2)=\Phi^c(-L/2)=Ac_h$. On the other hand, (\ref{Aodd}) and 
(\ref{Bodd}) correspond to odd solution, $\Phi^c(L/2)=-\Phi^c(-L/2)=Bs_h$. This
reflection symmetry is accidental one induced from the assumption of identical potentials at $z=\pm L/2$ branes. The parity under this reflection plays important roles in phenomenological discussions and is equivalent to the KK parity in the UED models. Detailed considerations will be given in the following sections.

\begin{figure}[t]
\hspace{4cm}(a)\hspace{7.7cm}(b)
\begin{center}
\includegraphics[scale = 0.9]{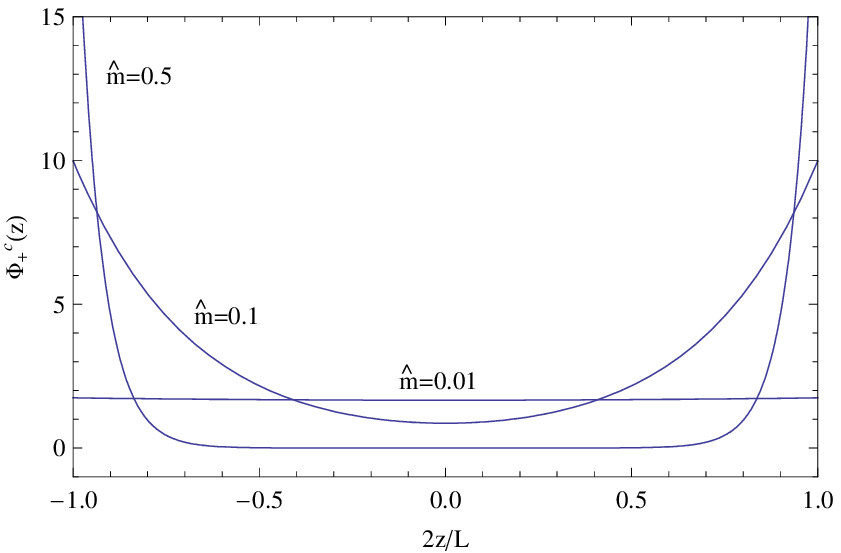}\hspace{4mm}
\includegraphics[scale = 0.91]{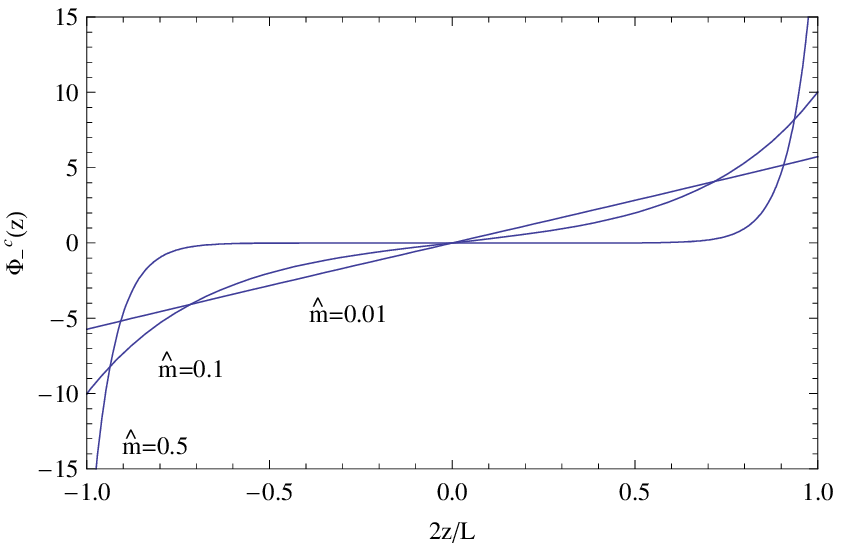}
\end{center}
\caption{The VEV profiles for various $\hat{m}$: (a) For $\Phi^c_+(z)$. (b) For
$\Phi^c_-(z)$. Input parameters are 
$(\hat{L},\hat{\lambda},\Lambda)=(20\pi,1,10\mbox{TeV})$ in both figures.}
\label{pfig1}
\end{figure}

\begin{figure}[t]
\hspace{3.85cm}(a)\hspace{7.7cm}(b)
\begin{center}
\hspace{-4mm}\includegraphics[scale = 0.9]{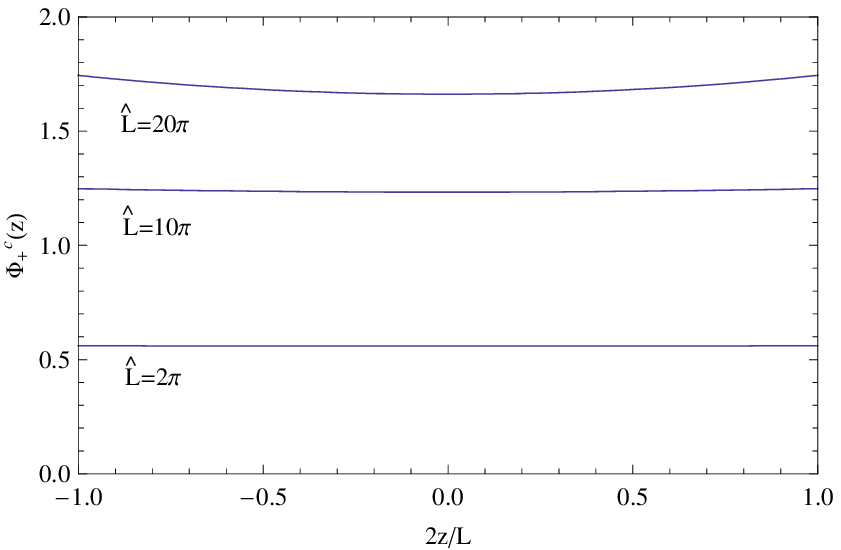}\hspace{4mm}
\includegraphics[scale = 0.91]{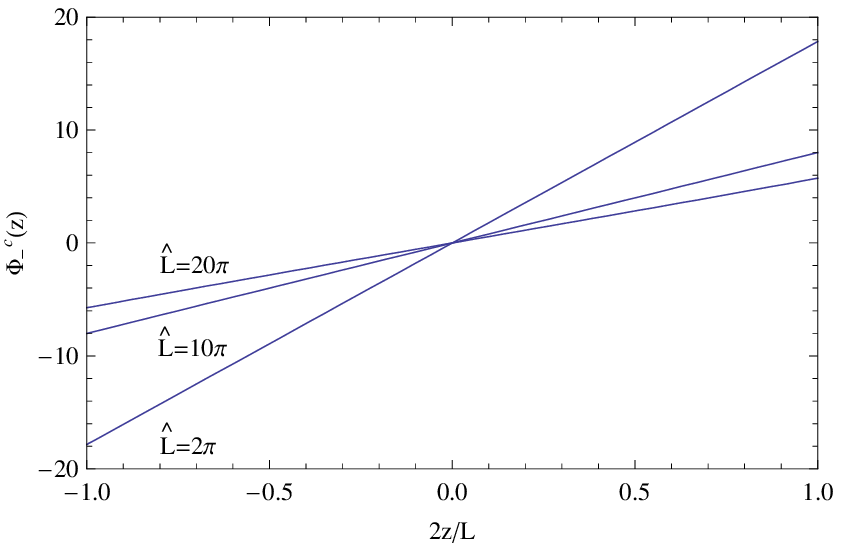}
\end{center}
\caption{The VEV profiles for various $\hat{L}$: (a) For $\Phi^c_+(z)$. (b) For
$\Phi^c_-(z)$. Input parameters are 
$(\hat{m},\hat{\lambda},\Lambda)=(0.01,1,10\mbox{TeV})$ in both figures.}
\label{fig1-2}
\end{figure}

\begin{figure}
\begin{center}
\hspace{1mm}(a)\hspace{7.8cm}(b)\vspace{3mm}

\hspace{-4mm}\includegraphics[scale = 0.9]{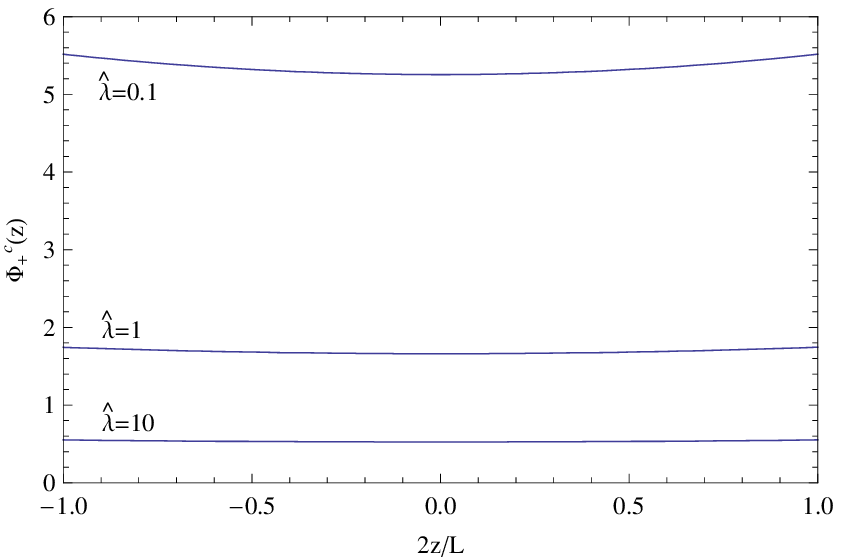}\hspace{4mm}
\includegraphics[scale = 0.9]{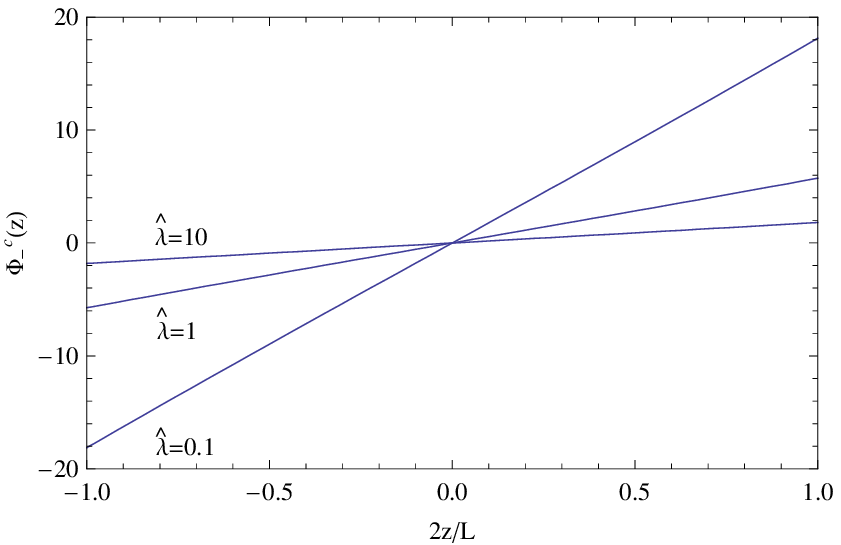}
\end{center}
\caption{The VEV profiles for various $\hat{\lambda}$: (a) For $\Phi^c_+(z)$. 
(b) For $\Phi^c_-(z)$. Input parameters are 
$(\hat{m},\hat{L},\Lambda)=(0.01,20\pi,10\mbox{TeV})$ in both figures.}
\label{pfig1-3}
\end{figure}

We note that if there is no bulk potential, only the flat VEV profile 
$\Phi^c=v$ is allowed under the presence of the reflection symmetry for 
the brane potentials. More generally, the following VEV profiles are 
given by the equation of motion with $\mathcal{V}=m^2|\Phi|^2$, with the BCs from the reflection symmetric 
boundary potentials
\begin{eqnarray}
 \Phi^c(z)=\left\{
  \begin{array}{ll}
   \pm\frac{\sqrt{\lambda v^2c_h-ms_h}}{c_h^{3/2}\sqrt{\lambda}}\cosh(mz) & [\mbox{even parity}],\\
   \pm\sqrt{\frac{\lambda v^2s_h-mc_h}{\lambda s_h(c_h-2)}}\sinh(mz)       & [\mbox{odd parity}],\\
   v & [\mbox{flat profile } (\mathcal{V}=0)],
  \end{array}\right.
\end{eqnarray}
The numerical calculations of the VEV profile for various $\hat{m}$, $\hat{L}$ 
and $\hat{\lambda}$ are shown in Figs.~\ref{pfig1}--\ref{pfig1-3}, where 
$\hat{m}$, $\hat{L}$, $\hat{v}$ and $\hat{\lambda}$ are dimensionless 
parameters corresponding to $m$, $L$, $v$ and $\lambda$, respectively defined as
\begin{eqnarray}
 m\equiv\hat{m}\Lambda,\hspace{5mm}
 L\equiv\frac{\hat{L}}{\Lambda},\hspace{5mm}
 v\equiv\hat{v}\Lambda^{3/2},\hspace{5mm}
 \lambda\equiv\frac{\hat{\lambda}}{\Lambda^2},
\end{eqnarray}
with the cutoff scale $\Lambda$.
We have also defined as
\begin{eqnarray}
 \Phi_+^c(z) &\equiv& \left|\frac{\sqrt{\lambda v^2c_h-ms_h}}
                                 {c_h^{3/2}\sqrt{\lambda}}\right|\cosh(mz),\\
 \Phi_-^c(z) &\equiv& \left|\sqrt{\frac{\lambda v^2s_h-mc_h}
                                 {\lambda s_h(c_h-2)}}\right|\sinh(mz).
\end{eqnarray} 
Figure \ref{pfig1} (a) shows that the VEV profile becomes flat when the bulk 
mass $m$ is small comparing with $\Lambda$. For simplicity, we will take $m=0$ 
and hence the flat VEV profile in phenomenological analyses throughout this 
paper, since it is the easiest case to reproduce $Z$ and $W$ 
masses.\footnote{There are generally quantum corrections that induce the bulk 
potential, whose sum with the tree one is assumed to be zero here. Note that, 
due to custodial symmetry, it could be that the resultant $Z$ and $W$ masses 
are correct ones even under the presence of bulk Higgs mass. We thank T. 
Yamashita on this point.}

%%%%%%%%%%%%%%%%%%%%%%%%%%%%%%%%%%%%%%%%%%%%%%%%%%%%%%
\subsection{Physical Higgs and other fields' profiles}
%%%%%%%%%%%%%%%%%%%%%%%%%%%%%%%%%%%%%%%%%%%%%%%%%%%%%%

In the previous subsection, the VEV profiles of the bulk scalar field and their general properties have been discussed. In this subsection, we investigate wave function profiles 
of quantum modes of the scalar field, which correspond to the Higgs, and 
Nambu-Goldstone (NG) modes. 

{}First, we expand the scalar field around the VEV as follows
\begin{eqnarray}
 \Phi(x,z)&=&\left(
 \begin{array}{c}
 \varphi^+(x,z)\\
 \Phi^c(z)+\frac{1}{\sqrt{2}}[\phi^q(x,z)+i\chi(x,z)] 
 \end{array}\right)\\
 &\equiv&\left(
 \begin{array}{c}
 \displaystyle\sum_{n=0}^\infty f_n^{\varphi^+}(z)\varphi_n^+(x)\\
 v(z)+\frac{1}{\sqrt{2}}\displaystyle\sum_{n=0}^\infty
 [f_n(z)\phi_n^q(x)+if_n^\chi(z)\chi_n(x)] 
 \end{array}\right),
 \label{higgs}
\end{eqnarray}
where we took $\Phi_R/\sqrt{2}=v(z)$ and $\Phi_I=0$ and the KK expansions are 
taken for $\phi^q(x,z)$, $\chi(x,z)$, and $\varphi^+(x,z)$. The $\phi_n^q(x)$, 
$\chi_n(x)$ and $\varphi_n^+(x)$ are the physical Higgs, NG, and charged Higgs 
bosons of the n-th KK modes in four dimensional spacetime, respectively. 
Hereafter the lowest modes (zero-mode) of each field are represented by 
$\phi^q(x)$, $\chi(x)$, and $\varphi^+(x)$, for simplicity. 

%%%%%%%%%%%%%%%%%%%%%%%%%%%%%%%%%%%%%%
\subsubsection{Physical Higgs profile}
%%%%%%%%%%%%%%%%%%%%%%%%%%%%%%%%%%%%%%

The action for the physical Higgs can be written down by\footnote{A detailed 
derivation is given in Appendix B.}
\begin{eqnarray}
 S_{\mbox{{\scriptsize free}},\phi^q}
 &=&\int d^4x\int_{-L/2}^{+L/2}dz
    \Bigg(\frac{1}{2}\phi^q(\Box+\partial_z^2
    -(v^2\mathcal{V}''{}^c+\mathcal{V}'{}^c))
     \phi^q\nonumber\\
 & &\phantom{\int d^4x\int_{-L/2}^{+L/2}dz
    \Bigg(}+\frac{\delta(z-L/2)}{2}\phi^q(
    -\partial_z-(v^2V_L''{}^c+V_L'{}^c))\phi^q
    \nonumber\\
 & &\phantom{\int d^4x\int_{-L/2}^{+L/2}dz
    \Bigg(}+\frac{\delta(z+L/2)}{2}\phi^q(
    \partial_z-(v^2V_0''{}^c+V_0'{}^c))\phi^q
    \Bigg)\nonumber\\
 &=&\int d^4x\int_{-L/2}^{+L/2}dz\Bigg[
    \frac{1}{2}\phi^q\left(\Box+\partial_z^2
    -\frac{\partial^2\mathcal{V}}{\partial
    \Phi_R^2}^c\right)\phi^q\nonumber\\
 & &+\frac{\delta(z-L/2)}{2}\phi^q\left(
    -\partial_z-\frac{\partial^2V_L}{
    \partial\Phi_R^2}^c\right)\phi^q     
    +\frac{
    \delta(z+L/2)}{2}\phi^q\left(\partial_z-
    \frac{\partial^2V_0}{\partial\Phi_R^2}^c
    \right)\phi^q\Bigg],\nonumber\\
 & &\label{hl}
\end{eqnarray}
where we used
\begin{align}
	{\partial V_\xi\over\partial\Phi_R}^c
		&=	v{V'}_\xi^c, &
	{\partial^2V_\xi\over\partial
       \Phi_R^2}^c
		&=	v^2{V''}_\xi^c
       +{V'}_\xi^c, &
	\left({\partial^2V_\xi\over\partial
       \Phi_R
       \partial\Phi_I}\right)^c
		&=	0.
 \end{align}

The KK equation and BCs for the Higgs are given
by
\begin{eqnarray}
 \left(\partial_z^2
 -\frac{\partial^2\mathcal{V}}{\partial
 \Phi_R^2}^c\right)f_n(z)=-\mu_n^2f_n(z),
 \label{kkh}
\end{eqnarray}
and
\begin{eqnarray}
 \left.\left(\mp\partial_z-\frac{
 \partial^2V_\xi}{\partial\Phi_R^2}^c
 \right)f_n(z)\right|_{z=\xi}=0.\label{p17}
\end{eqnarray} 
If we take the bulk and brane potentials as (\ref{bup}) and (\ref{bp}) again, 
the general solution of the KK equation (\ref{kkh}) and BCs become
\begin{eqnarray}
 f_n(z)=\alpha_n\cos(k_nz)+\beta_n\sin(k_nz),
\end{eqnarray}
and 
\begin{eqnarray}
 \mathcal{C}
 \left(
 \begin{array}{c} 
  \alpha_n \\
  \beta_n
 \end{array}
 \right)
 \equiv
 \left(
 \begin{array}{lr}
  k_ns_n-\tilde{\lambda}c_n & -k_nc_n-\tilde{\lambda}s_n \\
  k_ns_n-\tilde{\lambda}c_n & k_nc_n+\tilde{\lambda}s_n
 \end{array}
 \right)
 \left(
 \begin{array}{c} 
  \alpha_n \\
  \beta_n
 \end{array}
 \right)=0,\label{bcz}
\end{eqnarray}
where $k_n\equiv\sqrt{\mu_n^2-m^2}$. Furthermore, we defined the following 
parameters
\begin{eqnarray}
 s_n             
 &\equiv& \sin\left(\frac{k_nL}{2}\right), \\
 c_n             
 &\equiv& \cos\left(\frac{k_nL}{2}\right), \\
 \tilde{\lambda} 
 &\equiv& \frac{\partial^2V_\xi}{\partial
          \Phi_R^2}^c.
\end{eqnarray}

We look for non-trivial solutions of 
$\alpha_n$ and $\beta_n$, which correspond to $\det\mathcal{C}=0$,
leading to the following conditions~\footnote{A non-vanishing determinant of the coefficient matrix $\mathcal{C}$ in Eq.~(\ref{bcz}) corresponds to the trivial solution $(\alpha_n,\beta_n)=(0,0)$, which is of course
not of physical interest since this solution means that the wave function of the Higgs is zero everywhere. }
\begin{eqnarray}
 t_n=\text{$\frac{\tilde{\lambda}}{k_n}$, or $-\frac{k_n}{\tilde{\lambda}}$,}
 \label{tan}
\end{eqnarray}
where $t_n\equiv\tan(k_nL/2)$. 
The former condition $t_n=\tilde{\lambda}/k_n$ leads to $\alpha_n\neq0$ and $\beta_n=0$, and the wave function profile of the Higgs becomes even under the reflection between $\pm z$, 
that is, $f_n(z)=f_n(-z)$. On the other hand, the latter condition 
$t_n=-k_n/\tilde{\lambda}$ leads to odd profile, $f_n(z)=-f_n(-z)$. To summarize, we obtain
\begin{eqnarray}
 t_n=\left\{
 \begin{array}{llll}
 \tilde{\lambda}/k_n  & \Rightarrow & f_n(z)=\alpha_n\cos(k_nz) & 
 [\mbox{even}], \\
 -k_n/\tilde{\lambda} & \Rightarrow & f_n(z)=\beta_n\sin(k_nz)  & 
 [\mbox{odd}].
 \end{array}
 \right. \label{eo}
\end{eqnarray}
The appearance of this parity under the reflection between $\pm z$ is due to  the identical potentials on $z=\pm L/2$ branes. This accidental 
parity corresponds to the KK parity in the UED model. Hereafter, we also 
call it the even and odd KK parities:
\begin{eqnarray}
 f_n(z)=f_n(-z)\hspace{3mm}\mbox{[even]},\hspace{5mm}
 f_n(z)=-f_n(-z)\hspace{3mm}\mbox{[odd]}.
\end{eqnarray}
%If we assign a parity of the Higgs field, the wave function represents either even or odd profile in (\ref{eo}). Since the action should be $Z_2$ parity even, it is convenient to distinguish these parity by rewriting the wave function profile as,
So far, we have written everything in the ``downstairs'' picture at $-L/2\leq z\leq L/2$, or equivalently at $0\leq y\leq L$. When we go one step ``upstairs'' toward $-L< y\leq L$, we call the parity between $\pm z$ the KK parity, while the one between $\pm y$ the $Z_2$ parity:
\begin{eqnarray}
 &&f_n(z)=\alpha_n\cos(k_nz)~[\mbox{KK even}]\nonumber\\
 &&\hspace{3cm}\Leftrightarrow f_n(z)=\left\{
  \begin{array}{ll}
   \alpha_n[c_n\cos(k_ny)+s_n\sin(k_n|y|)]          & [Z_2\mbox{ even}],\\
   \alpha_n[\epsilon(y)c_n\cos(k_ny)+s_n\sin(k_ny)] & [Z_2\mbox{ odd}],
 \end{array}
 \right.\\
 &&f_n(z)=\beta_n\sin(k_nz)~[\mbox{KK odd}]\nonumber\\
 &&\hspace{3cm}\Leftrightarrow f_n(z)=\left\{
  \begin{array}{ll}
   \beta_n[c_n\sin(k_n|y|)+s_n\cos(k_ny)]          & [Z_2\mbox{ even}],\\
   \beta_n[c_n\sin(k_ny)+\epsilon(y)s_n\cos(k_ny)] & [Z_2\mbox{ odd}],
 \end{array}
 \right.
\end{eqnarray}
where we use $z=y-L/2$ and define $\epsilon(y)=\pm1$ for $\pm y>0$.

\begin{figure}[t]
\begin{center}
\includegraphics[scale = 1.0]{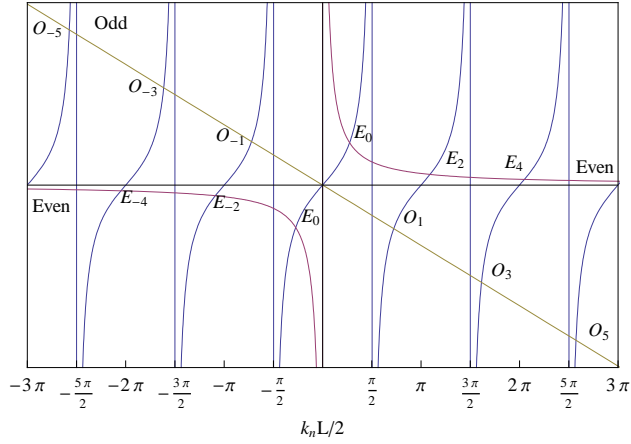}
\end{center}
\caption{Solutions of the KK equation with BCs.}
\label{figtan}
\end{figure}

Typical behaviours of Eq.~(\ref{tan}) is shown in Fig.~\ref{figtan}.
The figure shows functions $\tan(k_nL/2)$, $\tilde{\lambda}/k_n$, and 
$-k_n/\tilde{\lambda}$. Points of intersection $E_{2n}~(|n|=0,1,2,\ldots)$ are 
the solution of equation
\begin{eqnarray}
 t_n=\frac{\tilde{\lambda}}{k_n},\label{en}
\end{eqnarray}
and points of $O_{2n+1}$ correspond to the solution of 
\begin{eqnarray}
 t_n=-\frac{k_n}{\tilde{\lambda}}.\label{on}
\end{eqnarray}
We call $E_0$, $O_1$, $E_2,\ldots$ modes the zero, first, second, $\ldots$ modes, respectively. 
In particular, note that we always call the $E_0$ mode the zero-mode.

In the vanishing limit of $\tilde{\lambda}$, in which our model leads to the limit of the UED model without any bulk mass and brane potentials, 
we get $\tilde{\lambda}/k_n\rightarrow 0$ in Eq.~(\ref{en}) and 
$-k_n/\tilde{\lambda}\rightarrow-\infty\cdot k_n$ in Eq.~(\ref{on}). In the limit, 
solutions become
\begin{eqnarray}
 \frac{k_nL}{2}\simeq\frac{n\pi}{2}
\end{eqnarray}
for both even and odd $n$. Therefore, the wave function profiles are
\begin{eqnarray}
 f_n(z)=
  \left\{
   \begin{array}{ll}
    \alpha_n\cos\left(\frac{n\pi}{L}z\right) & n:\mbox{even}, \\
    \beta_n\sin\left(\frac{n\pi}{L}z\right) & n:\mbox{odd}.
   \end{array}
  \right.
 \label{wfp}
\end{eqnarray}
In the opposite infinity limit 
of $\tilde{\lambda}$, which means $\tilde{\lambda}/k_n\rightarrow\infty$ in 
Eq.~(\ref{en}) and $-k_n/\tilde{\lambda}\rightarrow0$ in Eq.~(\ref{on}),
\begin{eqnarray}
 \frac{k_nL}{2}\simeq\frac{(n+1)\pi}{2}
\end{eqnarray}
for both even and odd $n$. Then, the wave function profiles become
\begin{eqnarray}
 f_n(z)=
  \left\{
   \begin{array}{ll}
    \alpha_n\cos\left(\frac{(n+1)\pi}{L}z\right) & n:\mbox{even}, \\
    \beta_n\sin\left(\frac{(n+1)\pi}{L}z\right) & n:\mbox{odd}.
   \end{array}
  \right.
 \label{wfp1}
\end{eqnarray}

In Table~\ref{tab-pro}, we summarize the situation showing dependences of the Higgs profiles on the magnitude of boundary coupling. 
\begin{table}[t]
\begin{center}
\begin{tabular}{|c|}
\hline
KK parity / mode  \\
Typical Profiles \\
($\tilde{\lambda}\rightarrow0$: UED) \\
\hline 
\hline
$+$ / $n=0$ \\
\includegraphics[scale =0.47]{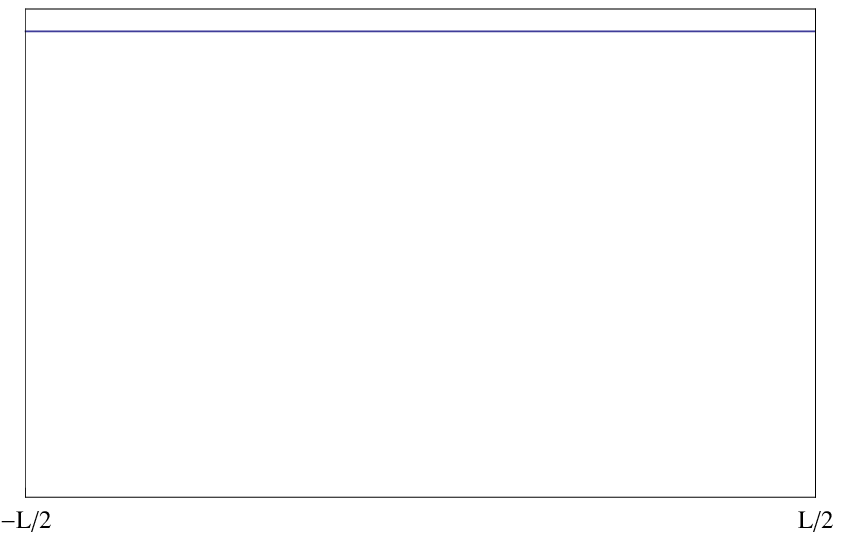}\\
\hline
$-$ / $n=1$ \\
\includegraphics[scale =0.47]{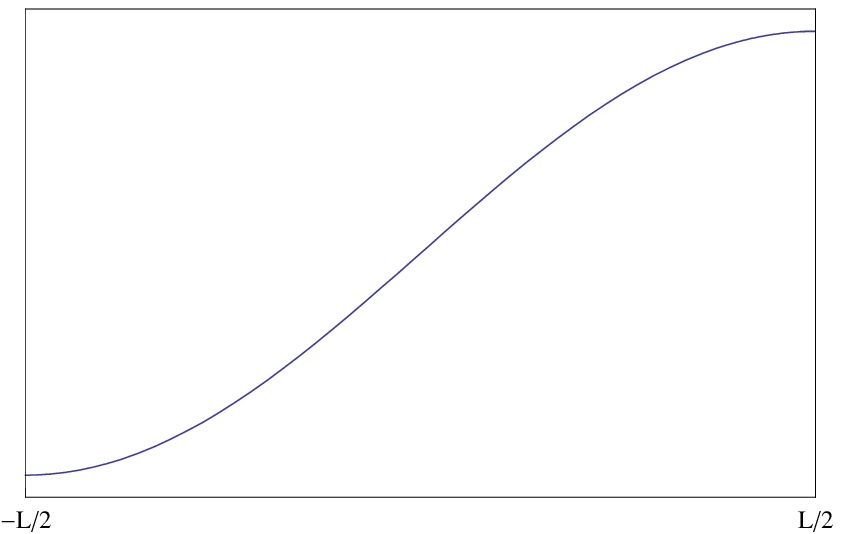}\\
\hline
$+$ / $n=2$ \\
\includegraphics[scale =0.47]{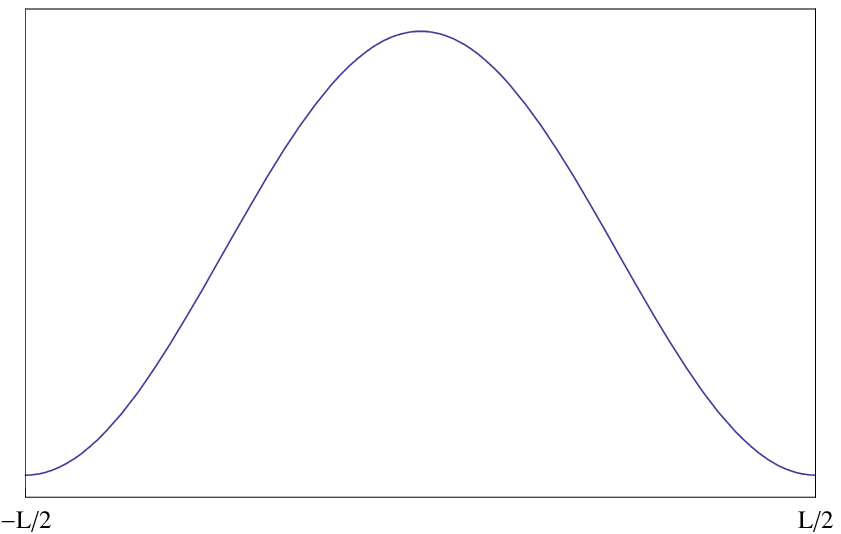}\\
\hline
$-$/  $n=3$ \\
\includegraphics[scale =0.47]{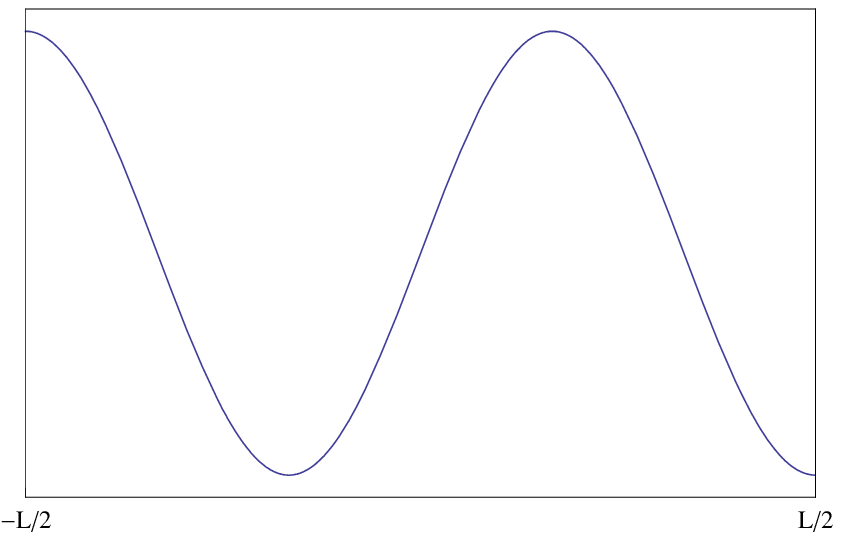}\\
$\vdots$ \\
\hline
\end{tabular}\hspace{1cm}
\begin{tabular}{|c|}
\hline
KK parity / mode \\
Typical Profile  \\
($\tilde{\lambda}\rightarrow\infty$) \\
\hline 
\hline
$+$ / $n=0$ \\
\includegraphics[scale =0.47]{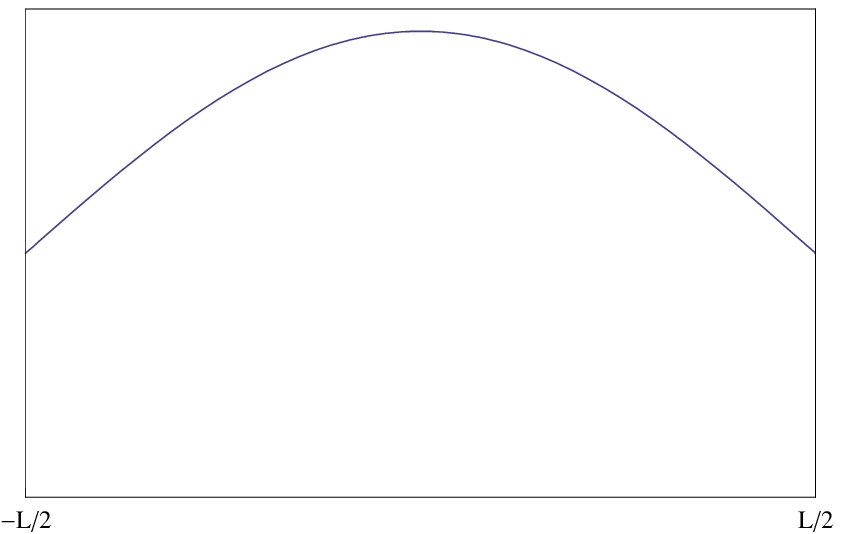}\\
\hline
$-$ / $n=1$ \\
\includegraphics[scale =0.47]{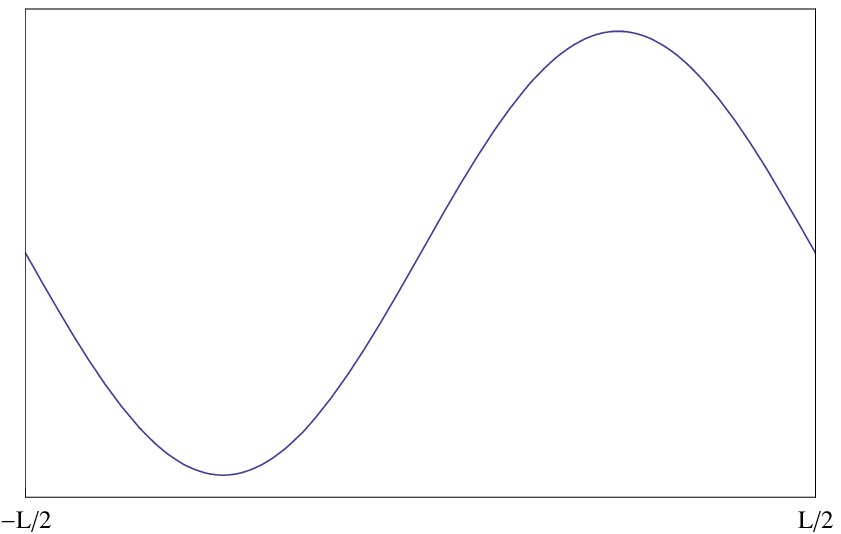}\\
\hline
$+$ / $n=2$ \\
\includegraphics[scale =0.47]{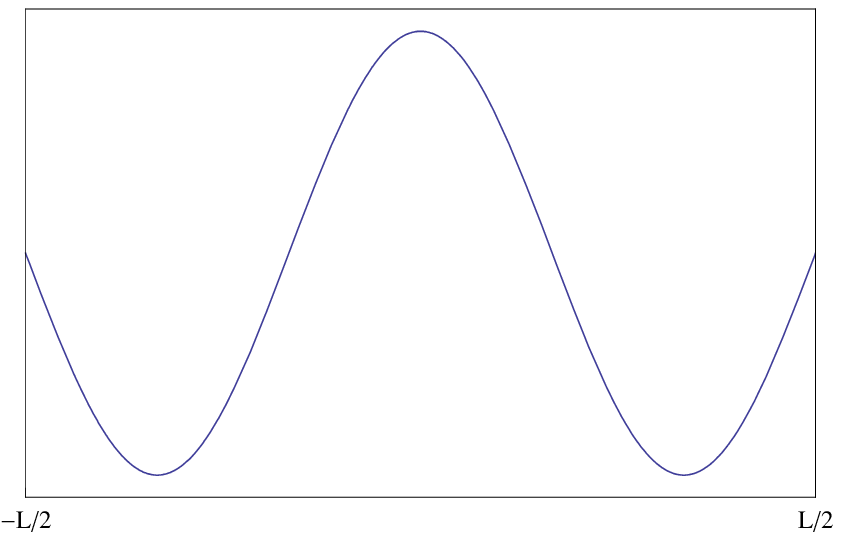}\\
\hline
$-$ / $n=3$ \\
\includegraphics[scale =0.47]{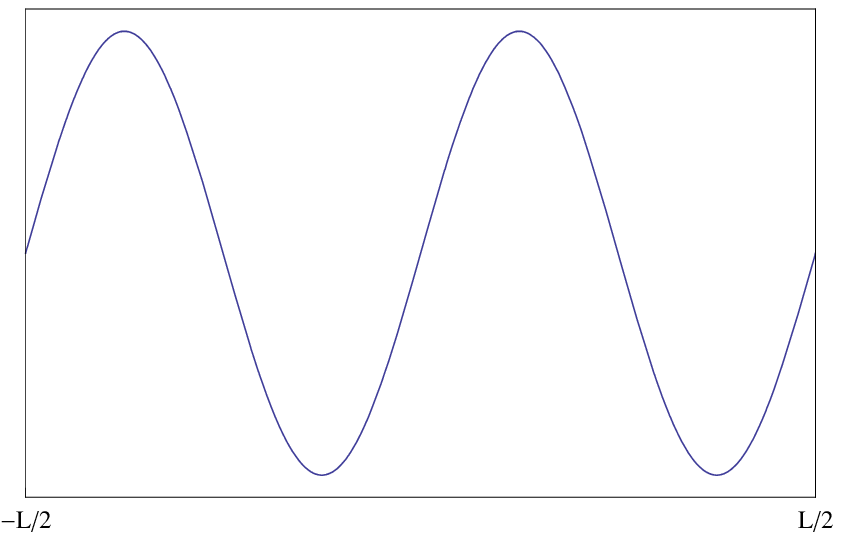}\\
$\vdots$ \\
\hline
\end{tabular}
\end{center}
\caption{Typical behaviours of wave function profile of the Higgs in this 
setup.}
\label{tab-pro}
\end{table}
The profile of zero mode Higgs in the small limit of the 
boundary coupling is obtained from Eq.~(\ref{wfp}) as 
\begin{eqnarray}
 f_0=\alpha_0,\label{flat}
\end{eqnarray}
which means that the Higgs has flat profile. It is shown in left upper figure 
of Table \ref{tab-pro}. On the other hand, the zero mode Higgs profile with a large boundary coupling is obtained from Eq.~(\ref{on}) as
\begin{eqnarray}
 f_0=\alpha_0\cos\left(\frac{\pi}{L}z\right),\label{non-flat}
\end{eqnarray}
which is given in right upper figure in Table \ref{tab-pro}.

Let us comment on a \emph{top Yukawa deviation}, which is one of the main points of this paper.
We find that the flat profile of Higgs~(\ref{flat}) continuously
changes to a non-flat one~(\ref{non-flat}) as 
the boundary coupling becomes larger. This means that the free quantum part of the Higgs tends to be expelled from the boundary when the boundary coupling is large. Since the VEV profile is flat in our setup of the vanishing bulk potential (even in this limit), the profiles of VEV and physical Higgs are generally different. 
This causes the top Yukawa deviation. If all the SM fermions
are localized on the brane, the magnitude of $f_0(\pm L/2)$ determines the Yukawa 
couplings. In this case, the top Yukawa deviation can occur drastically as the boundary coupling becomes large, while the top mass is being correctly reproduced.

In the vanishing boundary-coupling limit, the coefficients $\alpha_n$ and $\beta_n$ are determined by the normalization 
condition
\begin{eqnarray}
 \alpha_n\alpha_m\int_{-L/2}^{+L/2}dz\cos\left(\frac{n\pi}{L}z\right)
 \cos\left(\frac{m\pi}{L}z\right)=\delta_{n,m}\label{alnor}
\end{eqnarray}
for even $n,m$ and by 
\begin{eqnarray}
 \beta_n\beta_m\int_{-L/2}^{+L/2}dz\sin\left(\frac{n\pi}{L}z\right)
 \sin\left(\frac{m\pi}{L}z\right)=\delta_{n,m}\label{benor}
\end{eqnarray}
for odd $n,m$,
where $\delta_{n,m}$ is the Kronecker delta. 
Therefore in this limit, 
\begin{eqnarray}
 \alpha_n=\beta_n=\sqrt{\frac{2}{L}}
\end{eqnarray}
for $n\neq0$ and
\begin{eqnarray}
 \alpha_0=\sqrt{\frac{1}{L}}
\end{eqnarray}
for $n=0$.
In the opposite infinite limit of the boundary coupling, it can be easily checked that $\alpha_0=\alpha_n=\beta_n=\sqrt{2/L}$, which means that $n$ and $m$ are simply replaced with $n+1$ and $m+1$ in Eqs.~(\ref{alnor}) and (\ref{benor}), respectively. 

We should comment on a case of finite boundary coupling. It is 
seen from Fig.~\ref{figtan} that solutions of the KK equations $k_n$ are not 
represented by $n\pi/L$ or $(n+1)\pi/L$ with an integer $n$ generally. The 
deviation from $n$ or $n+1$ affects the normalization. We can generally 
parametrize the deviations as $n+\Delta_n$ by finite $\Delta_n$. In the case, 
the normalization conditions are written as
\begin{eqnarray}
 \alpha_{n+\Delta_n}\alpha_{m+\Delta_m}
 \int_{-L/2}^{+L/2}dz\cos\left(\frac{(n
 +\Delta_n)\pi}{L}z\right)\cos\left(\frac{(m
 +\Delta_m)\pi}{L}z\right)=\delta_{n
 +\Delta_n,m+\Delta_m},
\end{eqnarray}
for even $n$ and $m$, and 
\begin{eqnarray}
 \beta_{n+\Delta_n}\beta_{m+\Delta_m}
 \int_{-L/2}^{+L/2}dz
 \sin\left(\frac{(n+\Delta_n)\pi}{L}z\right)
 \sin\left(\frac{(m+\Delta_m)\pi}{L}z\right)
 =\delta_{n+\Delta_n,m+\Delta_m},
\end{eqnarray}
for odd $n$ and $m$. They lead to 
\begin{eqnarray}
 \alpha_n&=&\sqrt{\frac{2}{L\left[1+\frac{\sin((n+\Delta_n)\pi)}
                          {(n+\Delta_n)\pi}\right]}},\\
 \beta_n &=&\sqrt{\frac{2}{L\left[1-\frac{\sin((n+\Delta_n)\pi)}
                          {(n+\Delta_n)\pi}\right]}}.
\end{eqnarray}

To summarize, we show the wave function profile of the Higgs in this 
extradimensional setup as
\begin{eqnarray}
 f_n(z)=
  \left\{
   \begin{array}{l}
    \left\{
     \begin{array}{ll}
      \sqrt{\frac{1}{L}}                                 & 
      \hspace{3.35cm}0\mbox{ mode} \\
      \sqrt{\frac{2}{L}}\cos\left(\frac{n\pi}{L}z\right) & 
      \hspace{3.35cm}n:\mbox{even} \\
      \sqrt{\frac{2}{L}}\sin\left(\frac{n\pi}{L}z\right) & 
      \hspace{3.35cm}n:\mbox{odd}
     \end{array}
    \right.[\hat{\lambda}\rightarrow0],\\
    \left\{
     \begin{array}{ll}
      \sqrt{\frac{2}{L\left[1+\frac{\sin((n+\Delta_n)\pi)}{(n+\Delta_n)\pi}
      \right]}}\cos\left(\frac{(n+\Delta_n)\pi}{L}z\right) & n:\mbox{even} \\
      \sqrt{\frac{2}{L\left[1-\frac{\sin((n+\Delta_n)\pi)}{(n+\Delta_n)\pi}
      \right]}}\sin\left(\frac{(n+\Delta_n)\pi}{L}z\right) & n:\mbox{odd}
    \end{array}
  \right.[\mbox{finite }\hat{\lambda}],\\
  \left\{
   \begin{array}{ll}
    \sqrt{\frac{2}{L}}\cos\left(\frac{(n+1)\pi}{L}z\right)  & 
    \hspace{2.6cm}n:\mbox{even} \\
    \sqrt{\frac{2}{L}}\sin\left(\frac{(n+1)\pi}{L}z\right)  & 
    \hspace{2.6cm}n:\mbox{odd}
   \end{array}
  \right.[\hat{\lambda}\rightarrow\infty].
  \end{array}
  \right.
\end{eqnarray}
We repeat that at the large limit of the boundary coupling, the wave 
function profile of the zero mode Higgs is not flat but altered by a cosine function.

\begin{figure}[tbp]
\hspace{4.26cm}(a)\hspace{7.2cm}(b)
\begin{center}
\includegraphics[scale = 0.9]{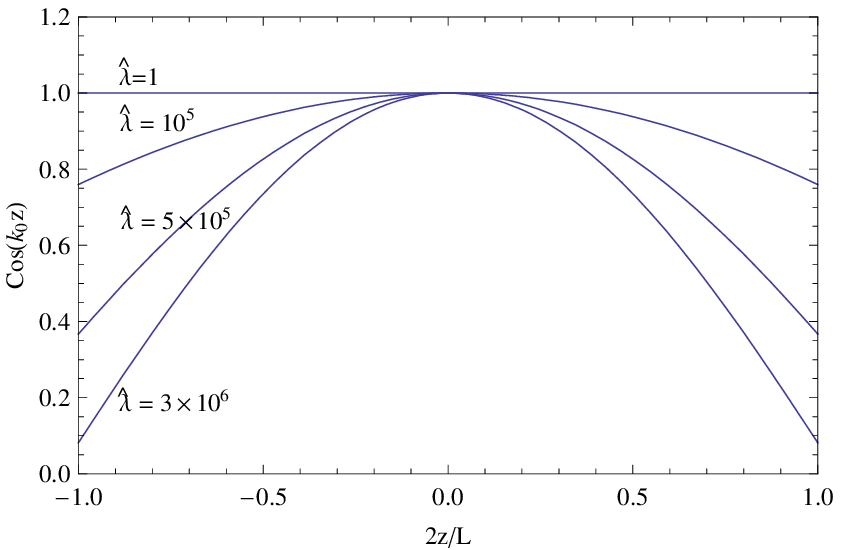}
\includegraphics[scale = 0.9]{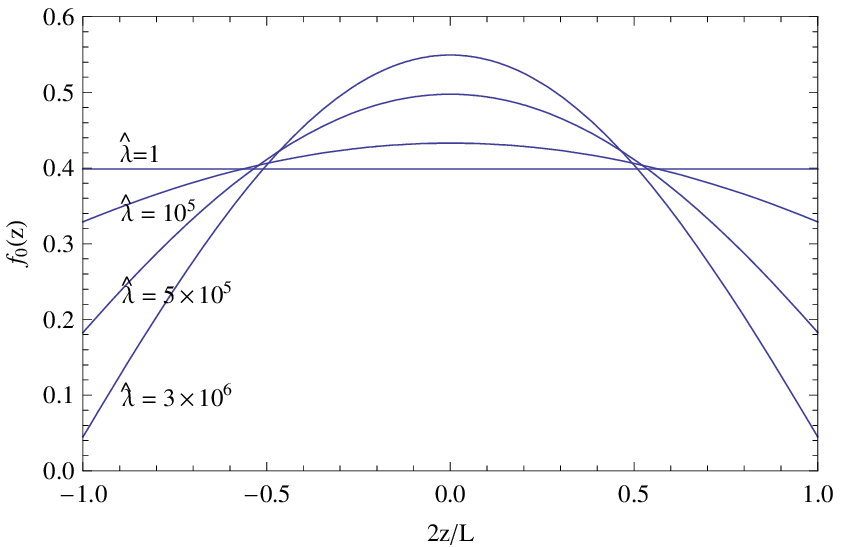}
\end{center}
\caption{The wave function profiles of the zero mode Higgs for various 
$\hat{\lambda}$: Input parameters are 
$(\hat{L},\hat{m},v,\Lambda)=(20\pi,0,174\mbox{GeV}/\sqrt{L},10\mbox{TeV})$ 
in both figures. (a) The vertical axis is $\cos(k_0z)$. (b) The vertical axis 
is $f_0(z)$.} 
\label{pfig2}
\end{figure}

Numerical calculations are given in Fig.~\ref{pfig2}.
These figures show changes of the wave function profile of zero mode Higgs when
$\hat{\lambda}$ becomes large, which confirm the change of flat profile for 
zero mode at small limit of $\tilde{\lambda}$ in left upper figure of 
Table~\ref{tab-pro} to non-flat one in corresponding right figure. The point in 
this feature is that there is parameter space where the magnitude of 
$|f_0(\pm L/2)|$ is tiny. The tiny $|f_0(\pm L/2)|$ means that the coupling 
among brane localized fields and a Higgs boson is also tiny. Such a small 
coupling leads to rich phenomenologies and would be observed at the LHC 
experiment. The investigations of these phenomenological implications are main 
purpose of this paper.

%%%%%%%%%%%%%%%%%%%%%%%%%%%%%%%%%%%%%%%%%%%%%%%%%%%%%%%%%%
\subsubsection{Nambu-Goldstone and other fields' profiles}
%%%%%%%%%%%%%%%%%%%%%%%%%%%%%%%%%%%%%%%%%%%%%%%%%%%%%%%%%%

Before turning to phenomenological discussions, we show the wave function 
profiles of NG mode and other fields in the SM. The action for the NG boson is 
given by\footnote{A detailed derivation is given in Appendix B.}
\begin{eqnarray}
 S_{\mbox{{\scriptsize free}},\chi}
 &=&\int d^4x\int_{-L/2}^{+L/2}dz\Bigg(
     \frac{1}{2}\chi(\Box+\partial_z^2
    -\mathcal{V}'{}^c)\chi+\frac{
    \delta(z-L/2)}{2}\chi(-\partial_z
    -V_L'{}^c)\chi\nonumber\\
 & &\phantom{\int d^4x\int_{-L/2}^{+L/2}dz
    \Bigg(\frac{1}{2}\chi(\Box+\partial_z^2
    -\mathcal{V}'{}^c)\chi}+\frac{
     \delta(z+L/2)}{2}\chi(+\partial_z
    -V_0'{}^c)\chi\Bigg)\nonumber\\
 &=&\int d^4x\int_{-L/2}^{+L/2}dz\Bigg[
    \frac{1}{2}\chi\left(\Box+\partial_z^2-
    \frac{\partial^2\mathcal{V}}{\partial
    \Phi_I^2}^c\right)\chi+\frac{
    \delta(z-L/2)}{2}\chi\left(-\partial_z-
    \frac{\partial^2V_L}{\partial\Phi_I^2}^c
    \right)\chi\nonumber\\
 & &\phantom{\int d^4x\int_{-L/2}^{+L/2}dz
    \Bigg[\frac{1}{2}\chi\left(\Box+
    \partial_z^2-\frac{\partial^2
    \mathcal{V}}{\partial\Phi_I^2}^c\right)
    \chi}+\frac{\delta(z+L/2)}{2}\chi\left(+
    \partial_z-\frac{\partial^2V_0}{\partial
    \Phi_I^2}^c\right)\chi\Bigg].\nonumber\\
 & &\label{46}
 \end{eqnarray}
where we used
\begin{align}
	{\partial V\over\partial\Phi_I}^c
		&=	0, &
	{\partial^2V\over\partial\Phi_I^2}^c
		&=	{V'}^c, &
	\left({\partial^2V\over\partial\Phi_R
       \partial\Phi_I}\right)^c
		&=	0.
 \end{align}
Thus the KK equation for the NG boson and BCs 
are written as 
\begin{eqnarray}
 \left(\partial_z^2-\frac{\partial^2
 \mathcal{V}}{\partial\Phi^2_I}
 \right)f_n^\chi(z)=-\mu_n^{\chi2}f_n^\chi(z) \label{gold1}
\end{eqnarray}
and
\begin{eqnarray}
 \left.\left(\mp\partial_z-\frac{1}{2}\frac{
 \partial^2V_\xi}{\partial\Phi_I^2}\right)
 f_n^\chi(z)\right|_{z=\xi}=0.\label{bc0}
\end{eqnarray}
The general solution of (\ref{gold1}) becomes 
\begin{eqnarray}
 f_n^\chi(z)=\alpha_n^\chi\cos(k_n^\chi z)+\beta_n^\chi\sin(k_n^\chi z),
\end{eqnarray}
where $k_n^\chi\equiv\sqrt{\mu_n^{\chi2}-m^2}$ as we take the bulk potential as Eq.~(\ref{bup}). The BCs~(\ref{bc0}) are also written down by
\begin{eqnarray}
 k_n^\chi
  \left(
   \begin{array}{rl}
    -s_n^\chi & c_n^\chi \\
    s_n^\chi  & c_n^\chi
   \end{array}
  \right)
  \left(
   \begin{array}{l}
   \alpha_n^\chi \\
   \beta_n^\chi
   \end{array}
  \right)=0,\label{bcchi}
\end{eqnarray}
where
\begin{eqnarray}
 s_n^\chi&\equiv&\sin\left(\frac{k_n^\chi L}{2}\right), \\
 c_n^\chi&\equiv&\cos\left(\frac{k_n^\chi L}{2}\right).
\end{eqnarray}
It is seen that non-trivial solutions are $k_n^\chi= n\pi/L$. Therefore, the 
wave function profile of NG boson is given as
\begin{eqnarray}
 f_n^\chi(z)=
  \left\{
  \begin{array}{ll}
   \sqrt{\frac{2}{L}}                                 & 0\mbox{ mode}, \\
   \sqrt{\frac{1}{L}}\cos\left(\frac{n\pi}{L}z\right) & n:\mbox{even}, \\
   \sqrt{\frac{1}{L}}\sin\left(\frac{n\pi}{L}z\right) & n:\mbox{odd},
  \end{array}
  \right.
\end{eqnarray}
where the normalization conditions are imposed to determine $\alpha_n^\chi$ and
$\beta_n^\chi$. We find that the profile of NG boson is the same as one in the
UED even if the brane potentials are introduced because the NG is a 
fluctuation in a flat direction along the potential minimum. 

We comment on wave function profiles of other fields in the SM such as the 
gauge bosons and fermions. If both the gauge bosons and fermions are bulk 
field, the Neumann BCs are simply written as 
$\mp\partial_zf_n^\ast(z)|_{z=\xi}=0$ because the brane potentials do not 
affect these fields, where $\ast$ distinguishes any field contents such as 
gauge bosons and fermions. Therefore, the BCs are essentially same ones as that
for the NG mode (\ref{bcchi}). 

We conclude discussions in this subsection 
that the presence of the brane potentials affects only the wave 
function profiles of the Higgs field because the Higgs VEV is flat, in our assumption of the vanishing bulk potential. 
The profiles of other bulk fields such as the NG and gauge bosons, and fermions, are given by
\begin{eqnarray}
 f_n^\ast(z)=
  \left\{
   \begin{array}{ll}
    \sqrt{\frac{2}{L}}                                 & 0\mbox{ mode} \\
    \sqrt{\frac{1}{L}}\cos\left(\frac{n\pi}{L}z\right) & n:\mbox{even} \\
    \sqrt{\frac{1}{L}}\sin\left(\frac{n\pi}{L}z\right) & n:\mbox{odd}.
   \end{array}
  \right.
 \label{owp}
\end{eqnarray}

%%%%%%%%%%%%%%%%%%%%%%%%%%%%%%%%
\subsection{Physical Higgs mass}
%%%%%%%%%%%%%%%%%%%%%%%%%%%%%%%%

We show the Higgs boson mass in this 
extradimensional setup. The Higgs mass is 
written from (\ref{hl}) as follows
\begin{eqnarray}
 m_H^2\equiv\int_{-L/2}^{+L/2}dzf_0(z)\left[-
 \partial_z^2+\frac{\partial^2\mathcal{V}}{
 \partial\Phi_R^2}^c\right]f_0(z)=\mu_0^2,
\end{eqnarray}
under the presence of a generic finite bulk potential. When the bulk potential is vanishing, the mass 
becomes
\begin{eqnarray}
 m_H^2=-\int_{-L/2}^{+L/2}dzf_0(z)\partial_z^2f_0(z)=\mu_0^2=k_0^2.
 \label{higgsmass}
\end{eqnarray}
The scale of the Higgs mass strongly depends on the boundary coupling $\lambda$
because both wave function profiles of the VEV and Higgs field are sensitive 
to the magnitude of this coupling, shown in Figs.~\ref{pfig1-3} and \ref{pfig2}. 
The coupling appears through the BCs. 

\begin{figure}
\begin{center}
\includegraphics[scale = 1.0]{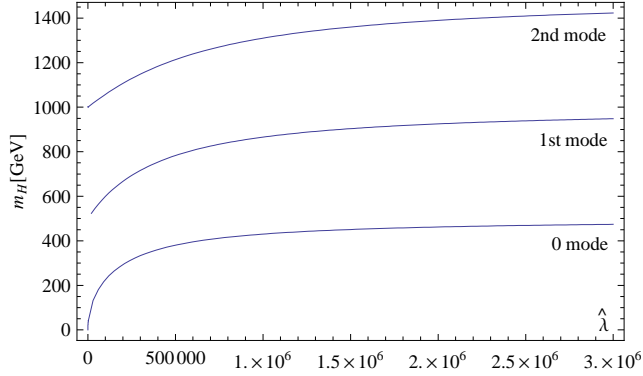}
\end{center}
\caption{$\lambda$ dependences of the Higgs 
masses for the zero, first, and second KK 
modes: Input parameters are 
$(\hat{m},\hat{L},\Lambda,v)=(0,20\pi,10
\mbox{TeV},174\mbox{GeV}/\sqrt{L})$.}
\label{higgsm}
\end{figure}

We consider the Higgs mass scale in three typical cases of the boundary coupling magnitude. {}First, at the vanishing limit of the coupling, the BCs 
(\ref{p17}) is rewritten as $\mp\partial_zf_0(z)|_{z=\xi}=0$. These BCs mean 
that the wave function profile of the Higgs dose not oscillate in an 
extradimensional direction. Therefore, it is seen from (\ref{kkh}) that the 
Higgs becomes massless at the vanishing limit of the boundary coupling 
$\lambda$. Secondly in the large limit of $\lambda$, since the solution of the 
KK equation for the zero mode $E_0$, shown in Fig.~\ref{figtan}, shifts to the 
point of $k_0\simeq\pi/L$, the Higgs mass is nearly equal to the KK scale 
defined as
\begin{eqnarray}
 m_{KK}\equiv\frac{\pi}{L}.
\end{eqnarray} 
Thirdly in the case of small $\lambda$, we can estimate 
as $k_0^2\simeq4\lambda v_{\mbox{{\scriptsize 
EW}}}^2m_{KK}^2/\pi^2$. It is seen from 
Fig.~\ref{figtan} that the Higgs mass 
continuously shift from vanishing $\lambda$ to 
large one. This feature is numerically 
confirmed in Fig.~\ref{higgsm}. 
To summarize,  in the case of the vanishing bulk potential,
the zero mode Higgs mass does not depend on the VEV and is given as
\begin{eqnarray}
 m_H^2=
  \left\{
   \begin{array}{ll}
    0                                                    & \mbox{for vanishing }\lambda,   \\
    4\lambda v_{\mbox{{\scriptsize EW}}}^2m_{KK}^2/\pi^2 & \mbox{for small }\lambda,\\
    m_{KK}^2                                             & \mbox{for large }\lambda.
   \end{array}
  \right.
\end{eqnarray}
The masses of higher KK mode is easily obtained 
by
\begin{eqnarray}
 m_{H^{(n)}}^2=k_n^2,
\end{eqnarray}
where we note that $k_n$ are points of intersection shown in Fig.~\ref{figtan} 
again.

% Note that when the bulk mass is vanishing, the Higgs mass is massless in the 
%case of disappearing $\lambda$, and is only determined by the KK scale in the 
%case of large $\lambda$, respectively. In the intermediate scale of $\lambda$ 
%with the vanishing bulk mass, the Higgs mass is approximated by
% \begin{eqnarray}
%  m_H\sim350\bigg(\frac{1}{\hat{L}}\bigg)
%            \bigg(\frac{\Lambda}{5\mbox{TeV}}\bigg)
%            \bigg(\frac{\hat{\lambda}}{1}\bigg)^{1/2}\mbox{ GeV}.
% \end{eqnarray}

Here we should comment on mode mixings of Higgs. Since this setup includes the 
brane potentials for the Higgs field such as 
$\delta(z\mp L/2)\lambda(|\Phi|^2-v^2)/2$, there are generally mode mixings 
in the Higgs sector. Therefore, the mass matrix of the Higgs in the basis of 
higher KK mode, $(H^{(0)},H^{(1)},H^{(2)},\cdots)^T$, is not diagonal, 
\begin{eqnarray}
 m_{HH}^2&\simeq&
  \left(
   \begin{array}{ccccc}
    \mu_0^2 & 0       & 0       & 0       & \cdots \\
    0       & \mu_1^2 & 0       & 0       & \cdots \\
    0       & 0       & \mu_2^2 & 0       & \cdots \\
    0       & 0       & 0       & \mu_3^2 & \cdots \\ 
    \vdots  & \vdots  & \vdots  & \vdots  & \ddots
   \end{array}
  \right)  
 +\lambda v^2
  \left(
   \begin{array}{ccccc}
    f_0f_0 & 0      & f_0f_2 & 0      & \cdots \\
    0      & f_1f_1 & 0      & f_1f_3 & \cdots \\
    f_0f_2 & 0      & f_2f_2 & 0      & \cdots \\
    0      & f_1f_3 & 0      & f_3f_3 & \cdots \\
    \vdots & \vdots & \vdots & \vdots & \ddots
   \end{array}
  \right),
 \label{hmm}
\end{eqnarray}
where $f_n\equiv f_n(\pm L/2)=\sqrt{1/L}$. It is found that there are generally
mode mixings among KK modes of Higgs field due to non-vanishing off-diagonal 
elements in the second term of Eq.~(\ref{hmm}). Furthermore, there survive only off-diagonal elements conserving the KK parity, that is, in $f_nf_m$ elements for zero and arbitrary positive integers $n$ and $m$, $(n,m)=(2n+1,2m+1)$ and $(2n,2m)$ are allowed but $(2n+1,2m)$ and vice versa are forbidden. Since we take the identical boundary potentials on $z=\pm L/2$ branes, mass terms from boundary potential which is proportional to 
$f_{2n+1}(L/2)f_{2m}(L/2)+f_{2n+1}(-L/2)f_{2m}(-L/2)$ is cancelled according to
its odd parity $f_{2n+1}(L/2)=-f_{2n+1}(-L/2)$.
While the UED conserves 
both the KK number and parity,\footnote{We should note that, if quantum corrections are taken into account on branes in the UED, the KK number conservation is broken.} our model with the 
same boundary potential on $z=\pm L/2$ branes conserves the KK number and 
parity except for the Higgs sector, and the Higgs sector conserves only the KK 
parity. 
That is just a grace of the presence of reflection symmetry between the Higgs 
potentials on $z=\pm L/2$ branes.

The magnitudes of mode mixing are roughly estimated as 
$\mathcal{O}(\sqrt{\lambda}v_{\mbox{{\scriptsize EW}}})$ for finite 
$\hat{\lambda}$, where we used $v\equiv 174\mbox{GeV}/\sqrt{L}\equiv 
v_{\mbox{{\scriptsize EW}}}/\sqrt{L}$. At the large limit of $\hat{\lambda}$, 
$f_n(\pm L/2)$ goes to zero. Therefore, mode mixings are suppressed in the 
case. Is is easy to see from Eq.~(\ref{hmm}) that they are also tiny in the case of
vanishing limit of $\hat{\lambda}$. Exactly speaking, the masses of each KK 
Higgs should be defined in the diagonal basis of each KK state. However, the 
above discussion for the Higgs mass of the zero mode gives a good approximation because mode mixings are generally suppressed.

%%%%%%%%%%%%%%%%%%%%%%%%%%%%%%%%%
\subsection{Higgs self-couplings}
%%%%%%%%%%%%%%%%%%%%%%%%%%%%%%%%%

In this subsection, we discuss the self-couplings in the Higgs sector of 
this setup. The 3 and 4-point couplings, and the Higgs decay to the 
longitudinal mode of the $W$ boson, $W_L$, are studied.

\subsubsection{3 and 4-point couplings of Higgs and NG bosons}

Let us consider 3 and 4-point couplings among the Higgs and NG bosons. These 
couplings are obtained from the boundary potential
\begin{eqnarray}
 S_{H,\chi}
 =-\int d^4x\int_{-L/2}^{+L/2}dz\delta(z\mp
 \xi)V_\xi,\label{hng}
\end{eqnarray}
where the brane potentials $V_\xi$ and the 
Higgs field $\Phi$ are defined in (\ref{bp}) 
and (\ref{higgs}). The Lagrangian for these 
couplings are written down as
\begin{eqnarray}
 -\mathcal{L}_{H,\chi}
 &=&\frac{\lambda}{4}\int_{-L/2}^{+L/2}dz\delta(z\mp\xi)
    \Bigg[\frac{1}{4}(f_0^\chi(z)\chi(x))^4
    +f_0^{\varphi^+}(z)f_0^{\varphi^-}(z)f_0^\chi(z)^2\varphi^+(x)\varphi^-(x)
    \chi(x)^2\nonumber\\
 & &+(f_0^{\varphi^+}(z)f_0^{\varphi^-}(z)\varphi^+(x)
    \varphi^-(x))^2+\frac{1}{4}(f_0(z)\phi^q(x))^4
    +\frac{1}{2}(f_0(z)f_0^\chi(z)\phi^q(x)\chi(x))^2\nonumber\\
 & &+f_0^{\varphi^+}(z)f_0^{\varphi^-}(z)f_0(z)^2\varphi^+(x)
    \varphi^-(x)\phi^q(x)^2+\sqrt{2}v(f_0(z)\phi^q(x))^3\nonumber\\
 & &+\sqrt{2}vf_0(z)f_0^\chi(z)^2
    \phi^q(z)\chi(z)^2
    +2\sqrt{2}vf_0^{\varphi^+}(z)f_0^{\varphi^-}(z)f_0(z)
    \varphi^+(x)\varphi^-(x)\phi^q(x)\Bigg]\nonumber\\
 & &\\
 &=&\frac{\lambda}{2L^2}\Bigg[\left(\frac{\chi(x)^2}{2}
    +\varphi^+(x)\varphi^-(x)\right)^2+c_\Delta^2\phi^q(x)^2[
    c_\Delta^2\phi^q(x)^2+\chi(x)^2+2\varphi^+(x)\varphi^-(x)]\nonumber\\
 & &\hspace{1cm}+2v_{\mbox{\scriptsize EW}}c_\Delta\phi^q(x)[
    c_\Delta^2\phi^q(x)^2+\chi(x)^2
    +2\varphi^+(x)\varphi^-(x)]\Bigg],\label{4ng}
\end{eqnarray}
where
\begin{eqnarray}
 c_\Delta\equiv\sqrt{\frac{1}{1+\sin(\Delta_0\pi)/\Delta_0\pi}}\cos
               \left(\frac{\Delta_0\pi}{2}\right).
\end{eqnarray}
We can read effective self-couplings from (\ref{4ng}) as 
\begin{eqnarray}
 &&\lambda_{\chi\chi\chi\chi}\equiv\frac{3\lambda}{L^2},\hspace{15mm}
   \lambda_{\chi\chi\varphi^+\varphi^-}
   =\lambda_{\varphi^+\varphi^-\varphi^+\varphi^-}
   \equiv\frac{2\lambda}{L^2},\label{ngself}\\
 &&\lambda_{HHHH}\equiv\frac{12\lambda}{L^2}c_\Delta^4,\hspace{5mm}
   \lambda_{HH\chi\chi}=\lambda_{HH\varphi^+\varphi^-}
   \equiv\frac{2\lambda}{L^2}c_\Delta^2,\label{HHHHandHHcc}
\end{eqnarray}
for 4-point couplings, $\chi(x)^4$, $\chi(x)^2\varphi^+(x)\varphi^-(x)$, 
$(\varphi^+(x)\varphi^-(x))^2$, $\phi^q(x)^4$, $\phi^q(x)^2\chi(x)^2$, and 
$\phi^q(x)^2\varphi^+(x)\varphi^-(x)$. Note that symmetric factors $1/N!$ 
have been taken into account for interactions $X^N$, where $X$ is the physical Higgs or
NG boson. The 3-point couplings for $\phi^q(x)^3$, $\phi^q(x)\chi(x)^2$, and 
$\phi^q(x)\varphi^+(x)\varphi^-(x)$ can also be obtained by
\begin{eqnarray}
 \lambda_{HHH}\equiv\frac{6v_{\mbox{{\scriptsize EW}}}\lambda}{L^2}
                    c_\Delta^3,\hspace{5mm}
 \lambda_{H\chi\chi}=\lambda_{H\varphi^+\varphi^-}
 \equiv\frac{2v_{\mbox{{\scriptsize EW}}}\lambda}{L^2}c_\Delta.
 \label{3self}
\end{eqnarray}

\begin{figure}
\begin{center}
\includegraphics[scale = 1.0]{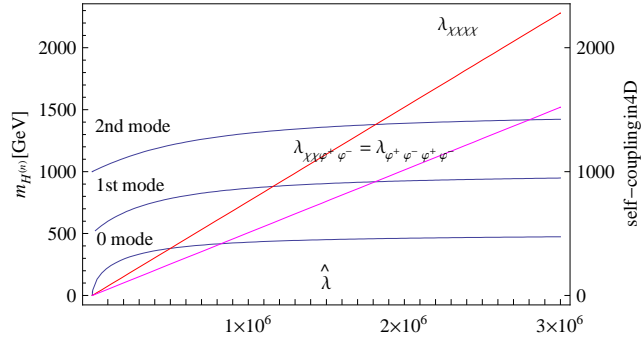}
\end{center}  
\caption{Dependences of NG self-couplings on the magnitude of boundary 
coupling. Input parameters are 
$(\hat{m},\hat{L},\Lambda,v)=(0,20\pi,10\mbox{TeV},174\mbox{GeV}/\sqrt{L})$.}
\label{self}
\end{figure}

\begin{figure}
\begin{center}
\includegraphics[scale = 1.0]{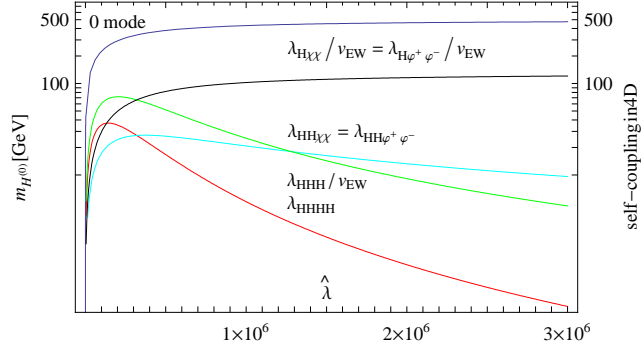}
\end{center}  
\caption{Dependences of Higgs self-couplings on the magnitude of boundary 
coupling. Input parameters are 
$(\hat{m},\hat{L},\Lambda,v)=(0,20\pi,10\mbox{TeV},174\mbox{GeV}/\sqrt{L})$. 
For references, the 0 mode Higgs mass is given.}
\label{hself}
\end{figure}

In the large limit of $\lambda$ i.e.\ $\Delta_0\rightarrow 1$, the coefficient $c_\Delta$ becomes
a suppression factor which is approximately proportional to $\pi/(2\lambda 
v^2L)$. 
We find that the $\lambda_{HHHH}$, $\lambda_{HHH}$, $\lambda_{HH\chi\chi}$, and
$\lambda_{HH\varphi^+\varphi^-}$ become small for larger value of $\lambda$ 
because $c_\Delta\rightarrow0$ in the limit and these couplings include higher 
power of $c_\Delta$ than $\lambda$. On the other hand the couplings 
$\lambda_{H\chi\chi}$ and $\lambda_{H\varphi^+\varphi^-}$ remain finite. 
In the limit of the large boundary coupling, the total potential energy would be lower when the Higgs profile is expelled from the boundary than when the Higgs stays at boundary. Therefore, the 3 and 4-point coupling for
the physical Higgs are suppressed while the 4-point couplings for the NG boson
are merely proportional to the boundary coupling. This difference between the 
Higgs and NG boson is just caused by the presence of the brane potentials as we mentioned. The large $\lambda$ modifies only the Higgs profile and the NG 
profile is not sensitive to the magnitude of $\lambda$. These features are 
confirmed by numerical calculations given in Figs.~\ref{self} and \ref{hself}. 
The masses of 0th, 1st, and 2nd KK Higgs are also presented in Figs.~\ref{self} and \ref{hself} for references. One should read left graduation for each Higgs mass, which is given in GeV unit, and right one for the self-couplings of Higgs and NG bosons in 4-dimensions. The horizontal axis corresponds to dimensionless coupling in 5-dimensions $\hat{\lambda}$. Note that 3-point couplings, which is dimensionful, is normalized to be  dimensionless by dividing by $v_{\mbox{{\scriptsize EW}}}$.

%%%%%%%%%%%%%%%%%%%%%%%%%%%
\subsubsection{Higgs decay}
%%%%%%%%%%%%%%%%%%%%%%%%%%%

Next, we consider decays of Higgs boson. When kinematically accessible and 
$m_H>140$GeV, the decay of the Higgs into $WW$ dominates over all other decay 
modes. The decay process of $H\rightarrow b\bar{b}$ dominates in the region 
of $m_H<140$GeV \cite{hpro}. 

Here, we argue the decay process especially for $H\rightarrow W_LW_L$. The 
decay width is given by
\begin{eqnarray}
 \Gamma(H\rightarrow W_LW_L)
 =\frac{\lambda_{H\varphi^+\varphi^-}v_{\mbox{\scriptsize EW}}G_F}{2\sqrt{2}}
  m_H\left(1-\frac{2m_W^2}{m_H^2}\right)^2
  \left(1-\frac{4m_W^2}{m_H^2}\right)^{1/2}.
\end{eqnarray}
Since $\lambda_{H\varphi^+\varphi^-}$ is given in (\ref{3self}) and $c_\Delta$ 
is proportional to $\pi/(2\lambda v^2L)$ at the large limit of $\lambda$, the 
coupling goes to $m_{KK}^2/\pi v_{\mbox{{\scriptsize EW}}}$ at this limit, 
where we used $m_{KK}=\pi/L$ and $v=v_{\mbox{{\scriptsize EW}}}/\sqrt{L}$. The 
decay width of $H\rightarrow W_LW_L$ at the large limit of the boundary 
coupling can be roughly estimated as
\begin{eqnarray}
 \Gamma(H\rightarrow W_LW_L)\simeq\frac{G_Fm_{KK}^2m_H}{2\sqrt{2}\pi}
                            \simeq164\left(\frac{m_{KK}}{500\mbox{GeV}}
                                  \right)^2\mbox{GeV},
\end{eqnarray} 
where we used $m_H\simeq m_{KK}$. We find that a peak in the cross section of 
this process appears even if the boundary coupling is large. 
\begin{figure}
\begin{center}
\includegraphics[scale = 1.0]{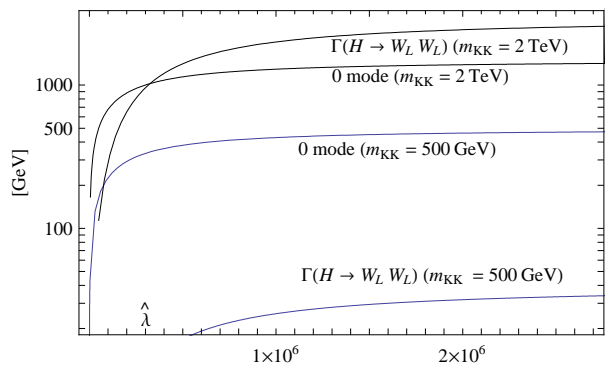}
\end{center}
\caption{Dependence of $\Gamma(H\rightarrow W_LW_L)$ on the boundary coupling 
and $m_{KK}$. Input parameters are 
$(\hat{m},\Lambda,v)=(0,10\mbox{TeV},174\mbox{GeV}/\sqrt{L})$. For references, 
the 0 mode Higgs masses for each $m_{KK}$ are shown. The horizontal axis is 
dimensionless coupling in 5-dimensions, $\hat{\lambda}$.}
\label{dec}
\end{figure}
However, the width is larger than the Higgs mass, $\Gamma_{HW_LW_L}>m_H$, when 
$m_{KK}$ is larger than few TeV. This means that the Higgs decays rapidly even 
if it can be produced, and thus, it is difficult to find the peak, equivalently
to capture the Higgs as a particle. These features for the cases of 
$m_{KK}=500$GeV and $2$TeV can be numerically checked in Fig.~\ref{dec}.
We can conclude that a particle description of the Higgs tends to get worse 
for larger boundary coupling and KK scale.

%%%%%%%%%%%%%%%%%%%%%%%%%%%%%%%
\subsection{Gauge boson masses}
%%%%%%%%%%%%%%%%%%%%%%%%%%%%%%%
We discuss the masses of $Z$ and $W$ bosons. They are derived from the 
Higgs kinetic term, which is given by 
\begin{eqnarray}
 S_{kin}&=&\int d^4x\int_{-L/2}^{+L/2}dz|D_M\Phi|^2,
\end{eqnarray}
where we take the gauge fields as bulk ones and define the covariant 
derivative as
\begin{eqnarray}
 D_M&\equiv&\partial_M+ig_5W_M^aT^a+ig_5'B_MY\label{cd}\nonumber\\
    &=&\partial_M+i\frac{g_5}{2}\left(
       \begin{array}{cc}
       W_M^3        & W_M^1-iW_M^2 \\
       W_M^1+iA_M^2 & -W_M^3 
       \end{array}\right)-i\frac{g_5'}{2}\left(
       \begin{array}{cc}
       B_M & 0   \\
       0   & B_M 
       \end{array}\right).
\end{eqnarray}
The $g_5$ and $g_5'$ are the gauge couplings in five dimensions. We have also 
taken the hyper charge as $Y\Phi=(1/2)\Phi$. The wave functions of gauge bosons
are defined by
\begin{eqnarray}
 Z_M\equiv\frac{g_5W_M^3-g_5'B_M}{\sqrt{g_5^2+g_5'{}^2}},\hspace{5mm}
 A_M\equiv\frac{g_5W_M^3+g_5'B_M}{\sqrt{g_5^2+g_5'{}^2}},\hspace{5mm}
 W_M^\pm\equiv\frac{W_M^1\mp iW_M^2}{\sqrt{2}}.
\end{eqnarray}

For estimation of the $Z$ boson mass, we show the profile of $Z$ in the extra 
dimensional direction. We take the KK expansion of $Z$ boson as 
\begin{eqnarray}
 Z_\mu=\sum_{n=0}^{\infty}f_n^Z(z)Z_\mu^{n}(x).
\end{eqnarray}
The KK equation and BCs are given as  
\begin{eqnarray}
 &&\left[\partial_z^2-\frac{v^2(z)}{2}(g_5^2+g_5'{}^2)\right]f_n^Z(z)
   =-\mu_{Z,n}^2f_n^Z(z),\label{81}\\
 &&\mp\partial_zf_n^Z(z)=0,\label{82}
\end{eqnarray}
where we have taken the Neumann type boundary conditions. Since $v(z)$ is defined as 
$v(z)\equiv\Phi^c(z)$, we consider the equation of motion for $\Phi(z)$ with 
BCs given in (\ref{l6z}) and (\ref{l7z}). We obtain the mass of $Z$ boson 
from (\ref{81}) as 
\begin{eqnarray}
 m_Z^2&\equiv&\int_{-L/2}^{+L/2}dzf_0^Z(z)
              \left[\frac{v^2(z)}{2}(g_5^2+g_5'{}^2)-\partial_z^2\right]
              f_0^Z(z),\label{zmass}\\
      &=&\mu_{Z,0}^2,
\end{eqnarray}
where we normalize the wave-function profile of the extra dimensional direction
for $Z$ boson field as
\begin{eqnarray}
 \int_{-L/2}^{L/2}dzf_n^Z(z)f_m^Z(z)=\delta_{n,m}.
\end{eqnarray}
The gauge coupling constants in five dimensions are related to the ones in 
four dimensions and the size of the extra dimension $L$ as
\begin{eqnarray}
 g_5 &=&g_4\sqrt{L},\\
 g_5'&=&g_4'\sqrt{L}.
\end{eqnarray} 
Therefore, the mass of Z boson is rewritten as
\begin{eqnarray}
 m_Z^2&=&\int_{-L/2}^{+L/2}dzf_0^Z(z)\left[\frac{v^2(z)}{2}(g_4^2+g_4'{}^2)L
         -\partial_z^2\right]f_0^Z(z),
 \label{zmass1}
\end{eqnarray}
with gauge couplings in four dimensions. Here a flat profile of $v(z)$ and zero
mode of the $Z$ boson, $f_0^Z(z)$, are required in order that the $Z$ boson 
mass can be reproduced in four dimensional SM where the mass is given as 
$m_Z^2=(g_4^2+g_4'{}^4)v_{\mbox{\scriptsize EW}}^2/2$ with 
$v_{\mbox{\scriptsize EW}}\simeq174$GeV. When the VEV profile $v(z)$ is enough 
flat, it is well approximated as $v(z)\simeq v$ where we have taken as
$v\equiv174\mbox{GeV}/\sqrt{L}$. Furthermore, when the profile $f_0^Z(z)$ is 
sufficiently flat, the magnitude of contribution to $Z$ boson mass from 
$f_0^Z(z)\partial_z^2f_0^Z(z)$ is negligibly small. Such situation of flat VEV 
profile is easily realized by taking a small bulk mass as discussed above. The 
reflection symmetry for the boundary potentials is important for these 
discussions. If the reflection symmetry is broken, the VEV profile is not 
generally flat, and thus, it is difficult to reproduce the gauge boson masses.
Actually, the $Z$ boson mass in Eq.~(\ref{zmass1}) cannot reproduce the 
experimental value $m_Z=91.1876\pm0.0021$GeV \cite{pdg}, when $m$ is larger 
than $2.65\times 10^{-3}$GeV and 
$(m_{KK},\Lambda,\hat{\lambda},v)=(500\mbox{GeV},10\mbox{TeV},3\times10^6,174\mbox{GeV}/\sqrt{L})$ are taken. 

The $W$ boson mass is similarly obtained to the 
case of $Z$ boson one. 
The KK equation for the $W$ boson with the BCs are given as  
\begin{eqnarray}
 &&\left[\partial_z^2-\frac{v^2(z)}{2}g_5^2\right]f_n^W(z)
   =-\mu_{W,n}^2f_n^W(z),\\
 &&\mp\partial_zf_n^W(z)=0.
\end{eqnarray}
The similar discussions give the mass of $W$ boson as 
\begin{eqnarray}
 m_W^2\equiv\int_{-L/2}^{+L/2}dzf_0^W(z)\left[\frac{v^{2}(z)}{2}g_4^2L
            -\partial_z^2\right]f_0^W(z).
\end{eqnarray}

As the results, the masses of gauge bosons at the limit of flat VEV profile are obtained 
\begin{eqnarray}
 m_{Z^{(n)}}^2&=&m_Z^2+\frac{n^2}{R^2},\\
 m_{W^{(n)}}^2&=&m_W^2+\frac{n^2}{R^2}.
\end{eqnarray}
The wave function profiles are represented by (\ref{owp}).

%%%%%%%%%%%%%%%%%%%%%%%%%%%%%%%%%
\subsection{KK number and parity}
%%%%%%%%%%%%%%%%%%%%%%%%%%%%%%%%%

We comment on the KK number and parity. It is known that the UED model has 
conservations of KK number and parity, which weaken constraints on the KK scale
from EW precision measurements. In our setup with
boundary Higgs potential, discussions for the conservations are modified. 
As we have mentioned in the considerations for the Higgs mass, the boundary 
potentials cause the mode mixing in the Higgs sector generally, and thus, the 
KK number conservation is completely broken in the Higgs sector. However, when 
we take the identical boundary potentials for both $z=\pm L/2$ branes, there is a reflection symmetry between the branes. This reflection symmetry guarantees the conservation of the KK parity in the gauge and Higgs sectors. The reflection symmetry breaking causes the KK parity violation in both sectors because a difference between the brane potentials generally gives non-flat VEV profile which does not guarantee the parity in the gauge sector. Since the gauge fields are bulk ones throughout this paper, the KK number is always conserved in the gauge sector. 

It is important for investigations of phenomenological implications of this 
setup whether the KK number and parity are conserved in the fermion sector or 
not. For these studies, we consider two kinds of scenario in the following 
sections. The first one is a case of Brane-Localized Fermion (BLF) scenario, 
and the other is Bulk Fermion (BF) one. Both KK number and parity in the 
fermion sector are broken in the BLF scenario, while both are conserved in the 
BF case. We will show KK number and parity conservations in gauge, Higgs, and 
fermion sectors in each model. We show the results in Table~\ref{tab1}.
\begin{table}[t]
\hspace{2.75cm}[Conserved Reflection Symmetry]
\hspace{9.5mm}[Broken Reflection Symmetry]\vspace{2mm}
\begin{center}
\begin{tabular}{|c||c|c|c|}
\hline
       & BLF      & BF   & UED  \\
\hline
\hline
Gauge   & N, P     & N, P & N, P \\
\hline
Higgs   & P        & P    & N, P \\
\hline
Fermion & $\times$ & N, P & N, P \\
\hline  
\end{tabular}\hspace{1cm}
\begin{tabular}{|c||c|c|c|}
\hline
       & BLF      & BF       & UED  \\
\hline
\hline
Gauge   & N        & N     & N, P \\
\hline
Higgs   & $\times$ & $\times$ & N, P \\
\hline
Fermion & $\times$ & N     & N, P \\
\hline  
\end{tabular}
\end{center}
\caption{The KK number and parity conservations: N and P means the presence of 
KK number and parity conservations, respectively.}
\label{tab1}
\end{table}

%%%%%%%%%%%%%%%%%%%%%%%%%%%%%%%%%%%%%%%%%%%%%%%%%%%
\section{Scenario I: Brane-Localized Fermion (BLF)}
%%%%%%%%%%%%%%%%%%%%%%%%%%%%%%%%%%%%%%%%%%%%%%%%%%%

We discuss phenomenological implications of the presence of brane potentials in
 the Brane-Localized Fermion (BLF) scenario where the gauge bosons are spreaded in the bulk and all the SM fermions are localized on brane. Especially, the deviation of the coupling between the top quark and Higgs boson from the naive standard-model expectation, the Higgs production at the LHC, and EW constraints are considered.

%%%%%%%%%%%%%%%%%%%%%%%%%%%%%%%%%
\subsection{Top Yukawa deviation}
%%%%%%%%%%%%%%%%%%%%%%%%%%%%%%%%%

The top Yukawa deviation generally occurs in the minimal supersymmetric standard model because there are two Higgs doublets. 
On the other hand, there is no such deviation in the UED model which contains only one Higgs doublet. 
We show that the top deviation can occur though there exists only one Higgs doublet in the BLF scenario.

The brane potentials with large boundary coupling may affect the Higgs couplings as we discussed in the previous section. Especially, it is expected that the effect on the Yukawa coupling of top quark with the Higgs is significant and we consider its deviation from the top Yukawa coupling in the four dimensional standard model.

In the BLF scenario, the Yukawa interaction for the top quark and the 
Higgs boson are written as
\begin{eqnarray}
 -\mathcal{L}_t
 &=&\int_{-L/2}^{+L/2}dz\delta(z-L/2)y_{t,5}\left[v(z)
    +f_0(z)\frac{\phi^q(x)}{\sqrt{2}}\right]\bar{t}(x)t(x)+\text{h.c.}
    \label{Yukawa1}\\
 &=&y_{t,5}\left[v(L/2)
    +f_0(L/2)\frac{\phi^q(x)}{\sqrt{2}}\right]\bar{t}(x)t(x)+\text{h.c.},
 \label{Yukawa}
\end{eqnarray}
where we take an even mode of the VEV profile.\footnote{
%Note that since the Yukawa coupling $y_{t,5}$ appearing in~(\ref{Yukawa1}) is 
%one in five dimensions, it is dimensionful parameter, whose mass dimension is 
%$[y_{t,5}]=-1/2$. 
Instead we can put fermions on both branes with exactly the equal 
brane-localized Lagrangians to each other. In that case, it would correspond to
 further halving the fundamental region to $0\leq y\leq L/2$, with appropriate 
redefinition of the effective Yukawa coupling by factor 2.} The mass of the top
 quark is defined as
\begin{eqnarray}
 m_t\equiv y_{t,5}v(L/2).
\end{eqnarray}
If the profile of the VEV is enough flat, 
$v(L/2)\simeq v=174\,\text{GeV}/\sqrt{L}$. Therefore, the effective top Yukawa
coupling in four dimensions $y_t$ is defined by
\begin{eqnarray}
 y_t\equiv\frac{m_t}{v\sqrt{L}}\simeq\frac{y_{t,5}}{\sqrt{L}}.
 \label{ysm}
\end{eqnarray}
On the other hand, the coupling between the top quark and Higgs boson in four 
dimensions is given as follow,
\begin{eqnarray}
 y_{t\bar{t}H}\equiv y_{t,5}f_0(L/2).
 \label{ytth}
\end{eqnarray}
Here we define the following ratio,
\begin{eqnarray}
 r\equiv\frac{y_{t\bar{t}H}}{y_t}.
\end{eqnarray}
This ratio corresponds to the deviation of the coupling between the top quark 
and Higgs boson from the top Yukawa coupling in the four dimensional SM. The ratio can be calculated from 
(\ref{ysm}) and~(\ref{ytth}) as
\begin{eqnarray}
 r\simeq f_0(L/2)\sqrt{L}=\hat{f}_0(L/2)\sqrt{L\Lambda}
  =\hat{f}_0(L/2)\sqrt{\hat{L}}.
\end{eqnarray}  

\begin{figure}[tbp]
\begin{center}
\includegraphics[scale = 1.0]{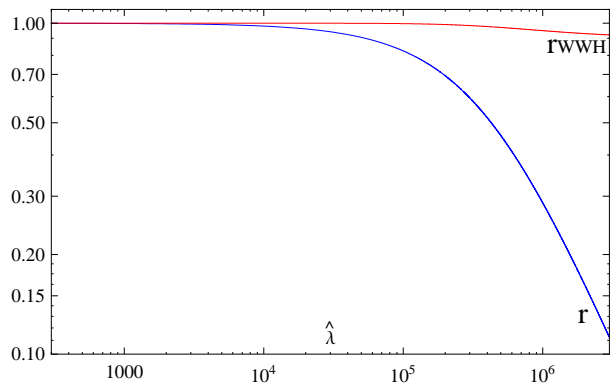}\hspace{4mm}
\end{center}
\caption{$\hat{\lambda}$ dependences of $r$ and $r_{WWH}$: Input parameters are
$(\hat{m},m_{\text{KK}})=(0,4\,\text{TeV})$. $r_{WWH}$ is defined in the next 
subsection.}
\label{fig5}
\end{figure}

Numerical calculation of this deviation is shown Fig.~\ref{fig5}. 
In the figure, $r=1$ means that there is no top deviation and $r=0.7$ 
corresponds to $30\%$ of deviation. It is seen that $50\%$ of deviation can be 
realized at $\hat{\lambda}\simeq5\times10^5$ and more larger $\hat{\lambda}$ 
leads to almost vanishing coupling between the top quark and Higgs boson, 
$r\ll1$. It would be achievable to find such large deviation at the LHC. We 
take the KK scale as $4\,\text{TeV}$ in the analysis. The magnitude of the deviation, 
equivalently the value of $|f_0(L/2)|$, is sensitive to only the size of the 
boundary coupling $\hat{\lambda}$ (see the discussions in Sec.2.2.1). The 
ref.\cite{Duhrssen:2004cv} gives a summary of extracting Higgs boson coupling 
from LHC data. Since various electroweak measurements constraint on the scale 
to be of order a few TeV~\cite{kk1}--\cite{kk5}, our predictions for the 
deviation of the coupling are very interesting ones which should be searched at the LHC.

We comment on a theoretical constraint from the Naive Dimensional Analysis 
(NDA) in higher dimensions~\cite{nda1,nda2}. According to the discussion of NDA, the coefficients of brane-localized terms are determined by requiring that all loop expansion parameters are of order one in four dimensional KK computation. That is applied to the coupling $\hat{\lambda}$, which mainly determines the magnitude of the 4-point coupling of NG boson. The NDA discussion constraints the four points coupling as 
\begin{eqnarray}
 \hat{\lambda}\leq\frac{8\pi^2\hat{L}^2}{3},
 \label{ndacon} 
\end{eqnarray}
where we referred to~(\ref{4ng}). When we take $\hat{L}=5\pi/2$ and 
$\Lambda=10\,\text{TeV}$, whose values are based on the numerical analyses 
given in Fig.~\ref{fig5} and which lead to $m_{\text{KK}}=4\,\text{TeV}$. By 
using these values, the NDA constraint gives $\hat{\lambda}<1.6\times10^3$. 
Therefore, if we respect with the NDA discussions, the top deviation is tiny. 
It is easy to understand that as follows. 

The BCs for the zero mode Higgs are given by
\begin{eqnarray}
 k_0\sin\left(\frac{k_0L}{2}\right)
 +4\lambda v^2\cos\left(\frac{k_0L}{2}\right)=0.
 \label{bcwd}
\end{eqnarray} 
At the small limit of $\lambda$, the value of $k_0$ is almost determined by the
 first term of~(\ref{bcwd}). It leads to $k_0\simeq0$, which corresponds to a 
flat profile of Higgs typically shown in left upper figure of 
Table~\ref{tab-pro}, and thus, there is no top deviation. On the other hand, a 
solution $k_0\simeq\pi/L$ is driven only by the second term of~(\ref{bcwd}) at 
the large limit of $\lambda$. This solution make the profile as a right upper 
figure of Table~\ref{tab-pro}, and the top deviation is significant. Therefore,
 the condition that there is a sizable top deviation is typically given by
\begin{eqnarray}
 \frac{\pi}{L}\ll 4\lambda v^2,
 \label{omosiro}
\end{eqnarray}
where we compare the magnitude of coefficient of the first term in~(\ref{bcwd}) with one of the second term, and substitute $k_0\simeq\pi/L$ after solving the equation at the large limit of $\lambda$. We can rewrite the condition~(\ref{omosiro}) as
\begin{eqnarray}
 1\ll\frac{1}{4\pi^2}\left(\frac{m_{\text{KK}}}{v_{\text{EW}}}\right)^2
 \ll\frac{\hat{\lambda}}{\hat{L}^2},
 \label{omosiro1}
\end{eqnarray} 
where we used $m_{\text{KK}}=\pi/L$ and $v=v_{\text{EW}}/\sqrt{L}$. It 
is easily seen that the condition~(\ref{omosiro}) for a top deviation cannot be consistent with the constraint from the NDA discussion~(\ref{ndacon}).

In the region of sizable top deviation, the 4-point coupling of the NG boson, 
which corresponds to that of the longitudinal gauge boson, becomes large as shown in~(\ref{4ng}). Since the large coupling is required for realization of top deviation, it seems that the perturbative unitarity does not work. However, a unitarity in a non-perturbative dynamics may be achieved because the EW symmetry breaking is caused by the usual Higgs mechanism. The BLF scenario shares the feature of strongly interacting model. An analysis of elastic scattering of the longitudinal mode through the vector-fusion and $Q\bar{Q}$ processes at the LHC has been discussed in a model of minimal strongly interacting EW symmetry breaking sector~\cite{Dobado:1999xb}.

%As mentioned above, a region where 
%$\hat{\lambda}$ is large, $\hat{\lambda}\sim500$, is phenomenologically 
%interesting, especially for a deviation of the coupling between the top quark and
% Higgs bosons from the top Yukawa coupling. Such a large $\hat{\lambda}$ 
%constraints on $\hat{L}$ to $\hat{L}\geq4.4$. This means that a typical KK 
%scale is less than $2.16$ TeV in this scenario.

%%%%%%%%%%%%%%%%%%%%%%%%%%%%%%%%%%%%%%%%
\subsection{Higgs production at LHC}
%%%%%%%%%%%%%%%%%%%%%%%%%%%%%%%%%%%%%%%%

\begin{figure}[t]
\begin{center}
\hspace{-5mm}(a) gluon fusion\hspace{2.6cm}(b) $WW$ fusion

\includegraphics[scale = 0.8]{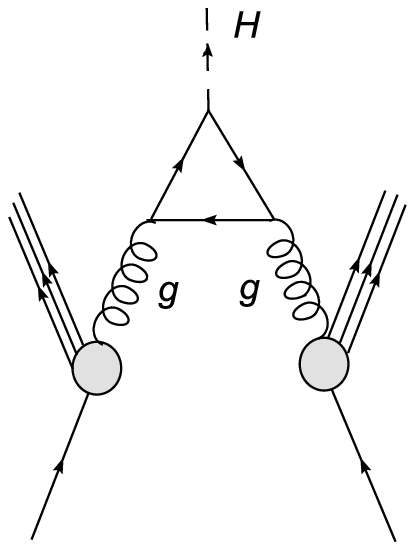}\hspace{2cm}
\includegraphics[scale = 0.8]{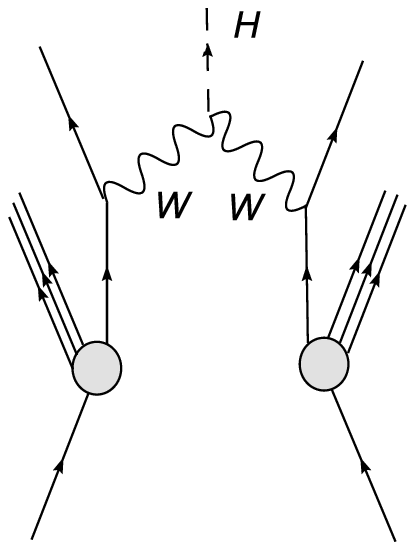}
\end{center}
\caption{Dominant process of Higgs production at the LHC}
\label{fig7}
\end{figure}

Next, we consider the Higgs production at the LHC experiment. The SM predicts 
that the gluon fusion through a one-loop diagram, which is the triangle heavy 
quark loop in Fig.~\ref{fig7}~(a), is the dominant process for the Higgs 
production when $m_H\leq600\,\text{GeV}$, $\sqrt{s}=14\,\text{TeV}$ and 
$m_t\simeq 175$ GeV and that the $WW$ fusion is a subdominant 
process~\cite{hpro}. The gluon fusion cross section is 10 times larger than the
 $WW$ fusion one when $m_H \leq 600\,\text{GeV}$. 

Since the coupling between top quark and physical Higgs boson is suppressed 
for large $\hat{\lambda}$ (see Fig.~\ref{fig5}), the processes for the Higgs production must be reconsidered in the case. The Higgs production cross section through the gluon fusion is determined by the magnitude of $|y_{t\bar{t}H}|^2$. The SM takes this coupling between the top quark and Higgs as the top Yukawa coupling $y_t$, that is, $r=1$. However, the BLF scenario induces deviations of $r$ from unity. This means that the cross section for the Higgs production though the gluon fusion are suppressed by $r^2$ in the case of large $\hat{\lambda}$. On the other hand, the cross section for the production through the $WW$ fusion also changes because the coupling between the $W$ and Higgs bosons is modified. The interaction is written by
\begin{eqnarray}
 -\mathcal{L}_{WWH}
 &=&\frac{em_W}{2\sin\theta_W}\frac{1}{2L}
    \int_{-L/2}^{+L/2}dzf_0(z)f_0^{W^+}(z)f_0^{W^-}(z)\phi^q(x)W^+(x)W^-(x)
    +\text{h.c.},\nonumber\\
 & &         
\end{eqnarray}
where $\theta_W$ is the Weinberg angle. When we write the ratio of the $WWH$ coupling in the five dimensional model to the one in four dimensional SM as $r_{WWH}$, the ratio is given by
\begin{eqnarray}
 r_{WWH}\equiv\frac{1}{\sqrt{L}}\int_{-L/2}^{+L/2}dzf_0(z),
 \label{rwwh}
\end{eqnarray}
where we have approximated that the extra dimensional profile of the Higgs VEV is flat and that, as a consequence, the zero mode profile of the $W$ boson is also flat $f_0^{W^\pm}(z)=1/\sqrt{L}$.
The cross section for the Higgs production through the $WW$ fusion in the BLF 
scenario is suppressed by $r^2_{WWH}$ from the one in the SM. The numerical 
evaluation of the ratio is given in Fig.~\ref{fig5}. The magnitude of this ratio is $\simeq1$ at a small $\hat{\lambda}$, and $\simeq0.92$ at a large boundary coupling. 

Let us summarize the above discussions for the Higgs production at the LHC. {}For $\hat{\lambda}$ of order one, the ratios of the production cross section through the gluon and $WW$ fusions to the corresponding ones in the SM are given by
\begin{eqnarray}
 r^\sigma_{g}&\equiv&\frac{\sigma_{5,g}}{\sigma_{g}}=r^2\simeq1,\\
 r^\sigma_{W}&\equiv&\frac{\sigma_{5,W}}{\sigma_{W}}=r_{WHH}^2\simeq1,
\end{eqnarray}
respectively, where $\sigma_{5,\ast}$ and $\sigma_\ast$ (with $\ast=g,W$) are the cross sections in five dimensions and SM, respectively. In this case, the discussions for the Higgs production at the LHC are the same as in the SM. On the other hand, these ratios in the case of a large boundary coupling, say $\hat{\lambda}=3\times10^6$, are estimated as
\begin{eqnarray}
 r^\sigma_{g}&=&10^{-2}\left(\frac{r}{0.1}\right)^2,\\
 r^\sigma_{W}&=&0.85\left(\frac{r_{WHH}}{0.92}\right)^2.
\end{eqnarray}
In the case, magnitudes of cross section for the Higgs production through the 
gluon and $WW$ fusion become $1\%$ and $85\%$ of the SM prediction. 
Furthermore, the dominant process of the production is not the gluon fusion but the $WW$ one. This is also important prediction of this setup in addition to the top deviation discussed in the previous subsection.

%%%%%%%%%%%%%%%%%%%%%%%%%%%%%%%%%%%%%%%%%
\subsubsection{Higher KK Higgs production}
%%%%%%%%%%%%%%%%%%%%%%%%%%%%%%%%%%%%%%%%%
Finally, let us consider higher KK Higgs production. 
In the above discussions, both the Higgs boson $H^{(0)}$ and the top quark $t^{(0)}$, circulating the triangle loop, have been zero mode. 
However, it is generally possible to consider a higher KK Higgs boson and 
heavy KK quarks in this process. In this case, generically the coupling $Q^{(n)}\bar{Q}^{(m)}\rightarrow H^{(l)}$ is employed at final 3-point vertex. These processes are KK-number and parity violating ones but they are possible in the BLF scenario. They would be important processes for future higher KK Higgs searches. The higher KK Higgs productions through KK quark loops in the BLF scenario with a large $\hat{\lambda}$ is suppressed by the same reason as that for the lowest mode Higgs production discussed above. In the case of the 
production through the $WW$ fusion, only the following processes are possible
since the KK parity is always conserved:
\begin{eqnarray}
 W^{(2n)}W^{(2m)}    &\rightarrow&H^{(2(l+1))},\\
 W^{(2n)}W^{(2m+1)}  &\rightarrow&H^{(2l+1)},\\
 W^{(2n+1)}W^{(2m+1)}&\rightarrow&H^{(2(l+1))}. 
\end{eqnarray}
These discussions are summarized in Table~\ref{comp}.

\begin{figure}[t]
\begin{center}
\includegraphics[scale = 1.0]{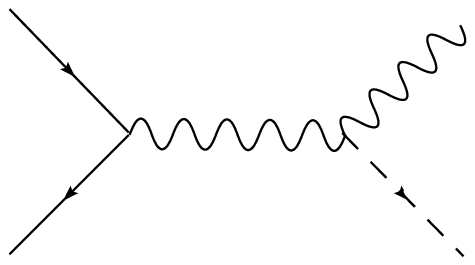}\hspace{2cm}
\includegraphics[scale = 1.0]{assoh.eps}
\end{center}\vspace{-3cm}

\hspace{1.4cm}$e^{-(0)}$\hspace{4.9cm}$Z^{(2m)}$
\hspace{2mm}$e^{-(0)}$\hspace{4.9cm}$Z^{(2m+1)}$

\hspace{4.2cm}$Z^{(2n)}$\hspace{6cm}$Z^{(2n)}$\vspace{1cm}

\hspace{1.4cm}$e^{+(0)}$\hspace{4.9cm}$H^{(2l)}$
\hspace{2mm}$e^{-(0)}$\hspace{4.9cm}$H^{(2l+1)}$
\begin{center}
\includegraphics[scale = 1.0]{assoh.eps}\hspace{2cm}
\includegraphics[scale = 1.0]{assoh.eps}
\end{center}\vspace{-3cm}

\hspace{1.4cm}$e^{-(0)}$\hspace{4.9cm}$Z^{(2m)}$
\hspace{1.5mm}$e^{-(0)}$\hspace{4.9cm}$Z^{(2m+1)}$

\hspace{4.2cm}$Z^{(2n+1)}$\hspace{5.5cm}$Z^{(2n+1)}$\vspace{1cm}

\hspace{1.4cm}$e^{+(0)}$\hspace{4.9cm}$H^{(2l+1)}$
\hspace{-1mm}$e^{-(0)}$\hspace{4.9cm}$H^{(2l)}$
\caption{Associated KK Higgs production in the BLF scenario with KK parity, where $e^{-(0)}$, $e^{+(0)}$, $Z^{(n)}$ and $H^{(l)}$ are the SM electron, positron, $n$ and $l$ modes of $Z$ and Higgs bosons, respectively, with $l,m,n=0,1,2,\cdots$.}
\label{pfigblf}
\end{figure}

\begin{figure}[t]
\begin{center}
\includegraphics[scale = 1.0]{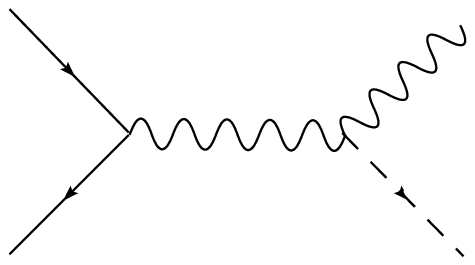}
\end{center}\vspace{-3cm}

\hspace{4.8cm}$e^{-(0)}$\hspace{4.9cm}$Z^{(m)}$

\hspace{7.6cm}$Z^{(n)}$\vspace{1cm}

\hspace{4.8cm}$e^{+(0)}$\hspace{4.9cm}$H^{(l)}$
\caption{Associated KK Higgs production in the BLF scenario without reflection 
symmetry.}
\label{pfig14}
\end{figure}

Higher KK Higgs can be also produced through an associated production process 
at a linear collider such as the International Linear Collider (ILC). 
In the BLF scenario with conserved KK parity, since the KK number is
violated in the fermion sector, the internal $Z$ boson can be higher KK mode, $Z^{(n)}$, shown in Fig.~\ref{pfigblf}. 
On the other hand, the KK parity is conserved in both gauge and Higgs sectors. 
Therefore, the following processes are allowed for the higher KK Higgs 
productions, 
\begin{eqnarray}
 Z^{(2n)}  &\rightarrow&Z^{(2m)}H^{(2l)},~Z^{(2m+1)}H^{(2l+1)},\\
 Z^{(2n+1)}&\rightarrow&Z^{(2m)}H^{(2l+1)},~Z^{(2m+1)}H^{(2l)}.
\end{eqnarray}
shown in Fig.~\ref{pfigblf}. If the reflection symmetry is broken in the BLF 
scenario, only the KK number of the gauge sector is conserved. All internal and
final states are possible,
\begin{eqnarray}
 Z^{(n)}\rightarrow Z^{(m)}H^{(l)},
\end{eqnarray} 
as given in Fig.~\ref{pfig14}.
Since the KK number is conserved in the UED, the internal $Z^{(n)}$ boson must 
be the zero mode $Z^{(0)}$ and final states $(Z^{(m)},H^{(l)})$ are limited to 
be equal $(Z^{(m)},H^{(m)})$. This could be checked at a future linear collider
 experiment.

%%%%%%%%%%%%%%%%%%%%%%%%%%%%%%%%%%%%%%%%%%%%%%%%%%%%%%%%%%%%%%%%
\subsection{Constraints from electroweak precision measurements}
%%%%%%%%%%%%%%%%%%%%%%%%%%%%%%%%%%%%%%%%%%%%%%%%%%%%%%%%%%%%%%%%

Next, let us consider constraints from EW precision measurements on the BLF 
scenario. The various EW measurements constrain the KK scale to be of order 
a few TeV~\cite{kk1}--\cite{kk5}. It is well known that a severe constraint on the scale in extra-dimensional model comes from the Fermi constant measured in the muon decay
\begin{eqnarray}
 G_F^{\text{SM}}=\frac{\pi\alpha}{\sqrt{2}m_W^{\text{(ph)}2}
                 \left(1-\frac{m_W^{\text{(ph)}2}}{m_Z^{\text{(ph)}2}}\right)
                 (1-\Delta r)},
\end{eqnarray}
where $\alpha$, $\Delta r$, $m_W^{\text{(ph)}}$, and $m_Z^{\text{(ph)}}$ are the fine 
structure constant, the SM radiative corrections, the measured masses of $W$ 
and $Z$ bosons, respectively. After compactification, integrating the $W$ 
boson and its KK states over extra-dimensional direction in this setup, the 
effective Fermi constant is affected as
\begin{eqnarray}
 G_F=G_F^{\text{SM}}\left[1-2\sum_{n=1}^\infty\frac{m_W^{\text{(ph)}2}}{(n/R)^2}\right],
\end{eqnarray}
where we identify $G_F$ with the experimentally measured value of the Fermi 
constant, $G_F=1.166367(5)\times10^{-5}$GeV$^{-2}$~\cite{Marciano:1988vm}--\cite{Barczyk:2007hp}. The experimental data puts a lower bound on the KK scale $m_{\text{KK}}>1.6\,\text{TeV}$. In addition to the discussion of the experimental bound from 
Fermi constant, some constraints from atomic parity violating experiments, 
leptonic width of the $Z$ boson, and $\rho$ parameter have been given in 
\cite{kk1}--\cite{kk5}. To conclude, the KK scale should be higher than a few TeV. 

It is well known that there is an useful parametrization to discuss 
constraints on a new physics such as the extra-dimensional model, which are $S$, 
$T$, and $U$ parameters~\cite{stu2,stu3}. $S$ and $T$ parameters are defined as
\begin{eqnarray}
 \alpha S&\equiv&4e^2\left[\Pi_{33}^{\text{new}}{}'(0)-\Pi_{3Q}^{\text{new}}{}'(0)\right],
 \label{spara}\\
 \alpha T&\equiv&\frac{e^2}{s^2c^2m_Z^2}[\Pi_{11}^{\text{new}}(0)
                     -\Pi_{33}^{\text{new}}(0)],\label{tpara}
\end{eqnarray}
where $c$, $s$, and $e$ are defined by
\begin{eqnarray}
 c  &\equiv&\cos\theta_W\equiv\frac{g_4}{\sqrt{g_4^2+g_4'{}^2}},\label{cos}\\
 s  &\equiv&\sin\theta_W\equiv\frac{g_4'}{\sqrt{g_4^2+g_4'{}^2}},\label{sin}\\
 e  &\equiv&\frac{g_4g_4'}{\sqrt{g_4^2+g_4'{}^2}}.\label{e}
\end{eqnarray}
Here $\Pi_{XY}(q^2)$ is the vacuum polarizations defined as
\begin{eqnarray}
 ig^{\mu\nu}\Pi_{XY}(q^2)+(q^\mu q^\nu\mbox{ terms})
 \equiv\int d^4e^{-iq\cdot x}\langle J_X^\mu(x)J_Y^\nu(0)\rangle,
\end{eqnarray}
and we define $\Pi'_{XY}(q^2)$ to be equal to $d\Pi_{XY}/dq^2$ only at $q^2=0$:
\begin{eqnarray}
 \Pi_{XY}(q^2)\equiv\Pi_{XY}(0)+q^2\Pi'_{XY}(q^2),
\end{eqnarray}
where $(XY)=(11),~(22),~(33),~(3Q)$, and $(QQ)$~\cite{stu2,stu3,stu1}. The 
$\Pi_{XY}^{\text{new}}(q^2)$ means the contribution of the new Higgs sector particles 
to $\Pi_{XY}(q^2)$. The $\Pi_{11}$ and $\Pi_{33}$ are represented by the 1PI 
self energies of the $W$, $Z$, photon and photon-$Z$ mixing as follows
\begin{eqnarray}
 \Pi_{11}&=&\frac{s^3}{e^2}\Pi_{WW},\\
 \Pi_{33}&=&\frac{s^3}{e^2}\left[c^2\Pi_{ZZ}+2sc\Pi_{ZA}+s^2\Pi_{AA}\right].
\end{eqnarray}

\begin{figure}[t]
\begin{center}
\hspace{3mm}(a)\hspace{5.8cm}(b)\vspace{1mm}

\hspace{3mm}$H$\hspace{6cm}$H$

$W$\includegraphics[scale = 1.0]{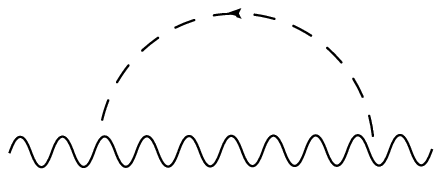}$W$\hspace{1cm}
$Z$\includegraphics[scale = 1.0]{fig38.eps}$Z$\vspace{2mm}

\hspace{4mm}(c)\hspace{5.8cm}(d)\vspace{1mm}

\hspace{4mm}$H$\hspace{6.cm}$H$

$W$\includegraphics[scale = 1.0]{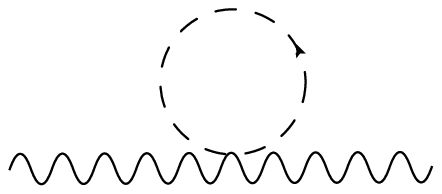}$W$\hspace{1.1cm}
$Z$\includegraphics[scale = 1.0]{fig39.eps}$Z$\vspace{2mm}

\hspace{4mm}(e)\hspace{5.8cm}(f)\vspace{2mm}

\hspace{4mm}$H$\hspace{6cm}$H$

\hspace{4mm}\includegraphics[scale = 1.0]{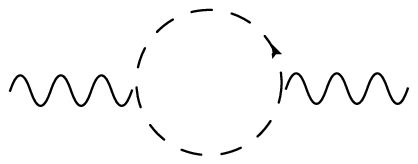}\hspace{2.1cm}
\includegraphics[scale = 1.0]{fig40.eps}\vspace{-8mm}

\hspace{3mm}$W$\hspace{4.1cm}$W$\hspace{1.5cm}$Z$\hspace{4.1cm}$Z$\vspace{1mm}

\hspace{4mm}$\chi$\hspace{6.1cm}$\chi$
\end{center}
\caption{Contributions to the $T$ parameter from one-loop diagrams, with 
the Higgs propagating in internal lines.}
\label{fig9}
\end{figure}

\begin{figure}
\begin{center}
(a)\hspace{3.3cm}(b)

\includegraphics[scale = 1.0]{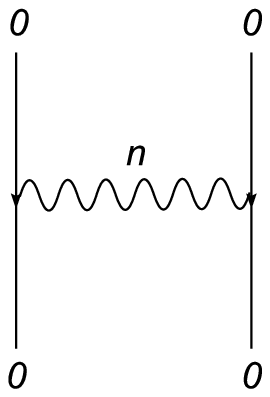}\hspace{1cm}
\includegraphics[scale = 1.0]{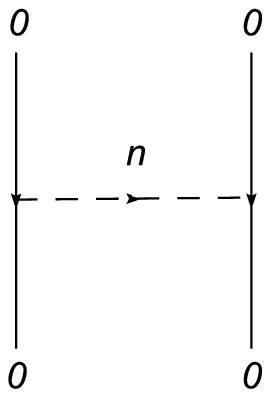}
\end{center}
\caption{Characteristic interactions in the BLF scenario: $0$ and $n$ stand for
the KK number.}
\label{blf-flow}
\end{figure}

Evaluations of these parameter is the most powerful method to consider whether or not a new physics is allowed by various electroweak precision measurements. 
Since the Higgs sector is generalized from the usual UED model in our setup, it should be carefully done to calculate diagrams including the Higgs loop shown in Fig.~\ref{fig9}.
However, there still exist tree level contributions to four Fermi interaction 
in the BLF scenario, shown in Fig.~\ref{blf-flow}~(a).
In Fig.~\ref{blf-flow}~(a), solid lines correspond to external currents and 
internal lines to gauge boson. $0$ and $n$ stand for the KK number. The KK 
number of the external currents should be $0$ because they are the standard 
model fermions. Such external currents can couple with the higher KK mode in 
the case of the extradimensional model without the KK number conservation such 
as our BLF scenario. These tree level couplings among $0$ and higher KK modes give severer constraints on the BLF scenario than the ones from one loop diagrams including Higgs. They give lower bound on the KK scale of the order of a few TeV~\cite{kk1}--\cite{kk5}. Therefore, we do not need to give tedious computations for $S$ and $T$ parameters in the BLF scenario. However, these constraints are modified in the BF scenario which has the KK parity conservation. Actual estimations of $S$ and $T$ parameter in the case will be presented in the next section.

There are also characteristic interactions in the BLF scenario shown in 
Fig.~\ref{blf-flow}~(b). An internal line is the Higgs. Since the conservation 
of KK number is violated in the fermion sector, the Higgs can also propagates 
on internal line. However, there is no such contribution to the muon decay processes because lepton flavor is conserved in our model and hence a flavor diagonal basis can always be taken without the loss of generality. 
Therefore, the diagram does not modify the constraints from the muon decay constant. Also the Flavor Changing Neutral Current (FCNC) process of Fig.~\ref{blf-flow}~(b) does not occur, by taking the flavor diagonal basis. 

\subsection{Summary of BLF scenario}
We have discussed the BLF scenario, where all standard model fermions are localized at a brane. The top Yukawa deviation have been quantitatively estimated, as the phenomenological results of the presence of brane potentials. We have pointed out that the deviation can be drastic in the case of a large boundary coupling.
We showed that the $50\%$ of deviation can be realized with large coupling and
more larger one leads to an enormous deviation. The discussions for the Higgs 
production at the LHC have also been presented. A large boundary coupling 
suppresses cross sections of the production through the gluon and $WW$ fusions 
to $1\%$ and $85\%$ of the SM prediction. Therefore, the $WW$ fusion becomes 
dominant process for the Higgs productions at the LHC. These predictions are 
consistent with the data of EW precision measurements and would be 
achievable to discover at the LHC.

%%%%%%%%%%%%%%%%%%%%%%%%%%%%%%%%%%%%%%%
\section{Scenario II: Bulk Fermion (BF)}
%%%%%%%%%%%%%%%%%%%%%%%%%%%%%%%%%%%%%%%
In this section, we investigate phenomenological implications of the 
presence of brane potentials to the BF scenario where all the SM fermions reside in the bulk and give consideration for the top Yukawa deviation and the Higgs production at the LHC experiment. 

%%%%%%%%%%%%%%%%%%%%%%%%%%%%%%%%%
\subsection{Top Yukawa deviation}
%%%%%%%%%%%%%%%%%%%%%%%%%%%%%%%%%

{}As a comparison with the BLF scenario, we also study the top Yukawa deviation 
in the BF scenario where all the SM fermions spread over the bulk.
The interactions among the top quark and Higgs boson in the BF scenario are given 
by
\begin{eqnarray}
 -\mathcal{L}_t=\frac{y_{t,5}}{L}\int_{-L/2}^{+L/2}dz\left[v(z)
                +f_0(z)\frac{\phi^q(x)}{\sqrt{2}}\right]\bar{t}(x,z)t(x,z).
\end{eqnarray}
The KK expansion is given by
\begin{eqnarray}
 t(x,z)\equiv\sum_{n=0}^\infty f_n^t(z)t_n(x),
 \label{topkk}
\end{eqnarray}
where $f_n^t(z)$ is normalized as
\begin{eqnarray}
 \int_{-L/2}^{+L/2}dzf_m^t(z)f_n^t(z)=\delta_{m,n}.
 \label{tnor}
\end{eqnarray}
We take $f_n^t(z)$ to be real so that complex phases reside in $t_n(x)$. Since 
the VEV $v(z)$ must be approximately flat in order not to have too large 
deviation of the $\rho$~parameter, we take the bulk Higgs mass to be zero, see 
Section~2.5 for detailed discussion. Then all the SM fermions have a flat 
zero-mode profile in the extra-dimension, that is, $f_0^t(z)$ in~(\ref{topkk}) 
is a constant. Therefore, the normalization~\eqref{tnor} gives 
$f_0^t={1/\sqrt{L}}$, and hence the final four dimensional coupling becomes
\begin{eqnarray}
 -\mathcal{L}_t=\left[m_t+\frac{y_{t,5}}{L}\int_{-L/2}^{L/2}dzf_0(z)
                \frac{\phi^q(x)}{\sqrt{2}}\right]t_0(x)\bar{t}_0(x).
                \label{5D_top_Yukawa}
\end{eqnarray}
{}Finally, the ratio $r$ defined in the section 3.1 is calculated in the BF 
scenario as follows
\begin{eqnarray}
 r=\frac{y_{t\bar{t}H}}{y_t}=\frac{1}{\sqrt{L}}\int_{-L/2}^{+L/2}dyf_0(z).
 \label{rtc}
\end{eqnarray}

We note that this ratio is just the same as that of the $WWH$ coupling in five 
dimensions to the one in four dimensions. It is because all the SM fermion in the 
BF scenario are similarly spreaded over the bulk as the gauge bosons in the BLF
scenario. Therefore, we can discuss the magnitude of the top deviation 
in the BF scenario by using a numerical plot given in Fig.~\ref{higgsm}, 
quantitatively. According to the numerical analyses, we can predict in the BF 
scenario that the top deviation becomes $8\%$ in the large boundary coupling limit but it is tiny when the boundary coupling is of order one. The deviation is smaller than the one in the BLF scenario.

%%%%%%%%%%%%%%%%%%%%%%%%%%%%%%%%%%%%%%%%%%%
\subsection{Higgs production at LHC}
%%%%%%%%%%%%%%%%%%%%%%%%%%%%%%%%%%%%%%%%%%%

Let us consider the Higgs production at the LHC in the BF scenario. The dominant 
process for the Higgs production at the LHC is the one through the gluon fusion. 
However, a large boundary coupling suppresses the cross sections through the gluon 
and $WW$ fusions to $1\%$ and $85\%$ from the SM prediction in the BLF scenario, respectively.
Therefore, the dominant process become the $WW$ fusion.

As is pointed out above, the coupling between the top quark and Higgs boson can 
slightly deviate from the top Yukawa coupling. The magnitudes of the deviation
are roughly estimated as $r\simeq0.92$ for the large boundary coupling in the BF
scenario. The ratio of the gluon-fusion cross section to the one 
predicted in the SM is proportional to $r^2$ as discussed above. The ratio 
for the $WW$ fusion process is also proportional to $r^2$ because the ratio to 
the SM prediction in~(\ref{rwwh}) is just one for the top Yukawa coupling 
given in~(\ref{rtc}). To summarize,
\begin{eqnarray}
 r_g^\sigma&=&r_W^\sigma=r^2=1
 \hspace{29.7mm}\mbox{for}\hspace{3mm}\hat{\lambda}=\mathcal{O}(1),\\
 r_g^\sigma&=&r_W^\sigma=r^2=0.85\left(\frac{r}{0.92}\right)^2
 \hspace{8mm}\mbox{for}\hspace{3mm}\hat{\lambda}=3\times10^{-6}.
\end{eqnarray}
Therefore, the cross sections for the Higgs production with a large boundary 
coupling are suppressed to be $85\%$ of the SM prediction in the BF scenario. 
Furthermore, other possible processes for the production such as 
$q\bar{q}\rightarrow HW$, $gg,q\bar{q}\rightarrow Ht\bar{t}$, and 
$q\bar{q}\rightarrow HZ$ are also suppressed simultaneously. It is seen that 
the dominant process is still the one through the gluon fusion. In the BLF 
scenario, the decay width also become large, the Higgs will decay rapidly when 
$m_{\text{KK}}$ is larger than a few TeV.

%%%%%%%%%%%%%%%%%%%%%%%%%%%%%%%%%%%
\subsubsection{Conserved KK parity}
%%%%%%%%%%%%%%%%%%%%%%%%%%%%%%%%%%%
Let us consider higher KK Higgs productions through the gluon and $W$ fusions. It is possible to consider higher KK productions through the processes like in the BLF scenario. Since the KK number conservation is violated only in the 
Higgs sector while the KK parity is conserved in all sectors, a higher KK Higgs 
can be produced through the gluon fusion as
\begin{eqnarray}
 Q^{(2n)}\bar{Q}^{(2m)}    &\rightarrow&H^{(2(l+1))},\\ 
 Q^{(2n)}\bar{Q}^{(2m+1)}  &\rightarrow&H^{(2l+1)},  \\ 
 Q^{(2n+1)}\bar{Q}^{(2m)}  &\rightarrow&H^{(2l+1)},  \\ 
 Q^{(2n+1)}\bar{Q}^{(2m+1)}&\rightarrow&H^{(2(l+1))}.
\end{eqnarray} 
On the other hand, processes
\begin{eqnarray}
 W^{(2n)}W^{(2m)}    &\rightarrow&H^{(2(l+1))},\\ 
 W^{(2n)}W^{(2m+1)}  &\rightarrow&H^{(2l+1)},  \\ 
 W^{(2n+1)}W^{(2m+1)}&\rightarrow&H^{(2(l+1))}
\end{eqnarray} 
are possible ones through the $W$ fusion. In the UED model, a Higher KK 
Higgs can be produced through the gluon fusion processes as 
\begin{eqnarray}
 Q^{(2n)}\bar{Q}^{(2|n-l-1|)}    &\rightarrow&H^{(2(l+1))},\\
 Q^{(2n)}\bar{Q}^{(2|n-l|+1)}    &\rightarrow&H^{(2l+1)},  \\ 
 Q^{(2n+1)}\bar{Q}^{(2|n-l|)}    &\rightarrow&H^{(2l+1)},  \\
 Q^{(2n+1)}\bar{Q}^{(|2(n-l)-1|)}&\rightarrow&H^{(2(l+1))},
\end{eqnarray} 
and $W$ fusion processes as 
\begin{eqnarray}
 W^{(2n)}W^{(2|n-l-1|)}    &\rightarrow&H^{(2(l+1))},\\ 
 W^{(2n)}W^{(2|n-l|+1)}    &\rightarrow&H^{(2l+1)},  \\ 
 W^{(2n+1)}W^{(2|n-l|)}    &\rightarrow&H^{(2l+1)},  \\ 
 W^{(2n+1)}W^{(|2(n-l)-1|)}&\rightarrow&H^{(2(l+1))},
\end{eqnarray}
because the both the KK parity and KK number are conserved in all sectors. 
We summarize the above consideration in Table~\ref{comp}.

\begin{figure}[t]
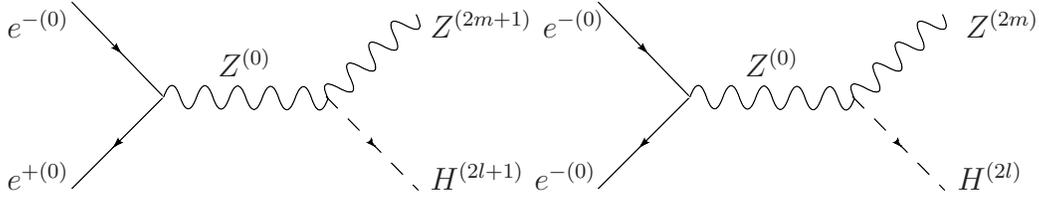

\begin{center}
\includegraphics[scale = 1.0]{assoh.eps}\hspace{2.1cm}
\includegraphics[scale = 1.0]{assoh.eps}
\end{center}\vspace{-3cm}

\hspace{1.4cm}$e^{-(0)}$\hspace{4.8cm}$Z^{(2m+1)}$
\hspace{0mm}$e^{-(0)}$\hspace{4.8cm}$Z^{(2m)}$

\hspace{4.2cm}$Z^{(0)}$\hspace{6.3cm}$Z^{(0)}$\vspace{1cm}

\hspace{1.4cm}$e^{+(0)}$\hspace{4.8cm}$H^{(2l+1)}$
\hspace{0mm}$e^{-(0)}$\hspace{4.8cm}$H^{(2l)}$
\caption{Associated KK Higgs production in the BF scenario with conserved KK parity.}
\label{pfigbf}
\end{figure}

Finally, we comment on higher KK Higgs productions at a linear collider. 
In the BF scenario with conserved KK parity, processes of 
\begin{eqnarray}
 Z^{(0)}\rightarrow Z^{(2m+1)}H^{(2l+1)},~Z^{(2m)}H^{(2l)},
\end{eqnarray}
are allowed because the KK parity is conserved in all sectors and the KK 
number is violated only in the Higgs sector shown in Fig.~\ref{pfigbf}. 

%%%%%%%%%%%%%%%%%%%%%%%%%%%%%%%
\subsubsection{Broken KK parity}
\begin{figure}
\begin{center}
\includegraphics[scale = 1.0]{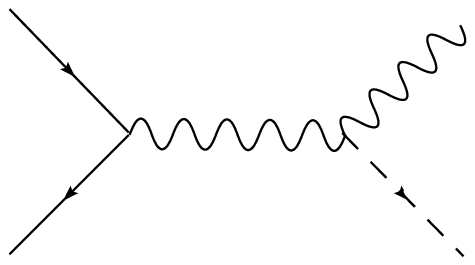}
\end{center}\vspace{-3cm}

\hspace{4.8cm}$e^{-(0)}$\hspace{4.9cm}$Z^{(m)}$

\hspace{7.6cm}$Z^{(0)}$\vspace{1cm}

\hspace{4.8cm}$e^{+(0)}$\hspace{4.9cm}$H^{(l)}$
\caption{Associated KK Higgs production in the BF scenario without KK parity.}
\label{pfig14bf}
\end{figure}

%%%%%%%%%%%%%%%%%%%%%%%%%%%%%%%
The reflection symmetry breaking in the BF scenario leads to the KK parity 
%conservation ?i?????????????????H?j
breaking in the gauge and Higgs sector. Therefore, the following 
productions are possible
\begin{eqnarray}
 Z^{(0)}\rightarrow Z^{(m)}H^{(l)},
\end{eqnarray} 
as given in Fig.~\ref{pfig14bf}.
These are the differences of BF scenario with the UED model. These higher KK Higgs productions could be searched in a future linear collider experiment. The discussions are summarized in Table~\ref{comp}.

%%%%%%%%%%%%%%%%%%%%%%%%%%%%%%%%%%%%%%%%%%%%%%%%%%%%%%%%%%%%%%%%
\subsection{Constraints from electroweak precision measurements}
%%%%%%%%%%%%%%%%%%%%%%%%%%%%%%%%%%%%%%%%%%%%%%%%%%%%%%%%%%%%%%%%

Let us consider constraints on this setup in the BF scenario from electroweak 
precision measurements. The radiative corrections to electroweak interaction 
observables can be summarized into three parameters $S$, $T$, and $U$ 
\cite{stu2,stu3}. Since the Higgs sector in this setup is generalized from the 
usual UED model, we especially consider contributions to the $S$ and $T$ 
parameters from the Higgs sector.

First, we estimate the $T$ parameter defined in~(\ref{tpara}) at one-loop 
level. As commented in the previous section, the Higgs sector is 
generalized from the UED model in this setup, and the calculation of diagrams 
including the Higgs loop shown in Fig.~\ref{fig9} should be reconsidered. The calculation of $T$ parameter in this setup can be divided into two parts systematically
\begin{eqnarray}
 T\simeq T_{\text{SM}}+T_{\text{BF}},\label{Ttot}
\end{eqnarray}
where the first and second terms are contributions from the SM and the BF 
scenario in this five dimensional setup. We also have another contribution 
to the parameter, which comes from effects of KK mode mixing. Since these mode
mixings must occur at least twice in internal lines of the Higgs field to 
conserve the KK parity in this setup, effects of mixings on physical quantities
are tiny enough to be neglected. In this sense, the estimation of $T$ parameter~(\ref{Ttot}) is a good approximation. 

The first term in~(\ref{Ttot}), the SM contribution, is given by 
\begin{eqnarray}
 T_{\text{SM}}\simeq\frac{1}{16\pi}\left[\frac{3m_t^2}{m_W^2}
             +2\left(1-\frac{s^2}{3c^2}\right)\log\frac{m_t^2}{m_W^2}\right]
             -\frac{s^2}{16\pi c^2}\left[3\left(\log\frac{m_H^2}{m_W^2}
             -\frac{5}{6}\right)\right],
 \label{smt}
\end{eqnarray}
where the first and second terms in~(\ref{smt}) are the dominant top quark contribution and the one from the Higgs sector, respectively~\cite{Vel,Jeg}. We 
should note that the Higgs mass appearing in the right hand side of~(\ref{smt})
is determined by $k_0$ in~(\ref{higgsmass}) which is different from the SM.
The characteristic contribution from the BF scenario is given in 
\begin{eqnarray}
 T_{\text{BF}}\simeq\frac{1}{4\pi}\left(\frac{m_t^2}{m_W^2}\right)
                     \left(\frac{m_t^2}{m_{\text{KK}}^2}\right)\zeta(2)
                     -\frac{s^2}{4\pi c^2}\frac{5}{12}\sum_{n=1}^\infty
                     \left(\frac{m_{H^{(n)}}-nm_{\text{KK}}}{nm_{\text{KK}}}\right)^2,
 \label{tkk}
\end{eqnarray}
where $m_{\text{KK}}\equiv1/R$ and $m_{H^{(n)}}$ is the mass of $n$th mode Higgs given 
by $k_n$ in this setup. The first and second terms correspond to contributions 
from KK top quark and KK Higgs, respectively. Since the Higgs sector in the BF
scenario is modified from the usual UED model, the contributions from the 
sector is somewhat different from calculations for the UED~\cite{App,Gogo}.

In the UED model, contribution from the Higgs sector is proportional to 
$(m_{H,\text{UED}}/nm_{\text{KK}})^2$, where $m_{H,\text{UED}}$ denotes the bulk Higgs mass in the UED. This effect in the UED model is from the universal shift of all the KK masses by the amount of the bulk mass. On the other hand, the effects is replaced by $(m_{H^{(n)}}-nm_{\text{KK}})^2$ in the BF scenario. These contributions become smaller for higher KK mode. Furthermore, the second term in~(\ref{tkk}) is negligibly small in a numerical estimation.

In the limit of vanishing boundary coupling, the Higgs and its KK masses become $m_{H^{(n)}}\rightarrow nm_{\text{KK}}$. Therefore, the contribution from the Higgs sector vanishes. In the opposite limit of the large boundary coupling, the Higgs KK masses are nearly equal to $(n+1)m_{\text{KK}}$. In this limit, the second term is estimated as $-(5s^2/48\pi c^2)\zeta(2)\simeq-0.016$ which is negligibly small. 
To summarize, the $T$ parameter in the BF scenario is given by
\begin{eqnarray}
 T&\simeq&\frac{1}{16\pi}\left[\frac{3m_t^2}{m_W^2}
          +2\left(1-\frac{s^2}{3c^2}\right)\log\frac{m_t^2}{m_W^2}\right]
          -\frac{s^2}{16\pi c^2}\left[3\left(\log\frac{m_H^2}{m_W^2}
          -\frac{5}{6}\right)\right]\nonumber\\
  &      &+\frac{1}{4\pi}\left(\frac{m_t^2}{m_W^2}\right)
          \left(\frac{m_t^2}{m_{\text{KK}}^2}\right)\zeta(2).
 \label{ft}
\end{eqnarray}

\begin{figure}[t]
\begin{center}
\includegraphics[scale = 0.8]{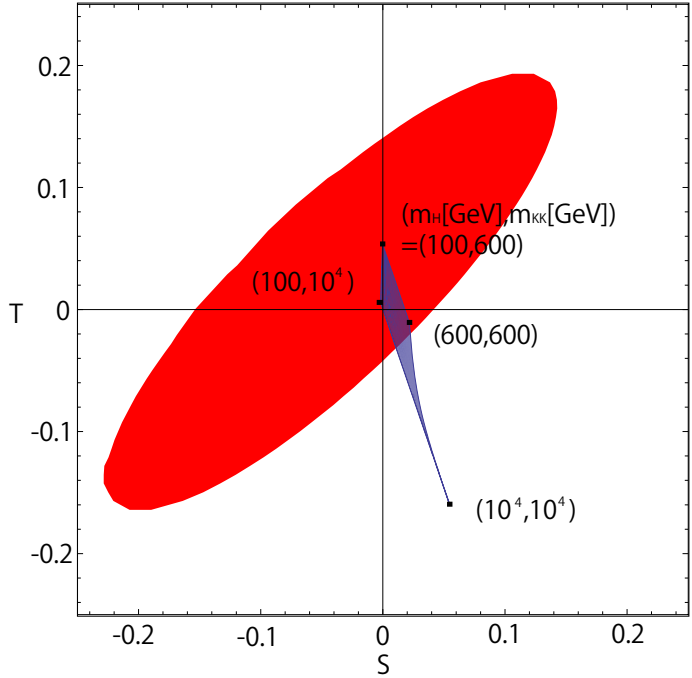}
\end{center}
\caption{$S$ and $T$ plot in this scenario: The ellipse is allowed region by all
 electroweak precision measurements at $90\%$ CL with $m_H^{\text{ref}}=117$ 
GeV \cite{pdg}. The four dots, $(m_H[\text{GeV}],m_{\text{KK}}[\text{GeV}])=(100,600),~(100,10^4),~(600,600),$ and $(10^4,10^4)$, represent typical predictions in 
the BF scenario.}
\label{fig12}
\end{figure}

Next, let us consider the $S$ parameter defined in~(\ref{spara}). The 
estimation of the parameter is also given by dividing into two parts,
\begin{eqnarray}
 S\simeq S_{\text{SM}}+S_{\text{BF}}.\label{Stot}
\end{eqnarray}
The SM contribution is as follows,
\begin{eqnarray}
 S_{\text{SM}}\simeq-\frac{3s^2}{2\pi}\log\frac{m_t}{m_Z}
             +\frac{s^2}{6\pi}\log\frac{m_H}{m_Z}.
\end{eqnarray}
The contribution to $S$ parameter from higher KK mode in the BF scenario is 
given by
\begin{eqnarray}
 S_{\text{BF}}\simeq\frac{s^2}{6\pi}\left(\frac{m_t^2}{m_{\text{KK}}^2}\right)\zeta(2)
             -\frac{s^2}{24\pi}\sum_{n=1}^\infty
             \left(\frac{m_{H^{(n)}}-nm_{\text{KK}}}{nm_{\text{KK}}}\right)^2.
 \label{sBF}
\end{eqnarray}
To summarize, we obtain the $S$ parameter in this scenario as 
\begin{eqnarray}
 S\simeq-\frac{3s^2}{2\pi}\log\frac{m_t}{m_Z}
        +\frac{s^2}{6\pi}\log\frac{m_H}{m_Z}
        +\frac{s^2}{6\pi}\left(\frac{m_t^2}{m_{\text{KK}}^2}\right)\zeta(2),
 \label{fs}
\end{eqnarray}
where we have neglected the second term in~(\ref{sBF}) because its value 
becomes $|(s^2/24\pi)\zeta(2)|\simeq4.9\times10^{-3}$, at maximum. The ($S$,$T$) plot 
is presented in Fig.~\ref{fig12} in range of 
$500\,\text{GeV}<m_{\text{KK}}(\equiv1/R)<10\,\text{TeV}$ and $100\,\text{GeV}<m_H<m_{\text{KK}}$.
\begin{figure}
\begin{center}
\includegraphics[scale = 1.0]{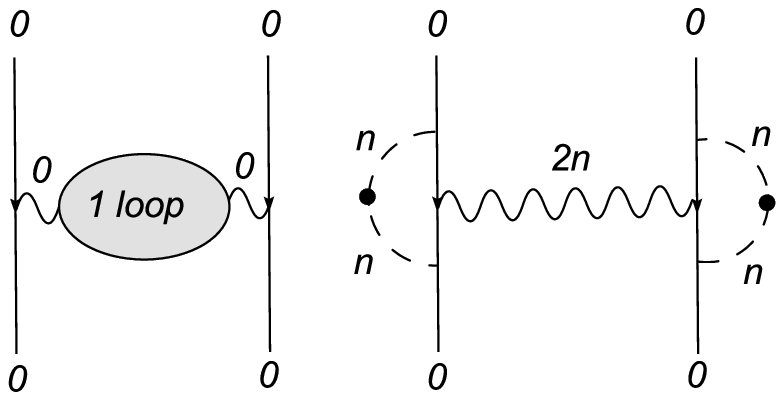}
\end{center}
\hspace{5.2cm}(a)\hspace{3.9cm}(b)
\caption{The $S$ and $T$ parameters are calculated based on diagram (a). (b) 
The BF scenario allow a contribution such a diagram from 2-loop level which is
forbidden in the usual UED model. However contributions are highly suppressed.}
\label{pfig13}
\end{figure}

The above calculations of BF scenario are similar to that of the UED model, where divergences cancels out in each KK level~\cite{App}. 
Let us see it more in detail.
Each diagram including Higgs loop shown in 
Fig.~\ref{fig9} leads to a divergence. However, it can be shown that each divergence from Fig.~\ref{fig9}~(c) and (d) are cancelled in the calculation of the $T$ parameter.
Furthermore, contributions to the parameter from diagrams in Fig.~\ref{fig9}~(e) 
and (f) lead to divergence depending only on the gauge boson masses but independent of the Higgs mass, which are cancelled between these two diagrams. 

There are also divergences 
depending only on the gauge boson masses from Fig.~\ref{fig9}~(a) and (b). In 
the SM and UED, these divergences are cancelled among all diagrams including 
the gauge and NG boson loops, after summing up all of them.
However, since the 
wave function profile of the Higgs is modified in the BF scenario, the 
additional factor is multiplied to three point couplings in Fig.~\ref{fig9}~(a), 
(b), (d), and (f), which are
\begin{eqnarray}
 \sum_{n,l}\frac{2}{L}\left(1+\frac{\sin(\pi\epsilon_l)}{\pi(l+\epsilon_l)}
 \right)^{-1}\int_{-L/2}^{+L/2}dz\sin\left(\frac{n}{L}\pi z\right)\sin\left(
 \frac{l+\epsilon_l}{L}\pi z\right),\label{faco}
\end{eqnarray}
for odd modes of the Higgs, gauge and NG bosons, and  
\begin{eqnarray}
 \sum_{n,l}\frac{2}{L}\left(1+\frac{\sin(\pi\epsilon_l)}{\pi(l+\epsilon_l)}
 \right)^{-1}\int_{-L/2}^{+L/2}dz\cos\left(\frac{n}{L}\pi z\right)\cos\left(
 \frac{l+\epsilon_l}{L}\pi z\right),\label{face}
\end{eqnarray}
for even modes, where $0\leq\epsilon_l\leq1$ determines the deviation of each KK mode of the BF scenario from that of the UED with vanishing bulk mass.
The $n$ stands for the KK number of gauge and NG boson, circulating in the loops in Fig.~\ref{fig9} (a), (b), (d), and (f), while $l$ is the one for the Higgs. 

We note that, in the KK expansion, the eigenfunction for each KK Higgs is different from the one for the gauge and NG bosons. In other words, the gauge/NG and Higgs are expanded by different complete set of eigenfunctions. Actually, in the limit of large boundary coupling, the profile of $n$-mode KK Higgs is given by  $\beta_n\sin((n+1)\pi z/L)$ for odd $n$ and $\alpha_n\cos((n+1)\pi z/L)$ for even $n$. Therefore, a complete KK summation can be done by summing over $n$ for the gauge/NG boson and $l$ for the Higgs separately. 

Let us see the $\epsilon_l\to0$ limit.
The magnitude of $\epsilon_l$ is correlated with the boundary coupling. 
The $\lambda\rightarrow0$ limit, which is equivalent to the UED with 
vanishing bulk mass, leads to $\epsilon_l\rightarrow0$. We find that both the
additional couplings in~(\ref{faco}) and (\ref{face}) can be expanded as
\begin{eqnarray}
 &&\sum_{n,l}\frac{2}{L}\left(1+\frac{\sin(\pi\epsilon_l)}{\pi(l+\epsilon_l)}
 \right)^{-1}\int_{-L/2}^{+L/2}dz\sin\left(\frac{n}{L}\pi z\right)\sin\left(
 \frac{l+\epsilon_l}{L}\pi z\right)\nonumber\\
 &&\hspace{8cm}\simeq 1+\sum_{n,l}F_o(n,l)\mathcal{O}(\epsilon_l^2)+\cdots,\\
 &&\sum_{n,l}\frac{2}{L}\left(1+\frac{\sin(\pi\epsilon_l)}{\pi(l+\epsilon_l)}
 \right)^{-1}\int_{-L/2}^{+L/2}dz\cos\left(\frac{n}{L}\pi z\right)\cos\left(
 \frac{l+\epsilon_l}{L}\pi z\right)\nonumber\\
 &&\hspace{8cm}\simeq 1+\sum_{n,l}F_e(n,l)\mathcal{O}(\epsilon_l^2)+\cdots,
\end{eqnarray}
where $F_o(n,l)$ and $F_e(n,l)$ are functions of $n$ and $l$. Therefore, we 
find that these additional couplings become $1$ at the leading order. 
In other words, the divergences cancels at the sub-leading $O(\epsilon_l)$ too.
That is, all the divergences from Fig.~\ref{fig9} (a), (b), (d), and (f) are cancelled among loop diagrams from the gauge and NG bosons at $O(\epsilon_l)$. 
%It may be that the divergences for all order are cancelled after summing over infinite modes, if there is no any phase transitions while evolution of $\epsilon_l$, equivalently the boundary coupling, to a finite value.

We give a brief comment on the KK parity which is related to our calculation of the EW constraints. In the calculations of $S$ and $T$ parameters, we considered radiative corrections from 1-loop diagrams, which are conceptually shown in Fig.~\ref{pfig13}, where $0$ and $n$ denote the KK numbers of propagating particles. As mentioned above, the gauge bosons that couple to external currents are zero mode at the 1-loop level, since the KK parity is always conserved on each vertex in the BF scenario. The situation is illustrated in Fig.~\ref{pfig13}~(a). 

Moreover, effects from longitudinal mode of gauge bosons are negligibly small. The reason is as follows. When the 1-loop corrected gauge boson propagator couples to the external fermion current in Fig.~\ref{pfig13}, the momentum of the gauge fields is replaced by the fermion mass of the current, which is typically the electron mass, due to the Dirac equation. 
Therefore, contributions from longitudinal mode of gauge bosons to the amplitude of the diagram in Fig.~\ref{pfig13}~(a) is suppressed by the factor $(m_e/m_W)^2$, where $m_e$ is the electron mass. They are negligibly small. 

If we consider corrections up to 2-loop level in the BF scenario, there are extra contributions forbidden in the UED. They are shown in Fig.~\ref{pfig13}~(b). In the figure, the black dot represents mode mixing of 
the Higgs. However, it is seen that effects from such a diagram are highly 
suppressed.

%%%%%%%%%%%%%%%%%%%%%
\section{Dark matter}
%%%%%%%%%%%%%%%%%%%%%
We comment on a dark matter candidate in the BLF and BF scenarios. 
In the UED model, since the KK parity is always conserved, the Lightest KK Particle (LKP) can be candidate for the dark matter. 
Especially, discussions for the KK photon as a candidate for the dark matter 
have been presented in many literatures, see~\cite{Hooper:2007qk} for an excellent review.

In the BLF scenario, the KK number and parity conservations are always 
broken in the fermion sector, and the LKP can decay to zero modes, which are the SM particles, through KK number and parity violating processes. Therefore, when we explain the dark matter in this scenario we need additional parity for the 
fermion sector or prepare other dark matter candidates. 

In the BF scenario with conserving reflection symmetry, the KK parity is conserved in all sectors, and the LKP whose KK parity is odd ($n=1$) is stable, 
which can be a candidate for dark matter. At tree level, all of the SM 
fields appear with towers of KK states with masses of
\begin{eqnarray}
 m_{X^{(n)}}^2=\frac{n^2}{R^2}+m_{X^{(0)}}^2,
\end{eqnarray}
where $X^{(n)}$ is the n-th KK excitation of the corresponding SM field $X^{(0)}$. 
Corrections to KK masses are generated by loop diagrams traversing around the 
extra dimension and brane localized kinetic terms. These contributions at 
1-loop level are given by~\cite{Cheng:2002iz}
\begin{eqnarray}
 \delta m_{B^{(n)}}^2&=&\frac{g_4'{}^2}{16\pi^2R^2}\left(-\frac{39}{2}
                        \frac{\zeta(3)}{\pi^2}-\frac{n^2}{3}\ln\Lambda R
                        \right),                                           \\
 \delta m_{W^{(n)}}^2&=&\frac{g_4^2}{16\pi^2R^2}\left(-\frac{5}{2}
                        \frac{\zeta(3)}{\pi^2}+15n^2\ln\Lambda R\right), %  \\
%  \delta m_{g^{(n)}}^2&=&\frac{g_3^2}{16\pi^2R^2}\left(-\frac{3}{2}
%                         \frac{\zeta(3)}{\pi^2}+23n^2\ln\Lambda R\right),   \\
%  \delta m_{Q^{(n)}}  &=&\frac{n}{16\pi^2R}\left(6g_3^2+\frac{27}{8}g_4^2
%                         +\frac{1}{8}g_4'{}^2\right)\ln\Lambda R,\\
%  \delta m_{u^{(n)}}  &=&\frac{n}{16\pi^2R}(6g_3^2+2g_4'{}^2)\ln\Lambda R,  \\
%  \delta m_{d^{(n)}}  &=&\frac{n}{16\pi^2R}\left(6g_3^2
%                         +\frac{1}{2}g_4'{}^2\right)\ln\Lambda R,           \\
%  \delta m_{L^{(n)}}  &=&\frac{n}{16\pi^2R}\left(\frac{27}{8}g_4^2
%                         +\frac{9}{8}g_4'{}^2\right)\ln\Lambda R,           \\
%  \delta m_{e^{(n)}}  &=&\frac{n}{16\pi^2R}\frac{9}{2}g_4'{}^2\ln\Lambda R,
\end{eqnarray}
for $U(1)$ and $SU(2)$ gauge bosons, respectively. After the KK modes of the 
$W$ and $Z$ bosons acquire masses by eating the fifth components of the gauge 
fields and Higgs KK modes (in a certain gauge), four scalar states remain at each KK level. The mass
of the neutral Higgs is written as
\begin{eqnarray}
 m_{H^{(n)}}^2   &=& \mu_n^2+\delta m_{H^{(n)}}^2, \label{kkh1} % \\
%  m_{H_n^\pm}^2 &=& \frac{n^2}{R^2}+m_W^2+\delta m_{H_n}^2, \\
%  m_{A_n^0}^2   &=& \frac{n^2}{R^2}+m_Z^2+\delta m_{H^{(n)}}^2,
\end{eqnarray}
where the radiative and boundary term corrections are given by
\begin{eqnarray}
 \delta m_{H^{(n)}}^2\simeq\frac{n^2}{16\pi^2R^2}
                           \left(3g_4^2+\frac{3}{2}g_4'{}^2
                           -2\lambda_H\right)\ln\Lambda R
                           +\frac{2\pi^2\hat{\lambda}
                            v_{\text{EW}}^2}{\Lambda^2R^2}.
 \label{kkhcor}
\end{eqnarray}
Here $\lambda_H$ is the Higgs quartic coupling and the second term in 
(\ref{kkhcor}) comes from the boundary mass term for the Higgs mode.

We note that there is no correction term like $m_H^2$ in~(\ref{kkh1}), where 
$m_H$ means the usual SM Higgs mass. Corresponding effects from $m_H^2$
are included into $\mu_n^2$. In the case of the vanishing bulk mass, $\mu_n$ 
equals to $k_n$, which is just the n-mode KK Higgs mass (see Fig.~\ref{figtan}).
 Since the mass of KK photon is written by
\begin{eqnarray}
 m_{A^{(n)}}^2=\frac{n^2}{R^2}+\delta m_{A^{(n)}}^2,
\end{eqnarray}
where
\begin{eqnarray}
 \delta m_{A^{(n)}}^2
 &\equiv&\frac{1}{2}\left[\delta m_{B^{(n)}}^2+\delta m_{W^{(n)}}^2
         +\frac{1}{2}(g_4^2+g_4'{}^2)v_{\text{EW}}^2\right.
         \nonumber\\
 &      &\hspace{5mm}\left.+\sqrt{\left\{\delta m_{B^{(n)}}^2
         -\delta m_{W^{(n)}}^2
         +\frac{1}{2}(g_4^2-g_4'{}^2)v_{\text{EW}}^2\right\}^2
         -(2g_4g_4'v_{\text{EW}}^2)^2}\right],
\end{eqnarray}
the LKP in the UED is the KK photon naively. The KK photon is also the LKP in 
this setup too, because the mass of the first KK Higgs is heavier than 
$\pi/L=1/R$ as shown in Fig.~\ref{figtan}. These discussions are summarized in 
Table~\ref{tab2}.
\begin{table}
\begin{center}
\begin{tabular}{|l||c|c|c|}
\hline
& BLF & BF & UED \\
\hline
\hline
Dark matter candidate & $\times$ & LKP & LKP \\
\hline
\end{tabular}
\end{center}
\caption{Dark matter candidate}
\label{tab2}
\end{table}

%%%%%%%%%%%%%%%%%%%%%%%%%%%%%%%%%
\section{Summary and discussions}
%%%%%%%%%%%%%%%%%%%%%%%%%%%%%%%%%

We have suggested a simple extra-dimension model which induces a deviation of the coupling between top and free physical Higgs fields from that of the SM top Yukawa, within one Higgs doublet scenario. Our setup is 5-dimensional flat spacetime, where one SM Higgs double field exists in the bulk. The wave function profile of the free physical Higgs becomes different from its VEV when brane potentials are introduced. We can consider two scenario depending on the location of matter fermions. One scenario is BLF and the other is BF. 

We found that the top-physical-Higgs coupling becomes small because of the brane potentials with large couplings in the BLF scenario, which could be checked at the LHC experiment. In this case dominant Higgs production channel at LHC becomes $WW$ fusion. Four-dimentional effective Higgs self-couplings do not become huge even in the large brane-coupling limit, since the wave-function profile of the free Higgs vanishes at the boundaries, and therefore the Higgs can be regarded as a particle (as long as the compactification scale is not higher than a few TeV). Reminding that the top Yukawa deviation can occur in multi-Higgs models, if we find only one Higgs particle as well as the top Yukawa deviation at LHC, our scenario become a realistic candidate beyond the SM. Note that our setup is similar to the universal extra dimension, but the latter has no top Yukawa deviation. 

In the BF scenario, the top Yukawa deviation is not significant even in a large boundary couplings, however a stable dark matter exists due to the existence of an accidental reflection symmetry. We also show that these two scenarios are consistent with the present electroweak precision measurements.

Finally, we comment on the tree level unitarity of $WW \to WW$ scattering. It 
is violated when the boundary-localized self-coupling of the Higgs $\lambda$ becomes 
large. In particular, the longitudinal coupling ($W^4_L$ contact coupling) becomes large as shown in Eq.~(\ref{ngself}) and in Fig.~\ref{self} and then we need calculation technique of strongly
interacting Higgs sector (see e.g.~\cite{Dobado:1999xb,Dobado:1990jy,Dobado:1989gr}). 
Notice that even in this limit, the coupling between physical Higgs $H$ and the longitudinal mode, corresponding to the NG-boson $\chi$, are saturated as in Fig.~\ref{hself}, since the four-dimensional Higgs effective couplings remain finite due to the suppression of Higgs profiles at boundaries, that is, due to the factor $c_\Delta\to 0$ in Eq.~\eqref{HHHHandHHcc}.
Furthermore even when the tree level unitarity is broken, the total unitarity of the model must be conserved because the EW symmetry breaking is caused by the ``Higgs mechanism'' on the boundary. The violation of the tree level unitarity is just a lack of a technique of treating a strong dynamics.

%Finally, we comment on the tree level unitarity of $WW\rightarrow WW$ 
%scattering. It is violated when the Higgs boundary localized self-couplings 
%become large. The longitudinal coupling ($W_L^4$ contact coupling) is also 
%large as shown in~(\ref{ngself}), and then we need calculation technique of strong interactions of Higgs sector (see e.g. 
%\cite{Dobado:1999xb,Dobado:1990jy,Dobado:1989gr}). Notice that even in this 
%case Higgs contributing diagrams are saturated as Fig.~\ref{hself} since the 
%4-dimensional Higgs effective couplings becomes saturated due to suppression of
% Higgs profiles at boundaries. 
%
%Furthermore even when the tree level unitarity is broken, the total unitarity 
%of the model must be conserved because the EW symmetry breaking is caused by 
%the ``Higgs mechanism'' on the boundary. The violation of the tree level 
%unitarity is just a lack of a technique of treating a strong dynamics.

\begin{table}
\begin{center}
\begin{tabular}{|c||l|l|}
\hline
& \hspace{2.5cm}BLF & \hspace{2.5cm}BF \\
\hline
\hline
Associated & $Z^{(2n)}\rightarrow Z^{(2m)}H^{(2(l+1))}$ & $Z^{(0)}\rightarrow Z^{(2m)}H^{(2(l+1))}$ \\
KK Higgs & $Z^{(2n)}\rightarrow Z^{(2m+1)}H^{(2l+1)}$ & $Z^{(0)}\rightarrow Z^{(2m+1)}H^{(2l+1)}$ \\
Productions & $Z^{(2n+1)}\rightarrow Z^{(2m)}H^{(2l+1)}$ & \\
           & $Z^{(2n+1)}\rightarrow Z^{(2m+1)}H^{(2(l+1))}$ & \\
\hline
Top Yukawa & Tiny in small $\hat{\lambda}$ & Tiny in small $\hat{\lambda}$ \\
Deviations & $\sim90\%$ in large $\hat{\lambda}$ & $\sim8\%$ in large $\hat{\lambda}$ \\
\hline
$H^{(0)}$ Production & Same as the SM in small $\hat{\lambda}$ & Same as the SM in small $\hat{\lambda}$ \\
by $gg$ fusion       & $1\%$ of SM in large $\hat{\lambda}$       & $85\%$ of SM in large $\hat{\lambda}$ \\
\hline
$H^{(0)}$ Production & Same as the SM in small $\hat{\lambda}$ & Same as the SM in small $\hat{\lambda}$ \\
by $WW$ fusion   & $85\%$ of SM in large $\hat{\lambda}$ & $85\%$ of SM in large $\hat{\lambda}$\\
\hline
KK Higgs & $Q^{(n+1)}\bar{Q}^{(m+1)}\rightarrow H^{(l+1)}$ in small $\hat{\lambda}$ & $Q^{(2n)}\bar{Q}^{(2m)}\rightarrow H^{(2(l+1))}$ \\
Production & & $Q^{(2n)}\bar{Q}^{(2m+1)}\rightarrow H^{(2l+1)}$  \\
by $gg$ fusion & & $Q^{(2n+1)}\bar{Q}^{(2m)}\rightarrow H^{(2l+1)}$ \\
& & $Q^{(2n+1)}\bar{Q}^{(2m+1)}\rightarrow H^{(2(l+1))}$ \\
\hline
KK Higgs & $W^{(2n)}W^{(2m)}\rightarrow H^{(2(l+1))}$ 
        & $W^{(2n)}W^{(2m)}\rightarrow H^{(2(l+1))}$\\
Production & $W^{(2n)}W^{(2m+1)}\rightarrow H^{(2l+1)}$ 
          & $W^{(2n)}W^{(2m+1)}\rightarrow H^{(2l+1)}$\\
by $WW$ fusion & $W^{(2n+1)}W^{(2m+1)}\rightarrow H^{(2(l+1))}$ 
              & $W^{(2n+1)}W^{(2m+1)}\rightarrow H^{(2(l+1))}$\\
\hline
\end{tabular}\vspace{1mm}

\hspace{-5.4cm}\begin{tabular}{|c||l|}
\hline
& \hspace{2.3cm}UED \\
\hline
\hline
Associated & $Z^{(0)}\rightarrow Z^{(m+1)}H^{(m+1)}$\\
KK Higgs     & \\
Productions & \\
\hline
Top Yukawa & $\times$\\
Deviations &\\
\hline
$H^{(0)}$ Production & Almost same as SM\\
by $gg$ fusion   & \\
\hline
$H^{(0)}$ Production & Almost same as SM\\
by $WW$ fusion   & \\
\hline
KK Higgs & $Q^{(2n)}\bar{Q}^{(2|n-l-1|)}\rightarrow H^{(2(l+1))}$ \\
Production & $Q^{(2n)}\bar{Q}^{(2|n-l|+1)}\rightarrow H^{(2l+1)}$ \\
by $gg$ fusion & $Q^{(2n+1)}\bar{Q}^{(2|n-l|)}\rightarrow H^{(2l+1)}$ \\
& $Q^{(2n+1)}\bar{Q}^{(|2(n-l)-1|)}\rightarrow H^{(2(l+1))}$ \\
\hline
KK Higgs &$W^{(2n)}W^{(2|n-l-1|)}\rightarrow H^{(2(l+1))}$ \\
Production &$W^{(2n)}W^{(2|n-l|+1)}\rightarrow H^{(2l+1)}$ \\
by $WW$ fusion & $W^{(2n+1)}W^{(2|n-l|)}\rightarrow H^{(2l+1)}$ \\
& $W^{(2n+1)}W^{(|2(n-l)-1|)}\rightarrow H^{(2(l+1))}$ ~~~\\
\hline
\end{tabular}
\end{center}
\caption{Comparisons of phenomenological consequences among the BLF, BF and UED
scenario. Note that it is difficult to find the peak in the cross section of 
$H^{(0)}$ production through the $WW$ process when $m_{\text{KK}}$ is larger 
than a few TeV with a large boundary coupling in the BLF and BF scenarios.} 
\label{comp}
\end{table}

%%%%%%%%%%%%%%%%%%%%%%%%%%%%%%
\subsection*{Acknowledgments}
%%%%%%%%%%%%%%%%%%%%%%%%%%%%%%

N.H.\ and  K.O.\ have been supported in part by scientific grants from the 
Ministry of Education, Science, Sports, and Culture of Japan Nos.\ 18204024, 20244028, 20025004, 20039006, and 205402272. The work of R.T.\ has 
been partially supported by the Japan Society of Promotion of Science. 
The authors are grateful to Toshifumi Yamashita and Shigeki Matsumoto for helpful and fruitful discussions, and also thank Kazunori Hanagaki, Yutaka Hosotani, Nobuchika Okada, Yutaka Sakamura, Minoru Tanaka, and Nobuhiro Uekusa for useful comments.

%\newpage
\appendix
\section*{Appendix}
%%%%%%%%%%%%%%%%%%%%%%%%%%%%%%
\section{Radion stabilization}
%%%%%%%%%%%%%%%%%%%%%%%%%%%%%%

In this section, we show solutions of the equation of motion~(\ref{l6z}) with 
BCs~(\ref{l7z}) and give potential analyses.

\begin{figure}
\begin{center}
\includegraphics[scale = 1.0]{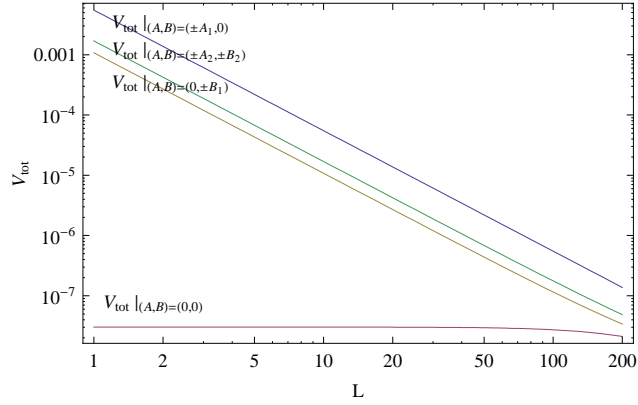}
\end{center}
\caption{Effective potentials for solutions, $(A,B)=(0,0),$ $(\pm A_1,0),$ 
$(0,\pm B_1),$ $(\pm A_2,\pm B_2)$. Input values are 
$(\hat{m},\hat{\lambda},\Lambda,v)=(10^{-3},3\times10^6,10\,\text{TeV},174\mbox{GeV}/\sqrt{L})$.}
\label{radion}
\end{figure}

The bulk equation of motion~(\ref{l6}) has a general solution,
\begin{eqnarray}
 \Phi^c(y)=A\cosh(mz)+B\sinh(mz),
\end{eqnarray}
where we take the bulk potential as~(\ref{bup}). The BCs~(\ref{l7}) with the 
brane potentials~(\ref{bp}) determine 
$(A,B)$ as 
\begin{eqnarray}
 (A,B)=(0,0),~(\pm A_1,0),~(0,\pm B_1),~(\pm A_2,\pm B_2),~(\pm A_2,\mp B_2),
 \label{AB}
\end{eqnarray}
where 
\begin{eqnarray}
 A_1&\equiv&\frac{\sqrt{\lambda v^2c_h-ms_h}}{c_h^{3/2}\sqrt{\lambda}},\\
 A_2&\equiv&\frac{\sqrt{\lambda v^2s_hc_h-m(2c_h^2+1)}}
                 {2c_h^{3/2}\sqrt{2\lambda s_h}},                      \\
 B_1&\equiv&\sqrt{\frac{\lambda v^2s_h-mc_h}{\lambda s_h(c_h-2)}},     \\
 B_2&\equiv&\sqrt{\frac{c_h[-\lambda v^2s_h+m(c_h-2)]}
                 {2\lambda s_h(1-c_h^2)}}.                             
\end{eqnarray}
We have defined $s_h$ and $c_h$ as in~(\ref{ssh}) and~(\ref{cch}), 
respectively.  Furthermore,~(\ref{lambdadef}) and~(\ref{vv}) have been taken.
We can write the bulk potential energy, the bulk kinetic energy, and 
effective potential energy in four dimension as follows,
\begin{eqnarray}
 V_p&\equiv&\int_{-L/2}^{+L/2}dz\mathcal{V},\\
 V_k&\equiv&\int_{-L/2}^{+L/2}dz|\partial_z\Phi^c|^2,\\
 V_{\mbox{{\scriptsize tot}}}&\equiv&V_{+L/2}+V_{-L/2}+V_p+V_k.
\end{eqnarray}
We obtain the effective potential energy for each solution in~(\ref{AB}),
\begin{eqnarray}
 V_{\mbox{{\scriptsize tot}}}|_{(A,B)=(0,0)}&=&2\lambda v^4, \\
 V_{\mbox{{\scriptsize tot}}}|_{(A,B)=(\pm A_1,0)}
 &=&\frac{mt_h}{\lambda}(2\lambda v^2-mt_h),\\
 V_{\mbox{{\scriptsize tot}}}|_{(A,B)=(0,\pm B_1)}
 &=&\frac{m}{\lambda t_h}\left(2\lambda v^2-\frac{m}{t_h}\right),\\
 V_{\mbox{{\scriptsize tot}}}|_{(A,B)=(\pm A_2,\pm B_2)}
 &=&V_{\mbox{{\scriptsize tot}}}|_{(A,B)=(\pm A_2,\mp B_2)}\nonumber\\
 &=&\frac{4(\lambda^2v^4-m^2)s_h^2c_h^2+m[4\lambda v^2s_h^2c_h^2(2c_h^2-1)+m]}
         {8\lambda s_h^2c_h^2},
\end{eqnarray}
where $t_h\equiv s_h/c_h$. The typical value for each effective potential is 
shown Fig.~\ref{radion}.
In the figure, the horizontal axis is $L$. We can identify $L$ with a VEV of a scalar field, so-called radion. When we consider the setup with a finite bulk mass, the
solution $(A,B)=(0,0)$ leads to a global minimum of the potential. However, 
the radion cannot be stabilized within the setup. The vanishing bulk mass has 
been taken in phenomenological discussions of this paper, and the effective 
potential become $V_{\mbox{{\scriptsize tot}}}|_{\Phi^c=v}=0$ which does not 
depend on $L$. Therefore, the radion cannot be stable. The study for the 
radion stabilization will be a future work.

%%%%%%%%%%%%%%%%%%%%%%%%%%%%%%%%
\section{Action of Higgs sector}
%%%%%%%%%%%%%%%%%%%%%%%%%%%%%%%%

In this section, we give derivations of the action for the Higgs (\ref{hl}) and
for the NG boson (\ref{46}). 

The free action for the physical Higgs and NG is written down as 
\begin{eqnarray}
 S_{\mbox{{\scriptsize free}},\phi^q,\chi}
 &=&\int d^4x\int_{-L/2}^{+L/2}dz
    \Bigg[-\frac{1}{2}(\partial_z\Phi_R+\partial_z\phi^q)^2
          -\frac{1}{2}(\partial_z\Phi_I+\partial_z\chi)^2
          -\mathcal{V}\nonumber\\
 & &-\frac{\partial\mathcal{V}}{\partial\Phi_R}^c\phi^q
    -\frac{\partial\mathcal{V}}{\partial\Phi_I}^c\chi
    -\frac{1}{2}\frac{\partial^2\mathcal{V}}{\partial\Phi_R^2}^c\phi^{q2}
    -\frac{1}{2}\frac{\partial^2\mathcal{V}}{\partial\Phi_I^2}^c\chi^2
     \nonumber\\
 & &-\delta(z-L/2)\Bigg\{V_{+L/2}
    +\frac{\partial V_{+L/2}}{\partial\Phi_R}^c\phi^q
    +\frac{\partial V_{+L/2}}{\partial\Phi_I}^c\chi\nonumber\\
 & &\phantom{-\delta(z-L/2)\Bigg\{V_{+L/2}
    +\frac{\partial V_{+L/2}}{\partial\Phi_R}^c\phi^q}
    +\frac{1}{2}\frac{\partial^2 V_{+L/2}}{\partial\Phi_R^2}^c\phi^{q2}
    +\frac{1}{2}\frac{\partial^2 V_{+L/2}}{\partial\Phi_I^2}^c\chi^2\Bigg\}
    \nonumber\\
 & &-\delta(z-L/2)\Bigg\{V_{-L/2}
    +\frac{\partial V_{-L/2}}{\partial\Phi_R}^c\phi^q
    +\frac{\partial V_{-L/2}}{\partial\Phi_I}^c\chi\nonumber\\
 & &\phantom{-\delta(z+L/2)\Bigg\{V_{-L/2}
    +\frac{\partial V_{-L/2}}{\partial\Phi_R}^c\phi^q}
    +\frac{1}{2}\frac{\partial^2 V_{-L/2}}{\partial\Phi_R^2}^c\phi^{q2}
    +\frac{1}{2}\frac{\partial^2 V_{-L/2}}{\partial\Phi_I^2}^c\chi^2\Bigg\}
    \Bigg],\nonumber\\ 
 \label{115}
\end{eqnarray}
The partial integrals in terms of $z$ for the first and second term in  
integrand give
\begin{eqnarray}
 S_{\mbox{{\scriptsize free}},\phi^q,\chi}
 &=&\int d^4x\int_{-L/2}^{+L/2}dz
    \Bigg[-\frac{1}{2}(\partial_z\phi^q)^2+\phi^q\partial_z^2\Phi_R^c
          -\frac{1}{2}(\partial_z\chi)^2+\chi\partial_z^2\Phi_I^c\nonumber\\
 & &-\frac{\partial\mathcal{V}}{\partial\Phi_R}^c\phi^q
    -\frac{\partial\mathcal{V}}{\partial\Phi_I}^c\chi
    -\frac{1}{2}\frac{\partial^2\mathcal{V}}{\partial\Phi_R^2}^c\phi^{q2}
    -\frac{1}{2}\frac{\partial^2\mathcal{V}}{\partial\Phi_I^2}^c\chi^2
    \nonumber\\
 & &-\delta(z-L/2)\Bigg\{+\phi^q\partial_z\Phi_R^c+\chi\partial_z\Phi_I^c
    \nonumber\\
 & &\phantom{-\delta(z-L/2)\Bigg\{}
    +\frac{\partial V_{+L/2}}{\partial\Phi_R}^c\phi^q
    +\frac{\partial V_{+L/2}}{\partial\Phi_I}^c\chi
    +\frac{1}{2}\frac{\partial^2 V_{+L/2}}{\partial\Phi_R^2}^c\phi^{q2}
    +\frac{1}{2}\frac{\partial^2 V_{+L/2}}{\partial\Phi_I^2}^c\chi^2\Bigg\}
    \nonumber\\
 & &-\delta(z+L/2)\Bigg\{-\phi^q\partial_z\Phi_R^c+\chi\partial_z\Phi_I^c
    \nonumber\\
 & &\phantom{-\delta(z+L/2)\Bigg\{}
    +\frac{\partial V_{-L/2}}{\partial\Phi_R}^c\phi^q
    +\frac{\partial V_{-L/2}}{\partial\Phi_I}^c\chi
    +\frac{1}{2}\frac{\partial^2 V_{-L/2}}{\partial\Phi_R^2}^c\phi^{q2}
    +\frac{1}{2}\frac{\partial^2 V_{-L/2}}{\partial\Phi_I^2}^c\chi^2\Bigg\}
    \Bigg].\nonumber\\
\end{eqnarray}
By using (\ref{l6z}) and (\ref{l7z}), we obtain
\begin{eqnarray}
 S_{\mbox{{\scriptsize free}},\phi^q,\chi}
 &=&\int d^4x\int_{-L/2}^{+L/2}dz
    \Bigg[-\frac{1}{2}(\partial_z\phi^q)^2-\frac{1}{2}(\partial_z\chi)^2
    -\frac{1}{2}\frac{\partial^2\mathcal{V}}{\partial\Phi_R^2}^c\phi^{q2}
    -\frac{1}{2}\frac{\partial^2\mathcal{V}}{\partial\Phi_I^2}^c\chi^2
    \nonumber\\
 & &\phantom{\int d^4x\int_{-L/2}^{+L/2}dz\Bigg[}-\delta(z-L/2)\Bigg\{
    +\frac{1}{2}\frac{\partial^2 V_{+L/2}}{\partial\Phi_R^2}^c\phi^{q2}
    +\frac{1}{2}\frac{\partial^2 V_{+L/2}}{\partial\Phi_I^2}^c\chi^2\Bigg\}
    \nonumber\\
 & &\phantom{\int d^4x\int_{-L/2}^{+L/2}dz\Bigg[}-\delta(z+L/2)\Bigg\{
    +\frac{1}{2}\frac{\partial^2 V_{-L/2}}{\partial\Phi_R^2}^c\phi^{q2}
    +\frac{1}{2}\frac{\partial^2 V_{-L/2}}{\partial\Phi_I^2}^c\chi^2\Bigg\}
    \Bigg].\nonumber\\
\end{eqnarray}
Finally, partial integral of the first and second term lead to
\begin{eqnarray}
 S_{\mbox{{\scriptsize free}},\phi^q,\chi}
 &=&\int d^4x\int_{-L/2}^{+L/2}dz
    \Bigg[\frac{1}{2}\phi^q\partial_z^2\phi^q+\frac{1}{2}\chi\partial_z^2\chi
    -\frac{1}{2}\frac{\partial^2\mathcal{V}}{\partial\Phi_R^2}^c\phi^{q2}
    -\frac{1}{2}\frac{\partial^2\mathcal{V}}{\partial\Phi_I^2}^c\chi^2
    \nonumber\\
 & &-\frac{\delta(z-L/2)}{2}\Bigg\{+\phi^q\partial_z\phi^q+\chi\partial_z\chi
    +\frac{\partial^2 V_{+L/2}}{\partial\Phi_R^2}^c\phi^{q2}
    +\frac{\partial^2 V_{+L/2}}{\partial\Phi_I^2}^c\chi^2\Bigg\}
    \nonumber\\
 & &-\frac{\delta(z+L/2)}{2}\Bigg\{-\phi^q\partial_z\phi^q-\chi\partial_z\chi
    +\frac{\partial^2 V_{-L/2}}{\partial\Phi_R^2}^c\phi^{q2}
    +\frac{\partial^2 V_{-L/2}}{\partial\Phi_I^2}^c\chi^2\Bigg\}\Bigg].
    \nonumber\\
\end{eqnarray}
This is just $S_{\mbox{{\scriptsize free}},\phi^q}+S_{\mbox{{\scriptsize 
free}},\chi}$.

%%%%%%%%%%%%%%%%%%%%%%%%%%%%%%%%%%%%%%%%%%%%
\section{Gauge interactions of Higgs fields}
%%%%%%%%%%%%%%%%%%%%%%%%%%%%%%%%%%%%%%%%%%%%

We show our notation in four dimensional spacetime. The extension to five 
dimensions is straightforward, that is, replacements $\mu(=0\sim3)\rightarrow 
M(=0\sim4)$ and $(g_4,g_4')\rightarrow(g_5,g_5')$ is taken and all fields have 
dependence on extradimensional space $y$.

%%%%%%%%%%%%%%%%%%%%%%%%%%%%%%%%%%%%%
\subsection{Gauge and Higgs coupling}
%%%%%%%%%%%%%%%%%%%%%%%%%%%%%%%%%%%%%

We expand the Higgs field as follows,
\begin{eqnarray}
 \Phi&=&\left(
 \begin{array}{c}
 \varphi^+\\
 v+\frac{1}{\sqrt{2}}[H+i\chi] 
 \end{array}\right).
\end{eqnarray}
The covariant derivative is defined as
\begin{eqnarray}
 D_\mu\equiv\partial_\mu+ig_4W_\mu^aT^a+ig_4'B_\mu Y.
\end{eqnarray}
Therefore, 
\begin{eqnarray}
 \sum_{a=1}^2W_\mu^aT^a=\frac{1}{2}
 \left(
  \begin{array}{cc}
   0                & W_\mu^1-iW_\mu^2 \\
   W_\mu^1+iW_\mu^2 & 0               
  \end{array}
 \right)\equiv\frac{1}{\sqrt{2}}
 \left(
  \begin{array}{cc}
   0       & W_\mu^+ \\
   W_\mu^- & 0               
  \end{array}
 \right),
\end{eqnarray}
and
\begin{eqnarray}
 ig_4W_\mu^3T^3+ig_4'B_\mu Y&=&\frac{i}{2}
  \left(
   \begin{array}{cc}
    g_4W_\mu^3+g_4'B_\mu & 0                     \\
    0                    & -g_4W_\mu^3+g_4'B_\mu
   \end{array}
  \right)\\
 &=&ie
  \left(
   \begin{array}{cc}
    \frac{1}{\tan2\theta_W}Z_\mu+A_\mu & 0                            \\
    0                                  & -\frac{1}{\sin2\theta_W Z_\mu}
   \end{array}
  \right),
\end{eqnarray} 
where $\theta_W$ is the Weinberg angle and we used
\begin{eqnarray}
 \left(
  \begin{array}{c}
   W_\mu^3 \\
   B_\mu
  \end{array}
 \right)=
 \left(
  \begin{array}{cc}
   \cos\theta_W  & \sin\theta_W \\
   -\sin\theta_W & \cos\theta_W
  \end{array}
 \right)\left(
  \begin{array}{c}
   Z_\mu \\
   A_\mu
  \end{array}
 \right)\equiv\left(
  \begin{array}{cc}
   c  & s \\
   -s & c
  \end{array}
 \right)\left(
  \begin{array}{c}
   Z_\mu \\
   A_\mu
  \end{array}
 \right),
\end{eqnarray}
and
\begin{eqnarray}
 c  &\equiv&\frac{g_4}{\sqrt{g_4^2+g_4'{}^2}},\\
 s  &\equiv&\frac{g_4'}{\sqrt{g_4^2+g_4'{}^2}},\\
 e  &\equiv&\frac{g_4g_4'}{\sqrt{g_4^2+g_4'{}^2}}.
\end{eqnarray}
Finally, we have
\begin{eqnarray}
 D_\mu\Phi&=&=\partial_\mu\Phi+\frac{ig_4}{\sqrt{2}}  
 \left(
  \begin{array}{cc}
   0       & W_\mu^+ \\
   W_\mu^- & 0               
  \end{array}
 \right)\Phi+ie
 \left(
  \begin{array}{cc}
    \frac{1}{\tan2\theta_W}Z_\mu+A_\mu & 0                            \\
    0                                  & -\frac{1}{\sin2\theta_W Z_\mu}
   \end{array}
  \right)\Phi\nonumber\\
 &=&
  \left(
   \begin{array}{ccc}
    \partial_\mu\varphi^+\\
    \frac{\partial_\mu H+i\partial_\mu\chi}{\sqrt{2}}
   \end{array}
  \right)+\frac{ie}{\sqrt{2}s}  
 \left(
  \begin{array}{c}
   W_\mu^+(v+\frac{H+i\chi}{\sqrt{2}}) \\
   W_\mu^-\varphi^+
  \end{array}
 \right)+ie
 \left(
  \begin{array}{c}
    (\frac{1}{\tan2\theta_W}Z_\mu+A_\mu)\varphi^+ \\
    -\frac{1}{\sin2\theta_W}Z_\mu(v+\frac{H+i\chi}{\sqrt{2}})
   \end{array}
  \right).\nonumber\\
 &&
 \label{dp}
\end{eqnarray}

The gauge boson mass and, 3 and 4-point couplings among the gauge and Higgs fields are obtained from the Higgs kinetic term,
\begin{eqnarray}
 \mathcal{L}_{kin}=\int d^4x|D_\mu\Phi|^2.
\end{eqnarray}
According to the~(\ref{dp}),
\begin{eqnarray}
 |D_\mu\Phi|^2&=&\left|\partial_\mu\varphi^+
                 +\frac{ie}{\sqrt{2}s}W_\mu^+
                  \left(v+\frac{H+i\chi}{\sqrt{2}}\right)
                 +ie\left(\frac{1}{\tan2\theta_W}Z_\mu+A_\mu\right)\varphi^+
                 \right|^2\nonumber\\
              & &+\left|\frac{\partial_\mu H+i\partial_\mu\chi}{\sqrt{2}} 
                 +\frac{ie}{\sqrt{2}s}W_\mu^-\varphi^+
                 -\frac{ie}{\sin2\theta_W}Z_\mu
                  \left(v+\frac{H+i\chi}{\sqrt{2}}\right)\right|^2.
\end{eqnarray}
The quadratic terms of $|D_\mu\Phi|^2$ are
\begin{eqnarray}
 \left.|D_\mu\Phi|^2\right|_{\mbox{quadratic}}
 &=&|\partial_\mu\varphi^+|^2
    +\left(\frac{ie}{\sqrt{2}s}(\partial_\mu\varphi^-)W^{+\mu}v+\text{h.c.}\right)
    +m_W^2W_\mu^+W^{-\mu}\nonumber\\
 & &+\frac{1}{2}((\partial_\mu H)^2+(\partial_\mu\chi)^2)+\frac{m_Z^2}{2}Z_\mu Z^\mu
    -m_Z(\partial_\mu\chi)Z^\mu.
\end{eqnarray}
The tri-linear terms are
\begin{eqnarray}
 \left.|D_\mu\Phi|^2\right|_{\mbox{tri-linear}}
 &=&\left[ie(\partial_\mu\varphi^--im_WW_\mu^-)
    \left\{\frac{1}{2s}W^{+\mu}(H+i\chi)
    +\left(\frac{Z^\mu}{\tan2\theta_W}+A^\mu\right)\varphi^+\right\}\right.
    \nonumber\\
 & &\hspace{5mm}\left.+\frac{ie}{s}
    \left(\frac{\partial_\mu H-i\partial_\mu\chi}{\sqrt{2}}
    +im_ZZ_\mu\right)\frac{W^{-\mu}\varphi^+}{\sqrt{2}}+\text{h.c.}\right]\nonumber\\
 & &+\frac{e}{\sin2\theta_W}Z^\mu
     \left\{\chi\partial_\mu H+H(-\partial_\mu\chi+m_ZZ_\mu)\right\}.
\end{eqnarray}
The quartic terms become
\begin{eqnarray}
 \left.|D_\mu\Phi|^2\right|_{\mbox{quartic}}
 &=&\frac{e^2}{4s^2}W_\mu^+W_\mu^{-\mu}(H^2+\chi^2)
    +e^2\left(\frac{Z_\mu}{\tan2\theta_W}+A_\mu\right)^2\varphi^+\varphi^-
    \nonumber\\
 & &+\left\{\frac{e^2}{2s}W_\mu^-(H-i\chi)\left(\frac{Z_\mu}{\tan2\theta_W}
    +A_\mu\right)\varphi^++\text{h.c.}\right\}\nonumber\\
 & &+\frac{e^2}{2s^2}W_\mu^+W^{-\mu}\varphi^+\varphi^-
    +\frac{e^2}{2\sin^22\theta_W}Z_\mu Z^\mu(H^2+\chi^2)\nonumber\\
 & &+\frac{1}{2}\left\{-\frac{e^2}{2s^2c}W_\mu^-Z^\mu\varphi^+(H-i\chi)
    +\text{h.c.}\right\}.
\end{eqnarray}

We take the following gauge fixing Lagrangian,
\begin{eqnarray}
 \mathcal{L}_\xi&=&-\frac{1}{2\xi}\left(\sum_{a=1}^3f^af^a+f^Bf^B\right)
                   \nonumber\\
                &=&-\frac{1}{2\xi}(f^+f^-+f^Zf^Z+f^Af^A),
\end{eqnarray}
where
\begin{eqnarray}
 f^\pm&\equiv&\frac{1}{\sqrt{2}}(f^1\mp if^2)
            =\partial_\mu W^{\pm\mu}\mp i\xi m_W\varphi^\pm,\\
 f^Z&\equiv&cf^3-sf^B=\partial_\mu Z^\mu-\xi m_Z\chi,\\
 f^A&\equiv&sf^3+cf^B=\partial_\mu A^\mu.
\end{eqnarray}
It is easily seen that
\begin{eqnarray}
 \mathcal{L}_{kin}|_{\mbox{quadratic}}+\mathcal{L}_{\xi}|_{\xi=1}
 &=&-\left[\partial_\mu\varphi^+\partial^\mu\varphi^-+m_W^2W_\mu^+W^{-\mu}
    +\frac{(\partial_\mu H)^2+(\partial_\mu\chi)^2}{2}\right.\nonumber\\
 & &\hspace{7mm}+\frac{m_Z^2}{2}Z_\mu Z^\mu+\partial_\mu W^{+\mu}\partial_\nu W^{-\nu}
    \nonumber\\
 & &\hspace{7mm}\left.+\frac{1}{2}\left\{\partial_\mu Z^\mu)^2
    +(\partial_\mu A^\mu)^2+m_Z^2\chi^2\right\}\right].
\end{eqnarray}

%%%%%%%%%%%%%%%%%%%%%%%
\section{Feynman rules}
%%%%%%%%%%%%%%%%%%%%%%%

%%%%%%%%%%%%%%%%%%%%%%%%%%%%%%%%%%%%%%%%%%%%%
\subsection{Feynman rules in four dimensions}
%%%%%%%%%%%%%%%%%%%%%%%%%%%%%%%%%%%%%%%%%%%%%

We give four dimensional Feynman rules in our notation.

Propagators:
\begin{eqnarray}
 &k&\nonumber\\\vspace{-15mm}
 &V_\mu\includegraphics[scale = 1.0]{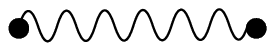}V_\nu&
 =\frac{-ig_{\mu\nu}}{k^2-m_V^2},\\
 &S\hspace{1mm}\includegraphics[scale = 1.0]{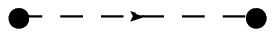}\hspace{1mm}S&
 =\frac{ig_{\mu\nu}}{k^2-m_S^2}.
\end{eqnarray}

Vector 4-point coupling:

\hspace{3cm}$V_{1,\mu}$\hspace{1.5cm}$V_{3,\rho}$

\hspace{3cm}\includegraphics[scale =1.0]{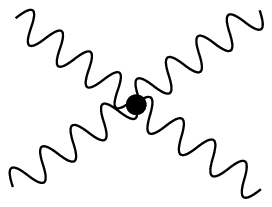}\vspace{-2.6cm}

\begin{eqnarray}
 \hspace{4cm}=-e^2C[2g_{\mu\nu}g_{\sigma\rho}-g_{\nu\rho}g_{\mu\sigma}
              -g_{\rho\mu}g_{\nu\sigma}]
\end{eqnarray}\vspace{-4mm}

\hspace{3cm}$V_{2,\nu}$\hspace{1.5cm}$V_{4,\sigma}$
\begin{eqnarray}\hspace{-3cm}
\begin{array}{lll}
W^+W^+W^-W^- & : & C=\frac{1}{s^2},\\
W^+W^-ZZ     & : & C=-\frac{c^2}{s^2},\\
W^+W^-AZ     & : & C=-\frac{c}{s},\\
W^+W^-AA       & : & C=-1.
\end{array}
\end{eqnarray}

\newpage
Vector 3-point coupling:

\hspace{5.5cm}$V_{2,\nu},k_2$

\hspace{2.5cm}\includegraphics[scale =1.0]{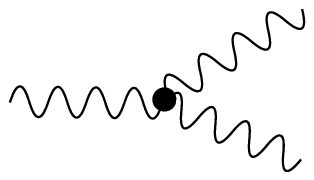}\vspace{-2.5cm}

\begin{eqnarray}
 \hspace{0.4cm}&&V_{1,\mu},k_1\hspace{2.6cm}=-eC[g_{\mu\nu}(k_2-k_1)_\rho
                 +g_{\nu\rho}(k_3-k_2)_\mu+g_{\rho\mu}(k_1-k_3)_\nu]
 \nonumber\\
\end{eqnarray}\vspace{-1.3cm}

\hspace{5.5cm}$V_{3,\rho},k_3$
\begin{eqnarray}\hspace{-4cm}
\begin{array}{lll}
AW^+W^- & : & C=1,\\
ZW^+W^- & : & C=\frac{c}{s}.
\end{array}
\end{eqnarray}

Vector and scalar 4-point coupling:

\hspace{2cm}$V_{1,\mu}$\hspace{2.5cm}$S_1$

\hspace{2.5cm}\includegraphics[scale =1.0]{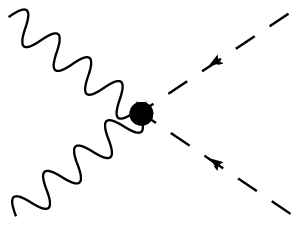}\vspace{-2.5cm}

\begin{eqnarray}
=-e^2g_{\mu\nu}C
\end{eqnarray}\vspace{-3mm}

\hspace{2cm}$V_{2,\nu}$\hspace{2.5cm}$S_2$
\begin{eqnarray}\hspace{-2cm}
\begin{array}{lll}
W^+W^+HH                 & : & C=\frac{1}{2s^2},              \\
W^+W^-\chi\chi           & : & C=\frac{1}{2s^2},              \\
W^+W^-\varphi^+\varphi^- & : & C=\frac{1}{2s^2},              \\
ZZ\varphi^+\varphi^-     & : & C=\frac{(s^2-c^2)^2}{2s^2c^2}, \\
ZA\varphi^+\varphi^-     & : & C=-\frac{s^2-c^2}{sc},         \\ 
AA\varphi^+\varphi^-     & : & C=2,                           \\
ZZHH                     & : & C=\frac{1}{2s^2c^2},           \\
ZZ\chi\chi               & : & C=\frac{1}{2s^2c^2},           \\
W^\pm Z\varphi^\mp H     & : & C=-\frac{1}{2c},               \\
W^\pm A\varphi^\mp H     & : & C=\frac{1}{2s},                \\
W^\pm Z\varphi^\mp\chi   & : & C=\mp\frac{i}{2c},             \\
W^\pm A\varphi^\mp\chi   & : & C=\pm\frac{i}{2c}. 
\end{array}
\end{eqnarray}

Vector and scalar 3-point coupling:

\hspace{5.5cm}$S_1,k_1$

\hspace{2.5cm}\includegraphics[scale =1.0]{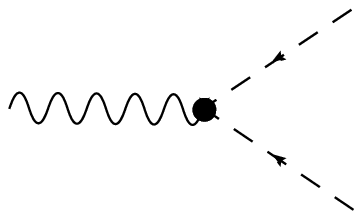}\vspace{-2.5cm}

\begin{eqnarray}
\hspace{-4cm}V_\mu\hspace{3cm}=eC(k_1-k_2)_\mu
\end{eqnarray}\vspace{-3mm}

\hspace{5.5cm}$S_2,k_2$
\begin{eqnarray}\hspace{-2cm}
\begin{array}{lll}
Z\chi H              & : & C=\frac{i}{2cs},       \\
A\varphi^+\varphi^-  & : & C=-1                   \\
Z\varphi^+\varphi^-  & : & C=\frac{s^2-c^2}{2sc}, \\
W^\pm\varphi^\mp H   & : & C=\pm\frac{1}{2s},     \\
W^\pm\varphi^\mp\chi & : & C=\frac{i}{2s},
\end{array}
\end{eqnarray}

\hspace{6cm}$V_{1,\mu}$

\hspace{2.5cm}\includegraphics[scale =1.0]{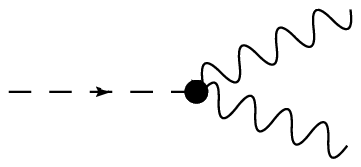}\vspace{-2.3cm}

\begin{eqnarray}
\hspace{-4.6cm}S\hspace{3.2cm}=eg_{\mu\nu}C
\end{eqnarray}\vspace{-4mm}

\hspace{6cm}$V_{2,\nu}$
\begin{eqnarray}\hspace{-2cm}
\begin{array}{lll}
HW^+W^-             & : & C=-\frac{m_W}{s},    \\
HZZ                 & : & C=-\frac{m_W}{sc^2}, \\
\varphi^\pm W^\mp Z & : & C=\frac{sm_W}{c},    \\
\varphi^\pm W^\mp A & : & C=-m_W.     
\end{array}
\end{eqnarray}

%%%%%%%%%%%%%%%%%%%%%%%%%%%%%%%%%%%%%%%%%%%%%%%%%%%%%%%%%%%
\subsection{Five dimensional Feynman rules in our scenario}
%%%%%%%%%%%%%%%%%%%%%%%%%%%%%%%%%%%%%%%%%%%%%%%%%%%%%%%%%%%

We give five dimensional Feynman rules in our two scenarios, the 
Brane-Localized Fermion (BLF) and Bulk Fermions (BF) scenarios.

Propagators:
\begin{eqnarray}
 &V_M^{(n)}\includegraphics[scale = 1.0]{gauge.eps}V_N^{(n)}&
 =\frac{-ig_{MN}}{k^2-m_V^2},\\
 &S^{(n)}\hspace{1mm}\includegraphics[scale = 1.0]{scalar.eps}\hspace{1mm}
 S^{(n)}&
 =\frac{ig_{MN}}{k^2-m_S^2}
\end{eqnarray}
where $n$ is the KK number. In the BLF and BF scenario, mode mixing can 
generally occur in the Higgs sector. Therefore, the Higgs propagator is 
effectively multiplied by 
$\mathcal{O}(\sqrt{\lambda}v_{\text{EW}})$ in the mass insertion 
method when corresponding amplitude is calculated. The propagator is given as
\begin{eqnarray}
 &H^{(n)}\hspace{1mm}\includegraphics[scale = 1.0]{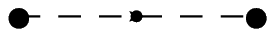}
 \hspace{1mm}
 H^{(m)}&
 =\frac{ig_{MN}}{k^2-m_S^2}\times
  \mathcal{O}(\sqrt{\lambda}v_{\text{EW}}).
\end{eqnarray}

Vector 4-point coupling:

\hspace{3cm}$V_{1,K}^{(l+m+n)}$\hspace{5mm}$V_{3,M}^{(m)}$

\hspace{3cm}\includegraphics[scale =1.0]{gauge4.eps}\vspace{-2.6cm}

\begin{eqnarray}
 \hspace{4cm}=-e^2C[2g_{KL}g_{NM}-g_{LM}g_{KN}-g_{MK}g_{LN}]
\end{eqnarray}\vspace{-4mm}

\hspace{3cm}$V_{2,L}^{(l)}$\hspace{1.5cm}$V_{4,N}^{(n)}$
\begin{eqnarray}\hspace{-3cm}
\begin{array}{lll}
W^+W^+W^-W^- & : & C=\frac{1}{s^2},\\
W^+W^-ZZ     & : & C=-\frac{c^2}{s^2},\\
W^+W^-AZ     & : & C=-\frac{c}{s},\\
W^+W^-AA     & : & C=-1.
\end{array}
\end{eqnarray}

\newpage
Vector 3-point coupling:

\hspace{5.5cm}$V_{2,M}^{(m)},k_2$

\hspace{2.5cm}\includegraphics[scale =1.0]{gauge3.eps}\vspace{-2.5cm}

\begin{eqnarray}
 \hspace{0.4cm}&&V_{1,L}^{(m+n)},
 k_1\hspace{2.6cm}=-eC[g_{LM}(k_2-k_1)_N+g_{MN}(k_3-k_2)_L+g_{NL}(k_1-k_3)_M]
 \nonumber\\
\end{eqnarray}\vspace{-1.3cm}

\hspace{5.5cm}$V_{3,N}^{(n)},k_3$
\begin{eqnarray}\hspace{-4cm}
\begin{array}{lll}
AW^+W^- & : & C=1,\\
ZW^+W^- & : & C=\frac{c}{s}.
\end{array}
\end{eqnarray}

Vector and scalar 4-point coupling:

\hspace{2cm}$V_{1,M}^{(2m)}$\hspace{2.5cm}$S_1^{(2l)}$\hspace{1.5cm}$V_{1,M}^{(2m)}$\hspace{2.5cm}$S_1^{(2l+1)}$

\hspace{2.5cm}\includegraphics[scale =1.0]{vvss.eps}
\hspace{2.5cm}\includegraphics[scale =1.0]{vvss.eps}\vspace{-2.5cm}

\begin{eqnarray}
\hspace{4cm}=-e^2g_{MN}CC_{eeee}\hspace{2.5cm}=-e^2g_{MN}CC_{eeoo}
\end{eqnarray}\vspace{-3mm}

\hspace{2cm}$V_{2,N}^{(2n)}$\hspace{2.5cm}$S_2^{(2|m+n-l|)}$
\hspace{0.3cm}$V_{2,N}^{(2n)}$\hspace{2.5cm}$S_2^{(|2(m+n-l)+1|)}$\vspace{2mm}

\hspace{2cm}$V_{1,M}^{(2m+1)}$\hspace{2cm}$S_1^{(2l)}$\hspace{1.6cm}$V_{1,M}^{(2m+1)}$\hspace{2.1cm}$S_1^{(2l+1)}$

\hspace{2.5cm}\includegraphics[scale =1.0]{vvss.eps}
\hspace{2.5cm}\includegraphics[scale =1.0]{vvss.eps}\vspace{-2.5cm}

\begin{eqnarray}
\hspace{4cm}=-e^2g_{MN}CC_{ooee}\hspace{2.5cm}=-e^2g_{MN}CC_{oooo}
\end{eqnarray}\vspace{-3mm}

\hspace{2cm}$V_{2,N}^{(2n+1)}$\hspace{2cm}$S_2^{(2|m+n+1-l|)}
$\hspace{2mm}$V_{2,N}^{(2n+1)}$\hspace{2.1cm}$S_2^{(|2(m+n-l)+1|)}$\vspace{2mm}

\hspace{2cm}$V_{1,M}^{(2m)}$\hspace{2.3cm}$S_1^{(2l)}$

\hspace{2.5cm}\includegraphics[scale =1.0]{vvss.eps}\vspace{-2.5cm}

\begin{eqnarray}
\hspace{-1.5cm}=-e^2g_{MN}CC_{eoeo}
\end{eqnarray}\vspace{-3mm}

\hspace{2cm}$V_{2,N}^{(2n+1)}$\hspace{2cm}$S_2^{(|2(m+n-l)+1|)}$\vspace{-4.5cm}

\begin{eqnarray}\hspace{-2cm}
\begin{array}{lll}
\hspace{9.5cm}W^+W^-HH                 & : & C=\frac{1}{2s^2},              \\
\hspace{9.5cm}W^+W^-\chi\chi           & : & C=\frac{1}{2s^2},              \\
\hspace{9.5cm}W^+W^-\varphi^+\varphi^- & : & C=\frac{1}{2s^2},              \\
\hspace{9.5cm}ZZ\varphi^+\varphi^-     & : & C=\frac{(s^2-c^2)^2}{2s^2c^2}, \\
\hspace{9.5cm}ZA\varphi^+\varphi^-     & : & C=-\frac{s^2-c^2}{sc},         \\ 
\hspace{9.5cm}AA\varphi^+\varphi^-     & : & C=2,                           \\
\hspace{9.5cm}ZZHH                     & : & C=\frac{1}{2s^2c^2},           \\
\hspace{9.5cm}ZZ\chi\chi               & : & C=\frac{1}{2s^2c^2},           \\
\hspace{9.5cm}W^\pm Z\varphi^\mp H     & : & C=-\frac{1}{2c},               \\
\hspace{9.5cm}W^\pm A\varphi^\mp H     & : & C=\frac{1}{2s},                \\
\hspace{9.5cm}W^\pm Z\varphi^\mp\chi   & : & C=\mp\frac{i}{2c},             \\
\hspace{9.5cm}W^\pm A\varphi^\mp\chi   & : & C=\pm\frac{i}{2c}. 
\end{array}\nonumber
\end{eqnarray}
where $C_{\ast\ast\ast\ast}$ $(\ast=e,o)$ are additional factors, which is 
caused by changing the wave function profile of the Higgs. We obtain\newpage
\begin{eqnarray}
 C_{eeee}&=&f_{2m}^{\ast e}(z)f_{2n}^{\ast e}(z)f_{2l}^e(z)f_{2|m+n-l|}^e(z),
            \label{eeee}\\
 C_{eeoo}&=&f_{2m}^{\ast e}(z)f_{2n}^{\ast e}(z)f_{2l+1}^o(z)
            f_{|2(m+n-l)+1|}^o(z),\\
 C_{ooee}&=&f_{2m+1}^{\ast o}(z)f_{2n+1}^{\ast o}(z)f_{2l}^e(z)
            f_{2|m+n-l+1|}^e(z),\\
 C_{oooo}&=&f_{2m+1}^{\ast o}(z)f_{2n+1}^{\ast o}(z)f_{2l+1}^o(z)
            f_{|2(m+n-l)+1|}^o(z),\\
 C_{eoeo}&=&f_{2m}^{\ast e}(z)f_{2n+1}^{\ast o}(z)f_{2l}^e(z)
            f_{|2(m+n-l)+1|}^o(z),\label{eoeo}
\end{eqnarray}
for $W^+W^-HH$ and $ZZHH$ coupling, where 
\begin{eqnarray}
 f_n^{\ast e}(z)&=&\sqrt{\frac{1}{L}}\cos\left(\frac{n\pi}{L}z\right),\\
 f_n^{\ast o}(z)&=&\sqrt{\frac{1}{L}}\sin\left(\frac{n\pi}{L}z\right),\\
 f_n^e(z)       &=&\sqrt{\frac{2}{L\left[1+\frac{\sin((n+\Delta_n)\pi)}
                   {(n+\Delta_n)\pi}\right]}}\cos\left(\frac{(n+\Delta_n)\pi}
                                                            {L}z\right),\\
 f_n^o(z)       &=&\sqrt{\frac{2}{L\left[1-\frac{\sin((n+\Delta_n)\pi)}
                   {(n+\Delta_n)\pi}\right]}}\sin\left(\frac{(n+\Delta_n)\pi}
                                                            {L}z\right),
\end{eqnarray}
and $\Delta_n$ determines a deviation of the Higgs profile. At the limits of 
$\lambda\rightarrow0$ and $\infty$, $\Delta_n$ can be parametrized as 
$\epsilon_n$ and $1-\epsilon_n$, respectively, where $\epsilon_n\ll1$. For 
couplings, $W^\pm Z\varphi^\mp H$ and $W^\pm A\varphi^\mp H$, we have
\begin{eqnarray}
 C_{eeee}&=&f_{2m}^{\ast e}(z)f_{2n}^{\ast e}(z)f_{2l}^{\ast e}(z)
            f_{2|m+n-l|}^e(z),\\
 C_{eeoo}&=&f_{2m}^{\ast e}(z)f_{2n}^{\ast e}(z)f_{2l+1}^{\ast o}(z)
            f_{|2(m+n-l)+1|}^o(z),\\
 C_{ooee}&=&f_{2m+1}^{\ast o}(z)f_{2n+1}^{\ast o}(z)f_{2l}^{\ast e}(z)
            f_{2|m+n-l+1|}^e(z),\\
 C_{oooo}&=&f_{2m+1}^{\ast o}(z)f_{2n+1}^{\ast o}(z)f_{2l+1}^{\ast o}(z)
            f_{|2(m+n-l)+1|}^o(z),\\
 C_{eoeo}&=&f_{2m}^{\ast e}(z)f_{2n+1}^{\ast o}(z)f_{2l}^{\ast e}(z)
            f_{|2(m+n-l)+1|}^o(z),
\end{eqnarray}
Finally, for other four-point couplings among vectors and scalars without 
Higgs, the following additional factors are written down,
\begin{eqnarray}
 C_{eeee}&=&f_{2m}^{\ast e}(z)f_{2n}^{\ast e}(z)f_{2l}^{\ast e}(z)
            f_{2|m+n-l|}^{\ast e}(z),\label{eeee1}\\
 C_{eeoo}&=&f_{2m}^{\ast e}(z)f_{2n}^{\ast e}(z)f_{2l+1}^{\ast o}(z)
            f_{|2(m+n-l)+1|}^{\ast o}(z),\\
 C_{ooee}&=&f_{2m+1}^{\ast o}(z)f_{2n+1}^{\ast o}(z)f_{2l}^{\ast e}(z)
            f_{2|m+n-l+1|}^{\ast e}(z),\\
 C_{oooo}&=&f_{2m+1}^{\ast o}(z)f_{2n+1}^{\ast o}(z)f_{2l+1}^{\ast o}(z)
            f_{|2(m+n-l)+1|}^{\ast o}(z),\\
 C_{eoeo}&=&f_{2m}^{\ast e}(z)f_{2n+1}^{\ast o}(z)f_{2l}^{\ast e}(z)
            f_{|2(m+n-l)+1|}^{\ast o}(z).\label{eoeo1}
\end{eqnarray}

Scalar 4-point coupling:

\hspace{2cm}$S_1^{(2m)}$\hspace{2.5cm}$S_3^{(2l)}$\hspace{1.5cm}$S_1^{(2m)}$\hspace{2.3cm}$S_3^{(2l+1)}$

\hspace{2.5cm}\includegraphics[scale =1.0]{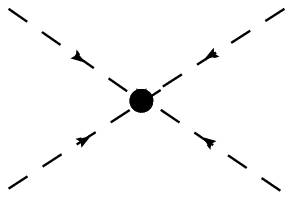}
\hspace{2.5cm}\includegraphics[scale =1.0]{scalar4.eps}\vspace{-2.5cm}

\begin{eqnarray}
\hspace{4cm}=CC_{eeee}\hspace{4cm}=CC_{eeoo}
\end{eqnarray}\vspace{-3mm}

\hspace{2cm}$S_2^{(2n)}$\hspace{2.5cm}$S_4^{(2|m+n-l|)}$
\hspace{0.3cm}$S_2^{(2n)}$\hspace{2.3cm}$S_4^{(|2(m+n-l)+1|)}$\vspace{2mm}

\hspace{2cm}$S_1^{(2m+1)}$\hspace{2cm}$S_3^{(2l)}$\hspace{1.6cm}$S_1^{(2m+1)}$\hspace{1.8cm}$S_3^{(2l+1)}$

\hspace{2.5cm}\includegraphics[scale =1.0]{scalar4.eps}
\hspace{2.5cm}\includegraphics[scale =1.0]{scalar4.eps}\vspace{-2.5cm}

\begin{eqnarray}
\hspace{4cm}=CC_{ooee}\hspace{4cm}=CC_{oooo}
\end{eqnarray}\vspace{-3mm}

\hspace{2cm}$S_2^{(2n+1)}$\hspace{2cm}$S_4^{(2|m+n+1-l|)}
$\hspace{2mm}$S_2^{(2n+1)}$\hspace{1.8cm}$S_4^{(|2(m+n-l)+1|)}$\vspace{2mm}

\hspace{2cm}$S_1^{(2m)}$\hspace{2.3cm}$S_3^{(2l)}$

\hspace{2.5cm}\includegraphics[scale =1.0]{scalar4.eps}\vspace{-2.5cm}

\begin{eqnarray}
\hspace{-1.5cm}=CC_{eoeo}
\end{eqnarray}\vspace{-3mm}

\hspace{2cm}$S_2^{(2n+1)}$\hspace{2cm}$S_4^{(|2(m+n-l)+1|)}$\vspace{-4cm}

\begin{eqnarray}\hspace{-2cm}
\begin{array}{lll}
\hspace{9.5cm}HHHH                       & : & C=\frac{12\lambda}{L^2}, \\
\hspace{9.5cm}\chi\chi HH                & : & C=\frac{2\lambda}{L^2},  \\
\hspace{9.5cm}\chi\chi\chi\chi           & : & C=\frac{3\lambda}{L^2},  \\
\hspace{9.5cm}HH\varphi^+\varphi^-       & : & C=\frac{2\lambda}{L^2},   \\
\hspace{9.5cm}\chi\chi\varphi^+\varphi^- & : & C=\frac{2\lambda}{L^2},  \\
\hspace{9.5cm}\varphi^+\varphi^+\varphi^-\varphi^- & : & C=\frac{2\lambda}{L^2},    
\end{array}\nonumber
\end{eqnarray}\vspace{0cm}

\noindent where for $HHHH$,
\begin{eqnarray} 
 C_{eeee}&=&f_{2m}^e(z)f_{2n}^e(z)f_{2l}^e(z)f_{2|m+n-l|}^e(z),\\
 C_{eeoo}&=&f_{2m}^e(z)f_{2n}^e(z)f_{2l+1}^o(z)f_{|2(m+n-l)+1|}^o(z),\\
 C_{ooee}&=&f_{2m+1}^o(z)f_{2n+1}^o(z)f_{2l}^e(z)f_{2|m+n-l+1|}^e(z),\\
 C_{oooo}&=&f_{2m+1}^o(z)f_{2n+1}^o(z)f_{2l+1}^o(z)f_{|2(m+n-l)+1|}^o(z),\\
 C_{eoeo}&=&f_{2m}^e(z)f_{2n+1}^o(z)f_{2l}^e(z)f_{|2(m+n-l)+1|}^o(z).
\end{eqnarray}
For $\chi\chi HH$ coupling, $C_{\ast\ast\ast\ast}$ is given 
by~(\ref{eeee})$-$(\ref{eoeo}), and for $\chi\chi\chi\chi$, 
$C_{\ast\ast\ast\ast}$ correspond to~(\ref{eeee1})$-$(\ref{eoeo1}), 
respectively.

\newpage
Vector and scalar 3-point coupling:

\hspace{7.5cm}$S_1^{(2n)},k_1$

\hspace{4.5cm}\includegraphics[scale =1.0]{vss.eps}\vspace{-2.5cm}

\begin{eqnarray}
V_M^{(2m)}\hspace{2.7cm}=eCC_{eee}(k_1-k_2)_M
\end{eqnarray}\vspace{-3mm}

\hspace{7.5cm}$S_2^{(2|m-n|)},k_2$\hspace{5.5cm}

\hspace{7.5cm}$S_1^{(2n+1)},k_1$

\hspace{4.5cm}\includegraphics[scale =1.0]{vss.eps}\vspace{-2.5cm}

\begin{eqnarray}
V_M^{(2m)}\hspace{2.7cm}=eCC_{eoo}(k_1-k_2)_M
\end{eqnarray}\vspace{-3mm}

\hspace{7.5cm}$S_2^{(|2(m-n)-1|)},k_2$

\hspace{7.5cm}$S_1^{(2n)},k_1$

\hspace{4.6cm}\includegraphics[scale =1.0]{vss.eps}\vspace{-2.5cm}

\begin{eqnarray}
 V_M^{(2m+1)}\hspace{2.7cm}=eCC_{oeo}(k_1-k_2)_M
\end{eqnarray}\vspace{-3mm}

\hspace{7.5cm}$S_2^{(2|m-n+1|)},k_2$

\hspace{7.5cm}$S_1^{(2n+1)},k_1$

\hspace{4.6cm}\includegraphics[scale =1.0]{vss.eps}\vspace{-2.5cm}

\begin{eqnarray}
 V_M^{(2m+1)}\hspace{2.7cm}=eCC_{ooe}(k_1-k_2)_M
\end{eqnarray}\vspace{-3mm}

\hspace{7.5cm}$S_2^{(2|m-n|)},k_2$

\begin{eqnarray}
\begin{array}{lll}
Z\chi H              & : & C=\frac{i}{2cs},       \\
A\varphi^+\varphi^-  & : & C=-1                   \\
Z\varphi^+\varphi^-  & : & C=\frac{s^2-c^2}{2sc}, \\
W^\pm\varphi^\mp H   & : & C=\pm\frac{1}{2s},     \\
W^\pm\varphi^\mp\chi & : & C=\frac{i}{2s},
\end{array}
\end{eqnarray}
where $C_{\ast\ast\ast}$ are also additional factors given as
\begin{eqnarray}
 C_{eee}&=&f_{2m}^{\ast e}(z)f_{2n}^{\ast e}(z)f_{2|m-n|}^e(z),\label{eee}\\
 C_{eoo}&=&f_{2m}^{\ast e}(z)f_{2n+1}^{\ast o}(z)f_{2|m-n-1|}^o(z),
           \label{eoo}\\
 C_{oeo}&=&f_{2m+1}^{\ast o}(z)f_{2n}^{\ast e}(z)f_{2|m-n+1|}^o(z),
           \label{oeo}\\
 C_{ooe}&=&f_{2m+1}^{\ast o}(z)f_{2n+1}^{\ast o}(z)f_{2|m-n|}^e(z),
\end{eqnarray}
for $Z\chi H$ and $W^\pm\varphi^\mp H$ couplings. And for 
$A\varphi^+\varphi^-$, $Z\varphi^+\varphi^-$, and $W^\pm\varphi^\mp\chi$ 
couplings, we obtain
\begin{eqnarray}
 C_{eee}&=&f_{2m}^{\ast e}(z)f_{2n}^{\ast e}(z)f_{2|m-n|}^{\ast e}(z),
           \label{eee1}\\
 C_{eoo}&=&f_{2m}^{\ast e}(z)f_{2n+1}^{\ast o}(z)f_{2|m-n-1|}^{\ast o}(z),
           \label{eoo1}\\
 C_{oeo}&=&f_{2m+1}^{\ast o}(z)f_{2n}^{\ast e}(z)f_{2|m-n+1|}^{\ast o}(z),
           \label{oeo1}\\
 C_{ooe}&=&f_{2m+1}^{\ast o}(z)f_{2n+1}^{\ast o}(z)f_{2|m-n|}^{\ast e}(z).
\end{eqnarray}

\hspace{5cm}$V_{1,M}^{(2m)}$\hspace{6cm}$V_{1,M}^{(2m)}$

\hspace{1.8cm}\includegraphics[scale =1.0]{svv.eps}
\hspace{3.3cm}\includegraphics[scale =1.0]{svv.eps}\vspace{-2.5cm}

\begin{eqnarray}
\hspace{2mm}S^{(2|m-n|)}\hspace{2.7cm}=eg_{MN}CC_{eee}
\hspace{3mm}S^{(2|m-n-1|)}\hspace{2.7cm}=eg_{MN}CC_{eoo}
\end{eqnarray}\vspace{-4mm}

\hspace{5cm}$V_{2,N}^{(2n)}$\hspace{6cm}$V_{2,N}^{(2n+1)}$\vspace{1mm}

\hspace{5cm}$V_{1,M}^{(2m+1)}$

\hspace{2.3cm}\includegraphics[scale =1.0]{svv.eps}\vspace{-2.5cm}

\begin{eqnarray}
 \hspace{-7cm}S^{(2|m-n+1|)}\hspace{2.7cm}=eg_{MN}CC_{oeo}
 \nonumber
\end{eqnarray}\vspace{-3mm}

\hspace{5cm}$V_{2,N}^{(2n)}$\vspace{-3.5cm}

\begin{eqnarray}\hspace{8cm}
\begin{array}{lll}
HW^+W^-             & : & C=-\frac{m_W}{s},    \\
HZZ                 & : & C=-\frac{m_W}{sc^2}, \\
\varphi^\pm W^\mp Z & : & C=\frac{sm_W}{c},    \\
\varphi^\pm W^\mp A & : & C=-m_W     
\end{array}
\end{eqnarray}\vspace{5mm}

\noindent where $C_{eee}$, $C_{eoo}$, and $C_{oeo}$ for $HW^+W^-$ $HZZ$ are 
given in~(\ref{eee}),~(\ref{eoo}), and~(\ref{oeo}), respectively. For 
$\varphi^\pm W^\mp Z$ and $\varphi^\pm W^\mp A$, $C_{eee}$, $C_{eoo}$, and 
$C_{oeo}$ for $HW^+W^-$ $HZZ$ are given in~(\ref{eee1}),~(\ref{eoo1}), and 
(\ref{oeo1}), respectively.

Scalar 3-point coupling:

\hspace{5cm}$S_2^{(2m)}$\hspace{6cm}$S_2^{(2m)}$

\hspace{1.8cm}\includegraphics[scale =1.0]{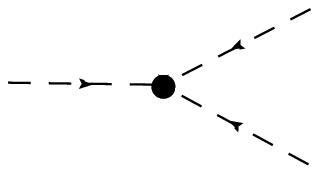}
\hspace{3.6cm}\includegraphics[scale =1.0]{scalar3.eps}\vspace{-2.5cm}

\begin{eqnarray}
\hspace{-0.8cm}S_1^{(2|m-n|)}\hspace{2.9cm}=CC_{eee}
\hspace{5mm}S_1^{(2|m-n-1|)}\hspace{2.7cm}=CC_{eoo}
\end{eqnarray}\vspace{-3mm}

\hspace{5cm}$S_3^{(2n)}$\hspace{6cm}$S_3^{(2n+1)}$

\hspace{5cm}$S_2^{(2m+1)}$\vspace{1mm}

\hspace{2.3cm}\includegraphics[scale =1.0]{scalar3.eps}\vspace{-2.5cm}

\begin{eqnarray}
 \hspace{-7.cm}S_1^{(2|m-n+1|)}\hspace{2.7cm}=CC_{oeo}
 \nonumber
\end{eqnarray}\vspace{-3mm}

\hspace{5cm}$S_3^{(2n)}$\vspace{-3.4cm}

\begin{eqnarray}\hspace{8cm}
\begin{array}{lll}
HHH       & : & C=\frac{6\lambda v_{\text{EW}}}{L^2}, \\
H\chi\chi & : & C=\frac{2\lambda v_{\text{EW}}}{L^2},  \\
H\varphi^+\varphi^- & : & C=\frac{2\lambda v_{\text{EW}}}{L^2},
\end{array}
\end{eqnarray}\vspace{5mm}

\noindent where for $HHH$,
\begin{eqnarray}
 C_{eee}&=&f_{2m}^e(z)f_{2n}^e(z)f_{2|m-n|}^e(z),\\
 C_{eoo}&=&f_{2m}^e(z)f_{2n+1}^o(z)f_{2|m-n-1|}^o(z),\\
 C_{oeo}&=&f_{2m+1}^o(z)f_{2n}^e(z)f_{2|m-n+1|}^o(z),
\end{eqnarray}
and for $H\chi\chi$, $C_{\ast\ast\ast}$ is given~(\ref{eee}),~(\ref{eoo}), and 
(\ref{oeo}), respectively.

Higgs and fermion coupling in the BLF scenario:

\hspace{5.5cm}$\bar{f}_i^{(m)}$

\hspace{2.3cm}\includegraphics[scale =1.0]{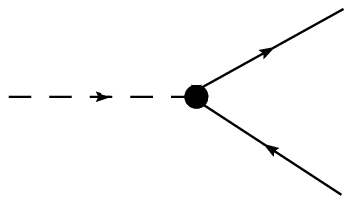}\vspace{-2.5cm}

\begin{eqnarray}
 \hspace{-4cm}H^{(l)}\hspace{2.5cm}
 =\delta_{ij}y_{f,5}\delta(\pm L/2)f_l(z),
\end{eqnarray}\vspace{-3mm}

\hspace{5.5cm}$f_j^{(n)}$

\noindent where $f_l(z)=f_l^e(z)$ for even $n$ and $f_l=f_l^o(z)$ for odd $n$.

Higgs and fermion coupling in the BF scenario:

\hspace{5cm}$\bar{f}_i^{(2m)}$

\hspace{1.4cm}\includegraphics[scale =1.0]{hff.eps}\vspace{-2.5cm}

\begin{eqnarray}
 \hspace{-4.5cm}H^{(2l+1)}\hspace{2.5cm}
 =\delta_{ij}y_{f,5}\frac{f_{2l+1}^o(z)f_{2m}^{\ast e}(z)f_{2n+1}^{\ast o}(z)}{L}
\end{eqnarray}\vspace{-5mm}

\hspace{5cm}$f_j^{(2n+1)}$

\hspace{5cm}$\bar{f}_i^{(2m+1)}$

\hspace{1.4cm}\includegraphics[scale =1.0]{hff.eps}\vspace{-2.5cm}

\begin{eqnarray}
 \hspace{-4.5cm}H^{(2l+1)}\hspace{2.5cm}
 =\delta_{ij}y_{f,5}\frac{f_{2l+1}^o(z)f_{2m+1}^{\ast o}(z)f_{2n}^{\ast e}(z)}{L}
\end{eqnarray}\vspace{-5mm}

\hspace{5cm}$f_j^{(2n)}$

\hspace{5cm}$\bar{f}_i^{(2m)}$

\hspace{1.4cm}\includegraphics[scale =1.0]{hff.eps}\vspace{-2.5cm}

\begin{eqnarray}
 \hspace{-5cm}H^{(2l)}\hspace{2.5cm}
 =\delta_{ij}y_{f,5}\frac{f_{2l}^e(z)f_{2m}^{\ast e}(z)f_{2n}^{\ast e}(z)}{L}
\end{eqnarray}\vspace{-5mm}

\hspace{5cm}$f_j^{(2n)}$

\hspace{5cm}$\bar{f}_i^{(2m+1)}$

\hspace{1.4cm}\includegraphics[scale =1.0]{hff.eps}\vspace{-2.5cm}

\begin{eqnarray}
 \hspace{-4.2cm}H^{(2l)}\hspace{2.5cm}
 =\delta_{ij}y_{f,5}\frac{f_{2l}^e(z)f_{2m+1}^{\ast o}(z)f_{2n+1}^{\ast o}(z)}{L}
\end{eqnarray}\vspace{-5mm}

\hspace{5cm}$f_j^{(2n+1)}$

%%%%%%%%%%%%%%%%%%%%%%%%%%%

\end{document}